\documentclass{./imammb}

\jno{dqnxxx}

\usepackage[round]{natbib}
\usepackage{caption}
\usepackage{subcaption}
\usepackage{tikz}
\usepackage[figuresleft]{rotating}

\usepackage{graphicx}
\usepackage{amsmath,amsthm,bm,mathrsfs,amssymb}
 \newtheoremstyle{theorem}{6pt}{6pt}{\rm}{}{\sffamily}{ }{ }{}
 \theoremstyle{theorem}

 \newtheoremstyle{algorithm}{6pt}{6pt}{\rm}{}{\sffamily}{ }{ }{}
 \theoremstyle{algorithm}

 \newtheoremstyle{lemma}{6pt}{6pt}{\rm}{}{\sffamily}{ }{ }{}
 \theoremstyle{lemma}

\newtheoremstyle{case}{6pt}{6pt}{\rm}{}{\sffamily}{. }{ }{}
 \theoremstyle{case}

 \newtheoremstyle{statement}{6pt}{6pt}{\rm}{}{\sffamily}{ }{ }{}
\theoremstyle{statement}

 \newtheoremstyle{corollary}{6pt}{6pt}{\rm}{}{\sffamily}{ }{ }{}
 \theoremstyle{corollary}

  \newtheoremstyle{definition}{6pt}{6pt}{\rm}{}{\sffamily}{ }{ }{}
 \theoremstyle{definition}

\newtheoremstyle{example}{6pt}{6pt}{\rm}{}{\sffamily}{ }{ }{}
\theoremstyle{example}

\newtheoremstyle{remark}{6pt}{6pt}{\rm}{}{\sffamily}{ }{ }{}
\theoremstyle{remark}

\newtheoremstyle{approximation}{6pt}{6pt}{\rm}{}{\sffamily}{ }{ }{}
\theoremstyle{approximation}

\newtheoremstyle{scheme}{6pt}{6pt}{\rm}{}{\sffamily}{ }{ }{}
\theoremstyle{scheme}

\newtheoremstyle{Algorithm}{6pt}{6pt}{\rm}{}{\sffamily}{ }{ }{}
\theoremstyle{Algorithm}

\newtheoremstyle{Assumption}{6pt}{6pt}{\rm}{}{\sffamily}{ }{ }{}
\theoremstyle{Assumption}

\newtheoremstyle{proposition}{6pt}{6pt}{\rm}{}{\sffamily}{ }{ }{}
\theoremstyle{proposition}

\newtheoremstyle{hypo}{6pt}{6pt}{\rm}{}{\sffamily}{ }{ }{}
 \theoremstyle{hypo}

  \newtheoremstyle{Step}{6pt}{6pt}{\rm}{}{}{ }{ }{}
 \theoremstyle{Step}

\newcommand\Real{\mbox{Re}} 
\newcommand\Pran{\mbox{\mbox{Pr}}} 

\numberwithin{equation}{section}

\def\t{\mathbf t}
\def\n{\mathbf n}
\def\u{\mathbf u}

\def\T{\mathbf T}
\newcommand{\eeas}{\end{eqnarray*}}
\newcommand{\bea}{\begin{eqnarray}}
\newcommand{\eea}{\end{eqnarray}}
\newcommand{\beq}{\begin{equation}}
\newcommand{\eeq}{\end{equation}}
\newcommand{\beqs}{\begin{equation*}}
\newcommand{\eeqs}{\end{equation*}}

\newcommand{\figWidth}{6.75}
\newcommand{\rowHeight}{3}  
\newcommand{\columnWidth}{7}  
\newcommand{\cbWidth}{6}

\newcommand{\trimfig}[2]{{\includegraphics[width=#2cm, clip, trim=0 760 0 760]{#1}}}
\newcommand{\getCB}[2]{{\includegraphics[height=#2cm, clip, trim=1800 237 162 231,angle=-90]{#1}}} 

\begin{document}

\title{Computed Flow and Fluorescence Over the Ocular Surface}
\author{ {\sc Longfei Li}\\[2pt]
Department of Mathematical Sciences, Rensselaer Polytechnic Institute, Troy NY, 12180, USA\\[6pt]
{\sc R. J. Braun$^{\ast}$} \\[2pt]
Department of Mathematical Sciences, University of Delaware, Newark, DE 19711, USA\\[2pt]
$^{\ast}${\rm Corresponding author: rjbraun@udel.edu}\\[6pt]
{\sc W. D. Henshaw}\\[2pt]
Department of Mathematical Sciences, Rensselaer Polytechnic Institute, Troy NY, 12180, USA\\[6pt]
{\rm AND}\\[6pt]
{\sc P. E. King-Smith}\\[2pt]
College of Optometry, The Ohio State University, Columbus, OH 43218, USA\\[6pt]
{\rm [Last modified: \today]}\vspace*{6pt}}
\pagestyle{headings}
\markboth{L. Li, R.J. Braun, T.A. Driscoll, W.D. Henshaw and P. E. King-Smith}{\rm Computed Flow and Fluorescence Over the Ocular Surface}
\maketitle

\begin{abstract}
{Fluorescein is  perhaps the most commonly used substance to visualize tear film thickness and dynamics; better understanding
of this process aids understanding of dry eye syndrome which afflicts millions of people.  We study a mathematical model for tear
film flow, evaporation, solutal transport and fluorescence over the exposed ocular surface during the interblink.   Transport of the fluorescein ion
by fluid flow in the tear film affects the intensity of fluorescence via changes in concentration and tear film thickness.  Evaporation causes increased osmolarity and potential irritation over the ocular surface; it also alters fluorescein concentration and thus fluorescence.  Using  thinning rates from in vivo measurements together with thin film equations for flow and transport of multiple solutes, we compute dynamic results for tear film quantities of interest.  We compare our computed intensity distributions with in vivo observations. A number of experimental features are recovered by the model.}
{tear film; thin film; tear osmolarity; fluorescence imaging; dry eye; tear breakup.}
\end{abstract}

\section{Introduction} \label{s:intro}
The purpose of the this paper is to construct and solve a model for the fluid flow and transport of solutes in the
tear film on an eye-shaped domain.  Using the computed solute distributions, we can calculate the intensity of
light emitted from instilled fluorescein dye, and compare the computed result with a typical experimental observation.
The results and comparison help understand tear film dynamics and imaging in vivo.

The tear film is critical for the eye's protection \citep{BronTiffRev04,Gipson10}
and proper optical function  \citep{MontesMico2010185,Tutt20004117}.
The tear film is a thin liquid film with multiple layers that establishes itself rapidly after a blink.
At the anterior interface with air is an oily lipid layer that primarily
retards evaporation \citep{BraunEtal15}, which helps to retain a smooth
well-functioning tear film \citep{Norn79}.  Posterior to the lipid layer is the
aqueous layer, which consists mostly of water \citep{Holly73}.
At the ocular surface, there is a region with transmembrane mucins protruding from the
epithelial cells of the cornea or conjunctiva.  This forest of glycosolated mucins, called
the glycocalyx, has been referred to as the mucus layer in past literature.  It is generally
agreed that the presence of the hydrophilic
glycocalyx on the ocular surface prevents the tear film from
dewetting \citep{Tiff90a,Tiff90b,Gipson04}.  The overall thickness of the
tear film is a few microns \citep{King-SmithFink04}, while the average thickness
of the lipid layer is on the order of tens to 100 nanometers \citep{Norn79,Yokoi96,GotoTseng03,KSOculSurfRev11,BraunEtal15}
and the thickness of the glycocalyx is a few tenths of a micron \citep{Gipson10}.
This overall structure is reformed on the order of a second after each blink in a
healthy tear film.

The aqueous part of tear fluid is primarily supplied from the lacrimal gland near the temporal canthus and the excess is drained through the puncta near the nasal canthus.  \citet{Doane81} proposed the mechanism of tear drainage {\it in vivo} whereby tear fluid is drained into the canaliculi through the puncta during the opening interblink phase.
Water lost from the tear film due to evaporation into air is an important process as well \citep{MishimaMaurice61,TomlinsonDoane09,KimballKing-Smith10}.  We believe that this is the primary
mechanism by which the osmolarity and other solute concentrations are increased in the tear film \citep{BraunEtal15}.
Some water is supplied from the ocular epithelia via osmosis
\citep{Braun12,CerretaniEtal14,BraunEtal15}.

The role of osmolarity on the ocular surface may be summarized as in \citet{BaudouinEtal13} and \citet{LiBraun15}.
The healthy tear film maintains homeostasis with the blood in the range 296-302 Osm/m$^3$ (or, equivalently,  mOsm/L or mOsM)
\citep{LempBron11,TomlinsonEtal06,VersuraEtal10}, while healthy blood is in the range 285-295 Osm/m$^3$  \citep{Tietz95}.  In dry eye syndrome (DES),
the lacrimal system may be unable to maintain this homeostasis and osmolarity values in the meniscus rise to 316-360 Osm/m$^3$ \citep{TomlinsonEtal06,GilbardFarris78,BSull10,DarttWillcox13}, and even higher values may be attained over the cornea.   Using in vivo experiment and sensory feedback, \citet{LiuBegEtal09}
estimated peak values of 800-900 Osm/m$^3$.  Similar or higher values were computed from mathematical
models of tear film break up in \citet{BraunEtal15} or
\citet{PengEtal14}, or for models of the whole ocular surface \citep{LiBraun15}.  The estimates from
these models are significant because the osmolarity in TBU, or in most locations over the ocular surface,
is higher than in the inferior meniscus, where the
only convenient method used in the clinic measures the osmolarity \citep{LempBron11}.

The ongoing supply and drainage of tear fluid affects the distribution and flow of the tear film.
A number of methods have been used to visualize and/or measure tear film thickness and flow, including
interferometry \citep{Doane89,King-SmithFink04,King-SmithFink09}, optical coherence tomography \citep{WangEtal03},
fluorescence imaging \citep{Begley13,King-SmithIOVS13b} and many others.

Fluorescein and other dyes have been used in a variety of ways, including:
estimation of tear drainage rates or turnover times \citep{Webber86}; assessment of the
condition of the ocular surface via staining of epithelial cells \citep[e.g.,]{BronEtal15};
to visualize overall tear film dynamics \citep{Benedetto84,Begley13,King-SmithIOVS13b,LiBraun14};
estimation of first breakup times of the tear film \citep{Norn69};
and the temporal progression of tear film breakup areas \citep{LiuBegEtal06}.
There are different ways that fluorescence may be used to visualize the tear film.
In the dilute limit, the fluorescein concentration is below the critical concentration, and the
intensity of the fluorescence from the tear film is approximately proportional to its thickness.
In the concentrated (or self-quenching) limit, the intensity drops as the tear film thins in response to
evaporation, and the thickness is roughly proportional to the square root of the intensity for a spatially
uniform (flat) tear film \citep{NicholsEtal12,BraunEtal14}.  In what follows, we will use FL to abbreviate
fluorescence imaging to FL imaging.  Fluorescein transport is difficult to measure, but we aim to
better understand that and FL imaging via the mathematical model we develop here.

A variety of mathematical models have incorporated various important effects of tear film
dynamics as recently reviewed by \citet{Braun12}.
The most common assumptions for these models are a Newtonian tear fluid and a flat cornea
\citep{BergerCorrsin74,BraunUsha12}.
Tear film models are often formulated on a one-dimensional (1D) domain oriented vertically through the center of the cornea with stationary ends corresponding to the eyelid margins.  We refer to models on this kind of domain as 1D models.
Surface tension, viscosity, gravity and evaporation are often incorporated into 1D models; wetting forces have been
included as well.

\citet{ZubkovBreward12} formulated and studied a mathematical model that
describes the spatial distribution of tear film
osmolarity that incorporates both fluid and solute (osmolarity) dynamics, evaporation,
blinking and vertical saccadic eyelid motion.
They found that both osmolarity was increased modestly
in the black line region of thinning near the meniscus \citep{MillerPolse02}
and that measurements of the solute concentrations within the inferior meniscus need
not reflect those elsewhere in the tear film.
This model gave smaller increases in osmolarity than spatially-uniform models
\citep{Braun12,BraunEtal15} because of the higher evaporation rates and lack of diffusion along
the tear film in those latter models.  Larger increases in osmolarity were reported in the eye-shaped domain
model studied by \citet{LiBraun15} because they used larger evaporation rates as may be seen
experimentally \citep{KSHinNic10}.

Compartment models of the tear film have also been developed.  \citet{GaffEtal10} developed a model
for the tear film with separate compartments for the central tear film, menisci and other locations with
an impermeable cornea (no osmosis).  The model allows water and osmolarity to pass between compartments so that
mass is conserved, but it does not directly take into account fluid flow.  From the model, they could estimate osmolarities obtained in the different compartments.  \citet{CerretaniEtal14} extended the model to include osmosis from the substrate.

Of interest in this paper are dynamics of the fluid flow and solute transport on an eye-shaped domain.
To our knowledge, \citet{MakiBraun10a}
were the first to extend models of fluid dynamics in the tear film to a geometry that approximates the
exposed ocular surface.
Besides specifying the tear film thickness at the boundary; they specified either the pressure  \citep{MakiBraun10a} or the flux \citep{MakiBraun10b}.
Their simulations recovered features seen in 1D models such as formation of the black line, and captured  some experimental observations of the tear film
dynamics around the lid margins.  \citet{MakiBraun10b} simplified the {\it in vivo} lacrimal supply and drainage
mechanisms and imposed a time-independent flux boundary condition; under some conditions, they recovered hydraulic connectivity as
seen experimentally.  \citet{LiBraun14} improved that model by adding evaporation and a wettable
ocular surface, as well as a time-dependent flux boundary condition that approximated the in- and out-flow of
the aqueous layer of the tear film.

\citet{LiBraun15} formulated a model that included osmolarity transport and osmosis from the tear/eye interface.  This model
provided, to our knowledge, the first global osmolarity distribution over the entire exposed eye.  The permeability of the ocular surface was treated as either constant over the whole surface, or as a space-dependent function with lower permeability over the cornea and higher permeability over the conjunctiva.   The location of the highest osmolarity varied with the permeability, and for the more realistic case with variability, the highest values were near the black lines over the cornea.  This model did not include local tear break up dynamics,
so they are an idealized case; however, the values found in that model are similar to recent calculations for tear breakup (TBU) models \citep{PengEtal14,BraunEtal15}.

A number of mathematical models have been developed for TBU.
Multi-layer models driven by dewetting van der Waals forces to rupture 
(corresponding to TBU in the eye literature) have appeared in linear \citep{SharmRuck86a},
nonlinear \citep{SharmRuck85,SharmRuck86b} and non-Newtonian forms \citep{ZMC03b,ZMC04}.
All of the theories could give reasonable TBU time (TBUT) ranges.
Related papers, which may apply to eyes via analogy with lung surfactants, include \citet{MatCraWar02}.
For a comprehensive review of related bilayer
work, see \citet{CrasMat09review}.

More recently, the DEWS report \citep{DEWSdef} has argued that TBU is driven, at least in many cases, by evaporation,
which causes increased osmolarity (concentration of ions) in the tear film, which may lead to irritation,
inflammation and damage to the ocular epithelium.  Currently, the authors are unaware of any method to directly
measure the osmolarity in regions of TBU due to the very small volumes of tear fluid involved, the rapid dynamics there,
and the extreme sensitivity of the cornea.
Theoretical efforts have responded to this situation by creating models that incorporate evaporation and osmosis, as well as other effects, into the mathematical models to better understand the dynamics of the process at a small scale.

In \citet{Braun12} and \citet{BraunEtal15}, a model for an evaporating, spatially uniform film was studied.
The model was a single ordinary differential
equation for the tear film thickness that included evaporation from the tear/air interface
at a constant rate and osmotic flow from the tear/cornea interface that was assumed to be a
semi-permeable boundary that allows water but not solutes to pass.
They found that the model predicted equilibration of the tear film thickness and that the osmolarity
could become quite large as the tear film thinned.
Similar conclusions were found from the model of \citet{Braun12}, which included van der Waals forces that
stopped thinning at the purported height of the glycocalix; this allowed the model to be used at zero
permeability at the tear/cornea interface.

These models were extended to include a specified space-dependent evaporation profile that varied in space by
\citet{PengEtal14} and \citet{BraunEtal15}.
In both models, the local thinning caused by locally increased evaporation led to increased osmolarity in the TBU
region, which could also be several times larger than the isotonic value.  \citet{BraunEtal15}
also included fluorescein transport in their model, and could compute intensity
distributions as well as fluorescein concentration.
They gave preliminary results that showed that FL imaging techniques for TBU may need to be interpreted
with care because fluorescein transport could alter the appearance of TBU from FL imaging relative to the actual thickness
distribution.  A detailed study that used an improved scaling has been been submitted \citep{BraunEtalTBU16}.
A model of TBU that incorporated an evaporation rate that depended on a insoluble surfactant concentration was studied
by \citet{SiddiqueBraun15}; they concluded that more physics was needed in that model to correctly capture TBU dynamics.

In this paper, we investigate a model for tear film flow, evaporation, osmolarity and
fluorescein transport, osmosis and fluorescence on an eye-shaped domain.  The emphasis in this paper is on understanding the transport of solutes inside the tear film, and how that transport affects FL imaging of the tear film.  Despite not specifically building in TBU
into the model, the model will show similar effects from thinning and flow particularly in the black lines over the cornea.
The paper is organized as follows.  We briefly describe a sample experiment using FL imaging.  We then develop the mathematical model for the fluid flow and solute transport on an eye-shaped domain.  We then move on to results, followed by discussion and conclusion.  The details of the mathematical formulation appear in the appendices.

\graphicspath{{./Figures/QuenchExpt/}}

\section{A Sample Experimental Result}

We begin by briefly describing the fluorescence imaging method used for visualizing tear film
dynamics \citep{NicholsEtal12,King-SmithIOVS13a,King-SmithIOVS13b} for the images shown here;
this description relies heavily on that used in \cite{LiBraun14}.  Video
recordings were made from normal and dry eye subjects for 60 s after instillation
of 1 $\mu$l of 5\% (sodium) fluorescein. Subjects were instructed to blink about 1 s after the start of
the recording and try to hold their eyes open for the remainder of the recording. The subjects’ eyes
were illuminated with blue light and a blocking interference filter was used to reduce the response to
reflected illumination light.  The horizontal illumination width was 15 mm, thus including the
cornea and part of the conjunctiva. The
research protocol was approved by an Institutional Review Board in accordance with the Declaration
of Helsinki. Informed consent was obtained from each subject at study enrollment.

Tear film thickness was studied by using the self-quenching of fluorescein, i.e., the reduction
of fluorescent efficiency with increasing fluorescein concentration \citep{NicholsEtal12}.
When diluted by about the assumed standard 7 $\mu$l of tears, fluorescein concentration
in the tears will be about 0.625\%. As the tear film thins from
evaporation, the fluorescent intensity is reduced inversely proportional to the square of concentration
and is therefore proportional to the square of tear thickness. Thus, tear thickness is proportional to
the square root of fluorescent intensity \citep{NicholsEtal12}. The method estimates relative changes in thickness given
an initial concentration, and does not directly visualize flow.
However, the lubrication model that we use computes the film thickness directly, so it is appropriate
to use this method to compare the theory with this type of experiment.  Comparisons for open eye shapes have
been made in our previous models \citep{MakiBraun10b,LiBraun14}.

In Figure 1, we show some images from the video recording of one subject (a different one than in our previous comparisons).
\begin{figure}[hp!]
\centering
        \begin{subfigure}[b]{0.47\textwidth}
                \centering
                \caption*{{\footnotesize$t=0.12$ s}}
                \includegraphics[width=\textwidth]{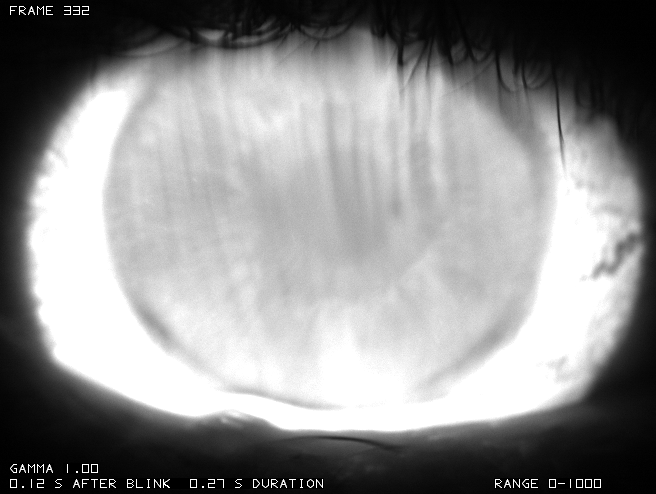}
        \end{subfigure}
        \begin{subfigure}[b]{0.47\textwidth}
                \centering
                \caption*{{\footnotesize$t=6.06$ s}}
                \includegraphics[width=\textwidth]{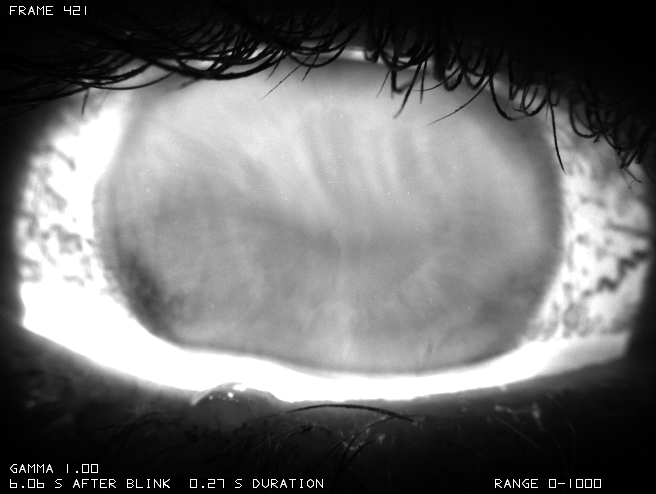}
        \end{subfigure}\\
        \begin{subfigure}[b]{0.47\textwidth}
                \centering
                \caption*{{\footnotesize$t=20.08$ s}}
                \includegraphics[width=\textwidth]{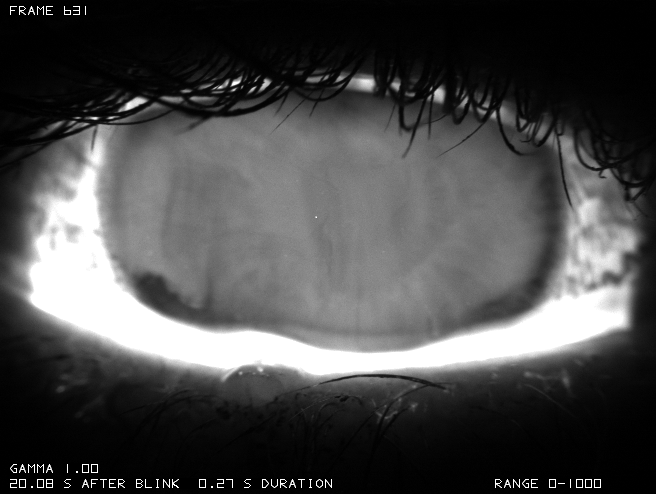}
        \end{subfigure}
        \begin{subfigure}[b]{0.47\textwidth}
                \centering
                \caption*{{\footnotesize$t=35.1$ s}}
                \includegraphics[width=\textwidth]{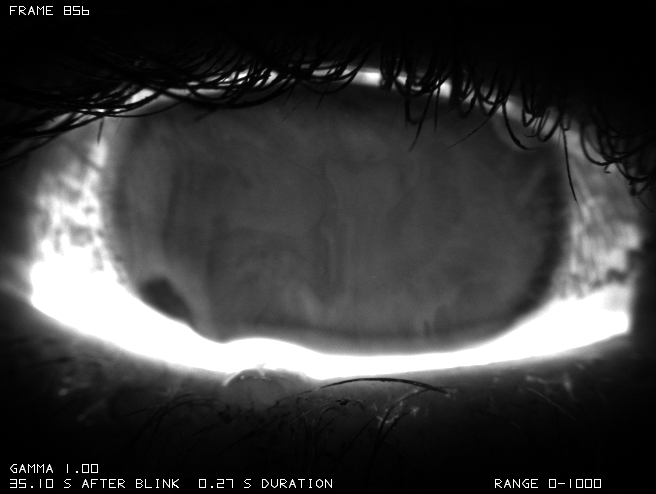}
        \end{subfigure}
\parbox{0.95\textwidth}{
\caption{An FL imaging experiment with the fluorescein concentration $f'$ in the quenching regime.  The times given are relative to a blink.}
}
\end{figure}
The images show four different times post-blink. The transparent cornea appears darker than the
surrounding reflective area corresponding to the conjunctiva; in these images, the contrast is enhanced to clearly show the
features of interest.  As time increases, the overall fluorescent intensity decreases away from the menisci.
This is due to evaporative loss of water with the corresponding increase in fluorescein concentration.
The self-quenching effect causes the intensity to decrease as the fluorescein concentration rises.  The menisci
remain bright throughout the experiment as they thin little and the fluorescein concentration changes little
there as a result.  Other sources for FL imaging results include, among others: \citet{LiuBegEtal06},
\citet{Begley13}, \citet{YokoiGeorgiev13}, and \citet{SuChang14}.
Localized effects such as TBU can occur as well; we don't deal with those dynamics in this paper, but FL imaging of TBU has begun to be
studied theoretically \citep{BraunEtal15,BraunEtalTBU16}.  In this work, we aim to capture the overall tear film image dynamics in the absence of localized TBU.

\graphicspath{{./Figures/}}
\section{Formulation}
\label{s:formulation}

\begin{figure}[htbp] 
   \centering
   \includegraphics[width=5in]{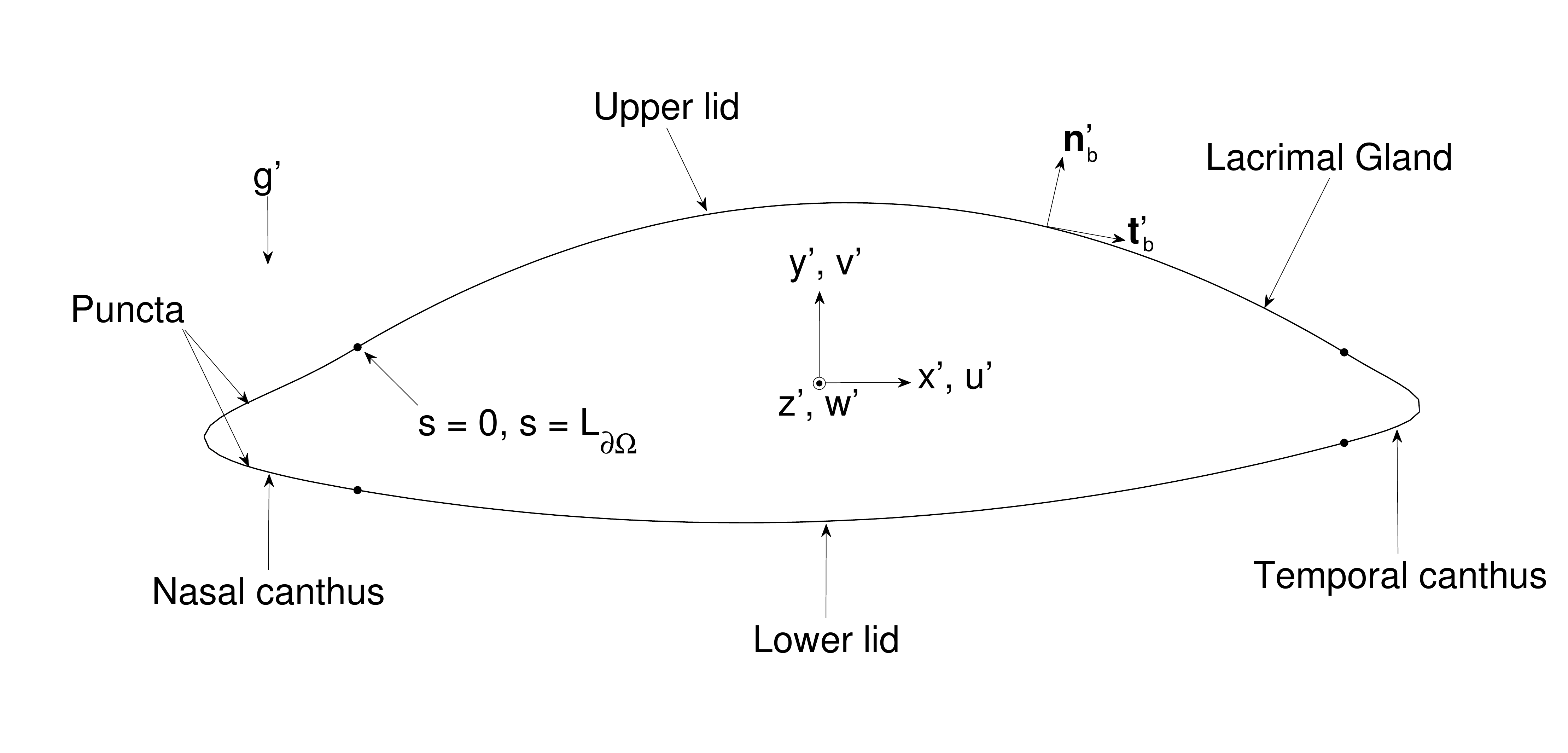}
   \caption{The coordinate system and eye-shaped domain. The $z'$ direction points out of the page.}
   \label{fig:coordEDIT}
\end{figure}

In this section, we present a mathematical model that incorporates fluid dynamics and transport of two solutes into a tear film model on a 2D eye shaped domain as shown in Figure \ref{fig:coordEDIT}.  In Figure \ref{fig:coordEDIT}, $(u',v',w')$ are the velocity components in the coordinate directions $(x',y',z')$; $z'$ is directed out of the page and primed variables are dimensional.  The gravitational acceleration $g'$ is specified in the negative $y'$ direction; we will neglect it in our computations in this paper.

\subsection{Eye-shaped domain}
The boundary curves of the eye-shaped domain are approximated from a digital photo by four polynomials. Two are parabolas in $x'$ and two are ninth-degree polynomials in $y'$, and $C_4$ continuity  is imposed where they join (indicated by dots) \citep{MakiBraun10a,MakiBraun10b,LiBraun14}.  $s'$ is the arc length of the boundary starting at the joint of the nasal canthus and upper lid, and is traversed in the counterclockwise direction as $s'$ increases.  The unit vectors tangential and normal to the boundary curves are given by $\mathbf{t'}_b$ and $\mathbf{n'}_b$, respectively. $z' = h'(x',y',t')$ denotes the free surface of tear film and $t'$ is the time.

\subsection{Governing Equations}
We assume that the tear fluid is incompressible and Newtonian with constant density $\rho$, viscosity $\mu$, specific heat $c_p$, and thermal conductivity $k$.  The tear fluid supplied from the lacrimal gland is more complex than water with salt ions, containing proteins and other organic molecules \citep{Dartt200521,Dartt2009155}.  Because these large molecules are present, whole tears have been found to be weakly shear thinning \citep{Tiffany91}.  Similar levels of shear thinning were incorporated by \citet{JossicLefevre09} and though some details change, the qualitative features of the flow and thickness distribution are, for our purposes here, similar to the Newtonian case.  More recently,  meibum (secreted from the glands in the eyelids) has been found to introduce elasticity into the tear film when temperature is below 32$^\circ$C  \citep{Leiske11}. Note that viscoelastic surfactants can affect film stability in a model system \citep{RosenfeldFuller12,BhamlaEtal14,BhamlaEtal16}.  However, it is not yet clear how these measurements affect overall tear film properties at typical in vivo operating temperatures that are often near 35$^\circ$C \citep{EfronYoung89}.  Because our current knowledge indicates that the shear thinning of the tear film is not strong \citep{Tiffany91} and because viscoelastic surface effects appear to be small at in vivo temperatures \citep{Leiske12}, we make the simplifying assumption of constant properties.

We also assume the ocular surface is flat due to the fact that the characteristic thickness of the human tear film is much less than the radius of curvature of the ocular globe \citep{BergerCorrsin74,BraunUsha12}.  The governing equations for the tear film thickness $h'(x',y',t')$, the osmolarity  $c'(x',y',t')$ and the fluorescein concentration $f'(x',y',t')$ are derived from the incompressible Navier-Stokes  and  convection-diffusion equations, respectively. The two equations are coupled by the osmotic flux, that is, the fluid that flows from the ocular surface at $z'=0$ into the tear film driven by the osmolarity difference between the tear film and the cornea \citep{LevVerk04}. In the model, water is lost to the air, treated as a passive gas outside the tear film ($z'>h'$),
and water is supplied due to the subsequent increased osmolarity via this osmotic flux.  It is assumed that fluorescein does not induce osmotic flux of water \citep[see Appendix, or][]{BraunEtal14}.

After nondimensionalization and simplification using lubrication theory (e.g. \cite{JenGrot93,Braun12}), we arrive at a system of PDEs for the dimensionless variables $h(x,y,t)$, $c(x,y,t)$ and $f(x,y,t)$:
\begin{align}
&\partial_t h +EJ +\nabla \cdot \mathbf{Q} - P_c(c-1) = 0,  \label{hPDE}\\
\nonumber\\
 &h\partial_t c +\nabla c\cdot \mathbf{Q}   = E c J + \frac{1}{\mathrm{Pe}_c}\nabla\cdot (h\nabla c)   -  P_c (c-1)c,  \label{cPDE} \\
  &h\partial_t f +\nabla f\cdot \mathbf{Q}   = E f J + \frac{1}{\mathrm{Pe}_f}\nabla\cdot (h\nabla f)   -  P_c (c-1)f.  \label{fPDE}
\end{align}
We define the pressure as
\begin{equation} \label{pPDE}
p = - S \Delta h + A h^{-3};
\end{equation}
the evaporative mass flux $J$ is given by
\begin{equation}
J =\frac{1 + \delta p}{\bar{K} + h},
\end{equation}
and the fluid flux $\mathbf{Q}$ across any cross-section of the film is  given by
\begin{equation} \label{flux_eqn}
 \mathbf{Q} = \frac{h^3}{12} \nabla \left( S\Delta h + Ah^{-3} \right).
\end{equation}
The conjoining pressure in modeling evaporation plays an important role in the tear film model.  It is meant to mimic the effect of the glycocalix, whose transmembrane mucins are strongly wet by water and we assume that they arrest the thinning of the tear film.  A secondary benefit is that the model allows solutions to be computed past the initial tear film breakup because the tear film thickness never reaches zero in this model.  These aspects of evaporation competing with conjoining pressure are discussed by \citet{WinterAnderson10} in the context of eyes, but the idea was developed by \citet{PotashWayner72} and \cite{MoosmanHomsy80}. More recent versions of the approach may be found in \citet{Morris01} and \citet{AjaevHomsy01}. The nondimensional parameters that arise are defined and given values in the following section and in Table~\ref{tab:DimensionlessParameters}.  The dimensional parameters used in those expression are given in Table~\ref{tab:DimensionalParameters}.  A detailed derivation of the governing equations (\ref{hPDE}) -- (\ref{fPDE}) can be found in  Appendix \ref{sect:FlowModelDerivation}.

Once the fluorescein concentration $f$ is computed, the FL intensity $I$ can be computed from
 the thickness $h$ and the fluorescein concentration $f$
via the nondimensional version of (\ref{e:dimensionalI}) \citep{Webber86,NicholsEtal12}
\begin{equation} \label{eqn:I-of-f}
I = I_0 \frac{ 1 - e^{- \phi f h} }{1 + f^2}.
\end{equation}
Here $\phi=\kappa d f_{cr}$ is the nondimensional Naperian extinction coefficient.
For fixed thickness $h$, $I$ decreases linearly for small concentrations; this is the dilute regime.
Sufficiently far above a critical concentration $f_{cr}=0.2\%$, the intensity
asymptotes to a quadratically decreasing function with increasing $f$; this is the self-quenching regime \citep{NicholsEtal12}.
For a tear film thinning by evaporation, the situation is slightly more complicated.
For a flat tear film, the product $hf$ is constant, and so in the dilute regime, $I$ is constant during thinning,
while in the self-quenching regime, the quadratic decrease for increasing $f$ still holds \citep{BraunEtal14}.
For a deforming tear film, mass conservation is more complicated, and it is a main purpose of this paper to compute
how images of the tear film appear subject to our model for flow and solute transport.

\subsection{Parameter Descriptions}
\label{s:parameters}

The nondimemsional parameters we use are shown in Table~\ref{tab:DimensionlessParameters}; the dimensional parameters are in Table~\ref{tab:DimensionalParameters}.
\begin{table}[htb]
  \begin{center}
\def~{\hphantom{0}}
  \begin{tabular}{cccccc}
\hline\hline
Parameter & Expression & Value & Parameter & Expression & Value\\
\hline
\vspace{.1in}
$\epsilon$ &$\displaystyle\frac{d'}{L'} $ & $1\times10^{-3}$ &
$S$   & $\displaystyle\frac{\sigma \epsilon^3}{\mu U_0}$            & 6.92$\times10^{-6}$             \\ \vspace{.1in}
$E$        & $\displaystyle\frac{k(T'_B-T'_s)}{d'\mathcal{L}_m \epsilon\rho U_0}$ & 118.3  &
$\bar{K}$  & $\displaystyle\frac{kK}{d'\mathcal{L}_m}$                            & 8.9$\times10^{3}$ \\ \vspace{.1in}
$\delta$   & $\displaystyle\frac{\alpha\mu U_0}{\epsilon^2L'(T'_B-T'_s)}$ & 4.66 &
$A$        & $\displaystyle\frac{A^*}{L'd\mu U_0}$                      & 2.14$\times10^{-6}$ \\ \vspace{.1in}
$\mathrm{Pe}_c $& $\displaystyle  \frac{U_0L'}{D_c}$   & $1.56\times10^4$ &
$\mathrm{Pe}_f $& $\displaystyle  \frac{U_0L'}{D_f}$   & $6.41\times10^4$ \\ \vspace{.1in}
$P_\text{corn}$        & $\displaystyle\frac{P_\text{corn}^{\text{tiss}}V_w c_0}{\epsilon U_0}$                      &0.013 &
$P_\text{conj}$        & $\displaystyle\frac{P_\text{conj}^{\text{tiss}}V_w c_0}{\epsilon U_0}$                      &0.06 \\
$\phi$                 & $\kappa d f_{cr}$                                                                           &0.466 & & & \\
\hline\\
  \end{tabular}
\parbox{5in}{ \caption{Dimensionless Parameters. Values and descriptions of the dimensional parameters that appear are given in Table \ref{tab:DimensionalParameters}. Note that the value of $\bar{K}$ corresponds to a thinning rate of $4 \mu$m/min.}\label{tab:DimensionlessParameters}
}
    \end{center}
\end{table}

\begin{table}[htb]
  \begin{center}
\def~{\hphantom{0}}
  \begin{tabular}{clll}
\hline\hline
Parameter & Description & Value & Reference\\
\hline
$\mu$    & Viscosity                              & 1.3$\times10^{-3}$Pa$\cdot$s & \cite{Tiffany91} \\
$\sigma$ & Surface tension                        & 0.045N$\cdot$m$^{-1}$&\cite{NagyovaTiffany99} \\
$k$      & Tear film thermal conductivity         & 0.68W$\cdot$m$^{-1}$$\cdot$K$^{-1}$& Water \\
$\rho$   & Density                                & $10^3$kg$\cdot$m$^{-3}$ & Water  \\
$\mathcal{L}_m$    & Latent heat of vaporization  & 2.3$\times10^6$J$\cdot$kg$^{-1}$&Water  \\
$T'_s$     & Saturation temperature               & 27$^\circ$C& Estimated  \\
$T'_B$     & Body temperature                     & 37$^\circ$C & Estimated\\
$g$        &  Gravitational acceleration          & 9.81m$\cdot$s$^{-2}$&  Estimated\\
$A^*$      & Hamaker constant                     & 3.5$\times10^{-19}$Pa$\cdot$m$^3$ & \cite{WinterAnderson10}\\
$\alpha$   & Pressure coefficient for evaporation & 3.6$\times10^{-2}$K$\cdot$Pa$^{-1}$& \cite{WinterAnderson10} \\
$K$        & Non-equilibrium coefficient          & 1.5$\times10^{5}$K$\cdot$m$^2$$\cdot$s$\cdot$kg$^{-1}$& Estimated\\
$d'$       & Characteristic thickness             & $5\times10^{-6}$m  & \cite{King-SmithFink04} \\
$L'$       & Half-width of palpebral fissure     & $5\times 10^{-3}$m  & Estimated  \\
$U_0$      & Characteristic speed                 & $5\times 10^{-3}$m/s  & \cite{King-SmithFink09}\\
$P_\text{corn}^{\text{tiss}}$  & Tissue permeability of cornea &  12.0$\mu$m/s & \cite{BraunEtal15}\\
$P_\text{conj}^{\text{tiss}}$  & Tissue permeability of conjunctiva &  55.4$\mu$m/s & \cite{BraunEtal15}\\
$V_w$      & Molar volume of water &  $1.8\times 10^{-5}$m$^3\cdot$mol$^{-1}$ &  Water\\
$D_c$      & Diffusivity of  osmolarity in water  &  $1.6\times 10^{-9}$m$^2$/s   & \citet{Riquelme07}  \\
$D_f$      & Diffusivity of  fluorescein in water &  $0.39\times 10^{-9}$m$^2$/s   & \citet{Casalini11} \\
$\kappa$   & Naperian extinction coefficient      & $1.75 \times 10^{7} $ m$^{-1}$M$^{-1}$ & \cite{Mota91} \\
\hline
  \end{tabular}
  \caption{Dimensional Parameters.  Here $K$ corresponds to a nominal thinning rate of $4 \mu$m/min; to obtain a $20\mu$m/min thinning rate,
  this quantity is reduced by a factor of 5.}
  \label{tab:DimensionalParameters}
  \end{center}
\end{table}

We note that when the ocular surface is permeable, we have a region corresponding to the cornea that is about 4.6 times less permeable than the
surrounding conjunctiva, with a smooth monotonic transition between them.  Details of the permeability function may be found in \citet{LiBraun15}.

\subsection{Initial Conditions}
The initial conditions are the same as those of \citet{LiBraun15} subject to the addition of fluorescein concentration.  For convenience we
give all the initial conditions here.
The initial thickness $h(x,y,0)$ is a numerically smoothed version of the function
\begin{equation} \label{eq:h-IC}
h(x,y,0)= 1+(h_0-1)e^{-\min\left(\mathrm{dist}\left(\left(x,y\right),\partial\Omega\right)\right)/x_0},
\end{equation}
where $x_0=0.06$  and $\mathrm{dist}\left(\mathbf{X},\partial\Omega\right)$ is the distance between a point with position vector $\mathbf{X}$ and a point on the boundary $\partial\Omega$ \citep{MakiBraun10a,MakiBraun10b}. It  specifies a dimensional initial volume of about 1.805$\mu$l. This value is within the experimental measurements of $2.23\pm2.5\mu$l by \citet{MathersDaley96}. The initial pressure $p(x,y,0)$ is calculated from equation (\ref{DAE2}) \citep{LiBraun14}. For the initial osmolarity, we assume the salt ions are well mixed and isotonic  (302 Osm/m$^3$, or 1 dimensionlessly), so that
\begin{equation}\label{eq:OsmoIC}
c(x,y,0)=1.
\end{equation}
In addition, we specify the initial fluorescence concentration to be a constant, i.e.
\begin{equation}\label{eq:FluorIC}
f(x,y,0)=f_0.
\end{equation}

\subsection{Boundary Conditions}
The boundary conditions follow those of \citet{LiBraun15} with the addition of no flux for fluorescein at the boundary.
Along the boundary (denoted by $\partial\Omega$), we prescribe constant tear film thickness:
\begin{equation} \label{eq:Dirichlet-h}
h\vert_{\partial{\Omega}} = h_0.
\end{equation}
We set $h_0 = 13$, which is in the range of experimental measurement from \citet{GoldingBruce97}.  In addition,
we specify the fluid flux normal to the boundary via
\begin{equation}
\mathbf{Q}\cdot  \mathbf{n}_b = Q_{lg}(s,t)+Q_{p}(s,t) \label{FluxBC}
\end{equation}
The details of this time dependent boundary condition are given in Appendix~\ref{sect:Time-dependentBC} and \citet{LiBraun15}.
For the osmolarity $c(x,y,t)$, we employ a Dirichlet boundary condition
\begin{equation}\label{DirichletBC}
c\vert_{\partial{\Omega}} =1.
\end{equation}
We use a  homogeneous Neumann boundary condition for $f(x,y,t)$
\begin{equation}\label{NeumannBC}
\nabla f\cdot \mathbf{n}_b\vert_{\partial{\Omega}} =0.
\end{equation}

\section{Numerical Methods}

To solve the problem numerically, we introduce the pressure $p(x,y,t)$ as a dependent variable and substitute for $J$ to arrive
at (\ref{DAE1})--(\ref{DAE4}).
The corresponding boundary conditions must be applied, (\ref{eq:Dirichlet-h})--(\ref{NeumannBC}).  Note that the flux condition (\ref{FluxBC}) is readily converted into a Neumann condition for $p$.  The initial conditions must be applied as well, using smoothed versions of (\ref{eq:h-IC}) and (\ref{pPDE}), as well as (\ref{eq:OsmoIC}) and (\ref{eq:FluorIC}).
We use the Overture computational framework (http://www.overtureframework.org. Contact: W.\ D.\ Henshaw, henshw@rpi.edu), which is a collection of C++ libraries for solving PDEs on complex domains \citep{ChesshireHenshaw90,Henshaw02}. The spatial discretization is carried out on five overlapping component grids, and the solution is found on the composite grid; further details are given elsewhere \citep{LiBraun15}.

To solve the equations (\ref{hPDE})--(\ref{pPDE}), we first discretize the spatial derivatives using the second-order accurate finite difference method for curvilinear and Cartesian grids from Overture.
Since  the model equations  (\ref{hPDE})--(\ref{pPDE}) are weakly coupled via the solutes, the hybrid time stepping scheme implemented by
\citet{LiBraun15} works well on this problem.  To solve the coupled system, equations (\ref{cPDE}) 
and (\ref{fPDE}) are solved using an RKC method \citep{SommeijerShampine97}. Then (\ref{hPDE}) \& (\ref{pPDE}) are updated with $c$ and $f$ values and solved using a variable step size BDF method with fixed leading coefficient based on \cite{BrenanCampbell96} and \cite{MakiBraun10a,MakiBraun10b}.  The resulting nonlinear system of the BDF method is solved using Newton's iteration method.  Further details are given in \citet{LiBraun15}.

\section{Results}
\label{s:results}

We begin with an impermeable ocular surface and a nominal thinning rate of $4\mu$m/min, which is at the upper end of normal rates \citep{NicholsMitchell05}.  We then proceed to a permeable ocular surface with either $4\mu$m/min or $20\mu$m/min nominal thinning rates, with different values for corneal and conjunctival permeability.

\subsection{Impermeable ocular surface $(P_c=0)$}

In this subsection, we study the dynamics where the ocular surface is impermeable to water and solutes,
$P_c=0$, and water evaporates via a uniform nominal (flat-film) thinning rate of $4 \mu$m/min.
This case will result in the highest solute concentrations that we could see in this model because
no water can be supplied by osmosis.  Figure~\ref{fig:hcf_Pc=0_4mpm} shows the dependent variables
$h$, $c$ and $f$ at times $t=5,10,15$.  We begin by discussing the change in the
thickness $h$, corresponding to the first column.
The initial thickness \citep[not shown; see][]{LiBraun14,LiBraun15} has a 
roughly flat interior with smoothly and rapidly increasing menisci
at the edges of the domains (corresponding to the lid margins).
The tear film menisci have a nondimensional thickness of 3 or more (15 $\mu$m or more dimensionally) in
the pink band adjacent to the lid margins.
Capillary-driven flow toward the menisci forms the so-called black line, as has been seen in many mathematical models \citep[e.g.][]{MillerPolse02,BraunFitt03}.  In this case, the thinning from evaporation cooperates with capillarity to thin this region a little faster than it may otherwise occur.  The interior of the tear film thins roughly at the specified thinning rate, while the thickness of the menisci may increase in places due to flow into that region inside the tear film.

Figure~\ref{fig:hcf_Pc=0_4mpm} shows the change in the osmolarity $c$ and fluorescein concentration $f$ in the second and third columns, respectively.  In both cases, the solute concentrations increase in the interior and black line regions, and change relatively little in the mensci.  A large relative change may be expected from mass conservation if the relative change in the thickness is large from evaporation and the relative change is larger away from the menisci.  For the osmolarity, the highest values are found near the black line regions, with a maximum value of $c_{\rm max}=1.83$ for $t=15$, similar to results found previously for osmolarity alone \citep{LiBraun15}.  Thus, for the impermeable case, the maximum osmolarity is increased more than 80\% from the isotonic value, to about 550 mOsM.  For $f$, the distribution at each time is similar to $c$, but the maximum value obtained is larger, with $f_{\rm max}=2.09$ being more than double the initial uniform concentration value.  The maximum in $f$ is larger because its diffusivity is about four times smaller than for the osmolarity.  The larger diffusivity for osmolarity allowed more diffusion out of the high concentration regions into the surrounding tear film, thus lowering the peak value.

{
\def\nRows{3}  
\def\nColumns{3}  
\begin{sidewaysfigure}[htb]
\begin{center}
\rotatebox{0}{
\begin{tikzpicture}[scale=1]
\pgfmathparse{\columnWidth*\nColumns} \let\xb\pgfmathresult
\pgfmathparse{\rowHeight*\nRows+2} \let\yb\pgfmathresult 
\useasboundingbox (0.0,0.0) rectangle (\xb,\yb);  
\def\row{2}
\def\column{0}
\def\xs{1.1} 
\def\ys{0.3}
\pgfmathparse{\xs+0.*\figWidth} \let\xcb\pgfmathresult 
\pgfmathparse{\xcb+0.*\cbWidth} \let\xx\pgfmathresult
\draw(\xx,\ys) node[anchor=south west] {\getCB{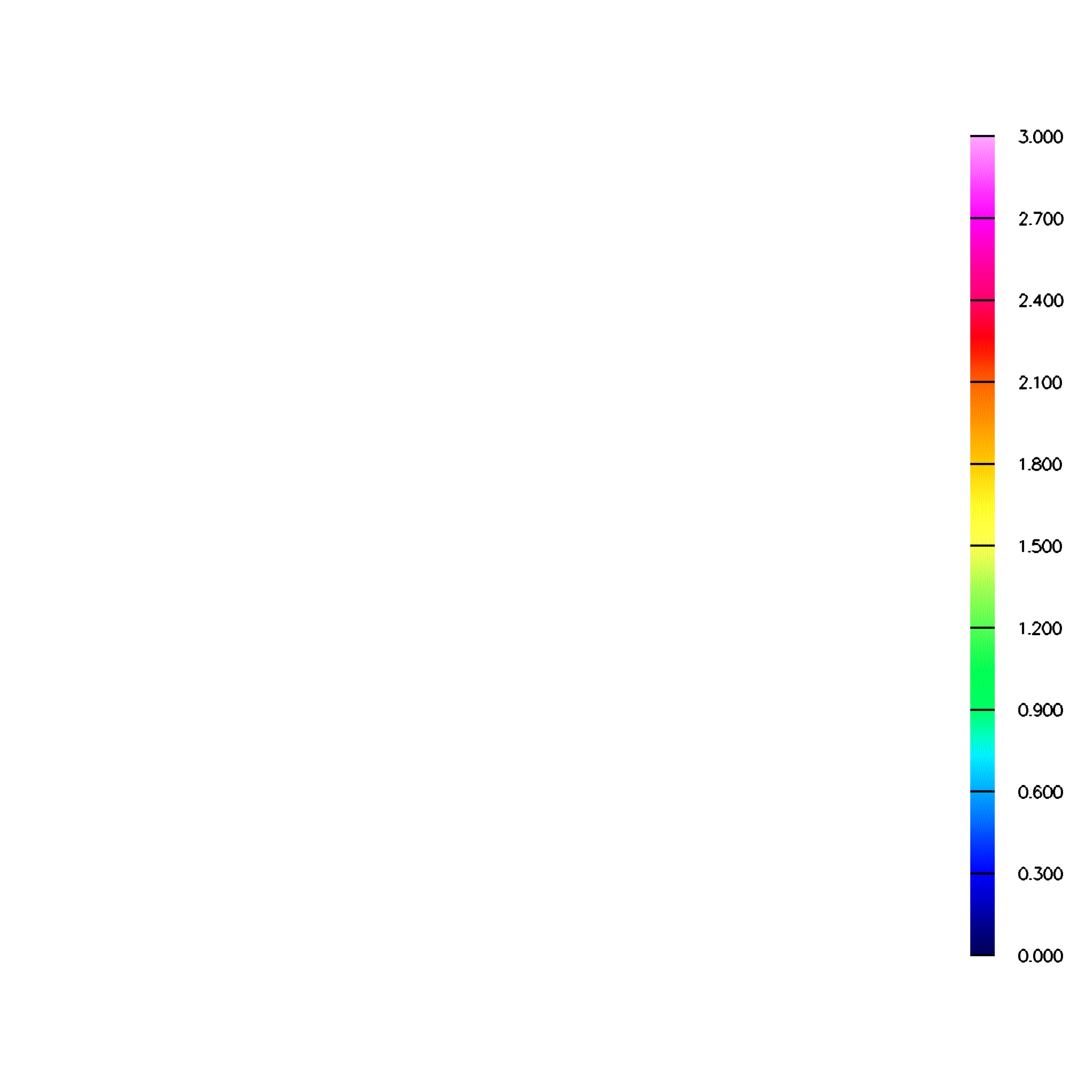}{\cbWidth}};
\draw(\xx,\ys) node[anchor= west]{$0$};
\pgfmathparse{\xcb+0.47*\cbWidth} \let\xx\pgfmathresult
\draw(\xx,\ys) node[anchor= west]{$1.5$};
\pgfmathparse{\xcb+0.97*\cbWidth} \let\xx\pgfmathresult
\draw(\xx,\ys) node[anchor= west]{$3$};
\pgfmathparse{\xs+1.*\figWidth} \let\xcb\pgfmathresult 
\pgfmathparse{\xcb+0.*\cbWidth} \let\xx\pgfmathresult
\draw(\xx,\ys) node[anchor=south west] {\getCB{colorbar_rainbow.pdf}{\cbWidth}};
\draw(\xx,\ys) node[anchor= west]{$0$};
\pgfmathparse{\xcb+0.47*\cbWidth} \let\xx\pgfmathresult
\draw(\xx,\ys) node[anchor= west]{$1$};
\pgfmathparse{\xcb+0.97*\cbWidth} \let\xx\pgfmathresult
\draw(\xx,\ys) node[anchor= west]{$2$};
\pgfmathparse{\xs+2.*\figWidth} \let\xcb\pgfmathresult 
\pgfmathparse{\xcb+0.*\cbWidth} \let\xx\pgfmathresult
\draw(\xx,\ys) node[anchor=south west] {\getCB{colorbar_rainbow.pdf}{\cbWidth}};
\draw(\xx,\ys) node[anchor= west]{$0$};
\pgfmathparse{\xcb+0.47*\cbWidth} \let\xx\pgfmathresult
\draw(\xx,\ys) node[anchor= west]{$1.1$};
\pgfmathparse{\xcb+0.97*\cbWidth} \let\xx\pgfmathresult
\draw(\xx,\ys) node[anchor= west]{$2.2$};
\def\xs{0.8}
\def\ys{0.}
\pgfmathparse{\xs+0.5*\figWidth} \let\xx\pgfmathresult
\pgfmathparse{\yb-1} \let\yy\pgfmathresult
\draw(\xx,\yy) node[anchor=south]{$h$};
\pgfmathparse{\xs+1.5*\figWidth} \let\xx\pgfmathresult
\draw(\xx,\yy) node[anchor=south]{$c$};
\pgfmathparse{\xs+2.5*\figWidth} \let\xx\pgfmathresult
\draw(\xx,\yy) node[anchor=south]{$f$};
\pgfmathparse{\row+1} \let\tP\pgfmathresult
\draw(0,\tP) node[anchor=west,xshift=-0.1cm] {$t=15$};
\pgfmathparse{0*\figWidth} \let\xf\pgfmathresult
\draw(\xf,\row) node[anchor=south west,xshift=\xs cm,yshift=\ys cm] {\trimfig{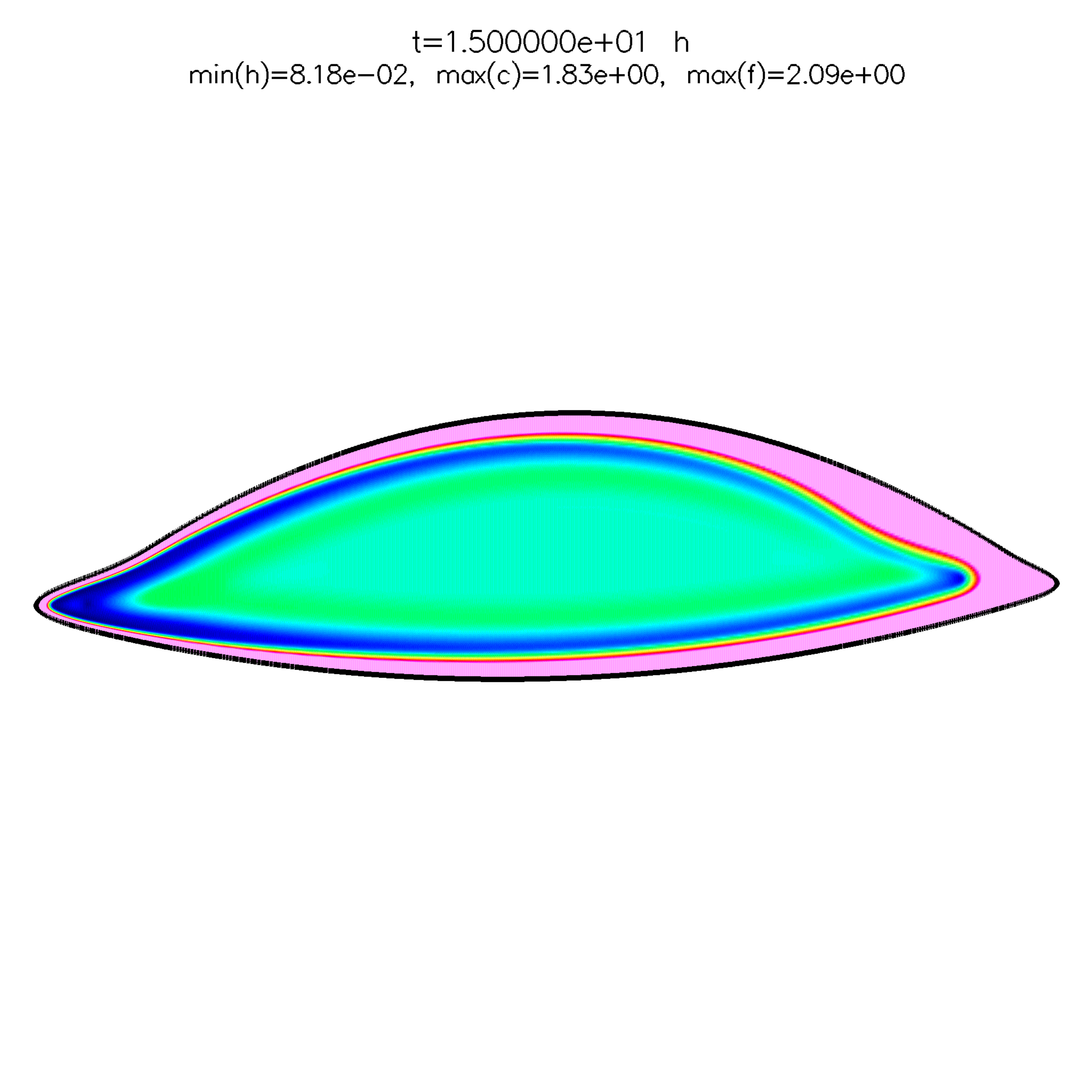}{\figWidth}};
\pgfmathparse{\xs+0.5*\figWidth} \let\xx\pgfmathresult
\draw(\xx,\row) node[anchor=north]{$\min(h)=$ 8.18e-2};
\pgfmathparse{1*\figWidth} \let\xf\pgfmathresult
\draw(\xf,\row) node[anchor=south west,xshift=\xs cm,yshift=\ys cm] {\trimfig{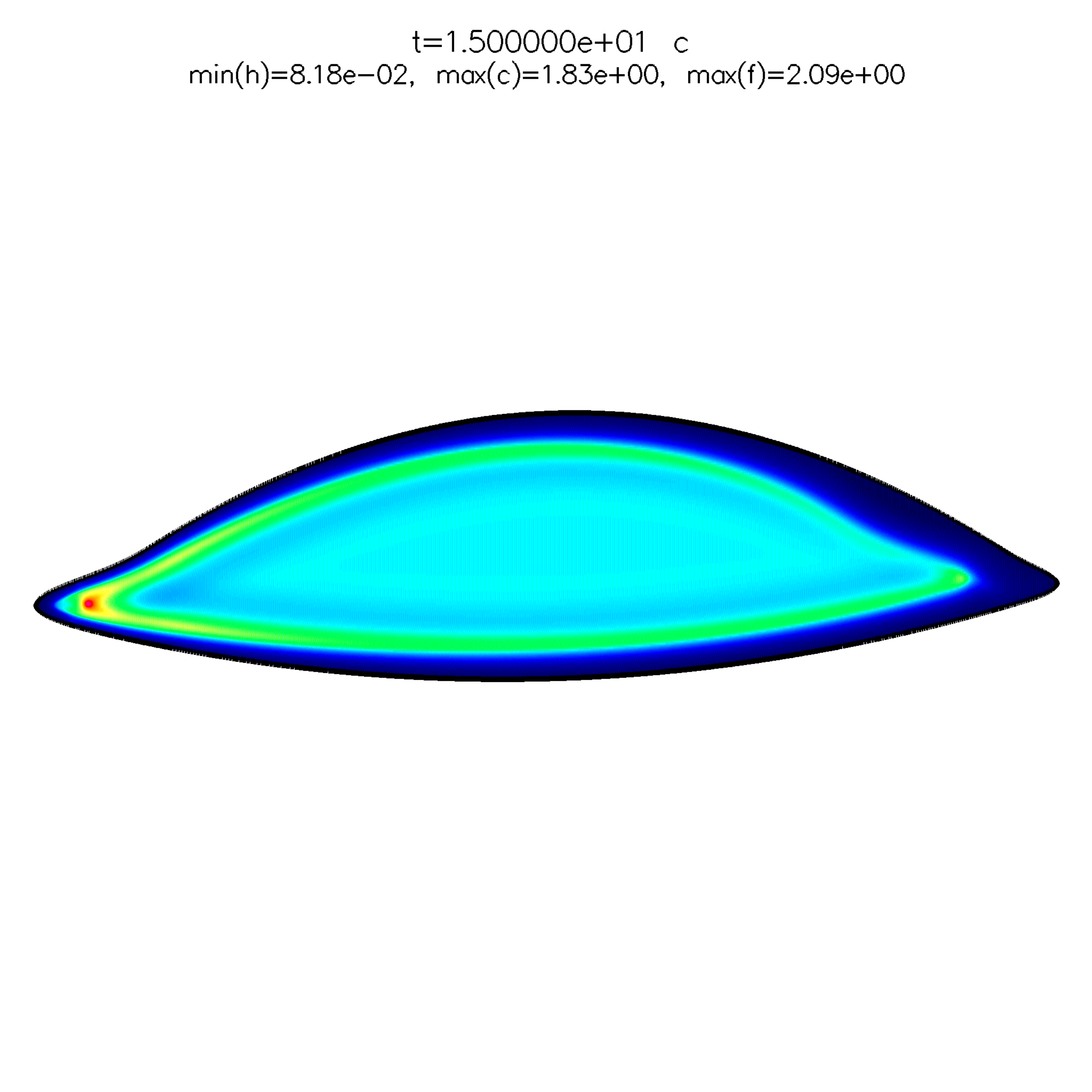}{\figWidth}};
\pgfmathparse{\xs+1.5*\figWidth} \let\xx\pgfmathresult
\draw(\xx,\row) node[anchor=north]{$\max(c)=$ 1.83};
\pgfmathparse{2*\figWidth} \let\xf\pgfmathresult
\draw(\xf,\row) node[anchor=south west,xshift=\xs cm,yshift=\ys cm] {\trimfig{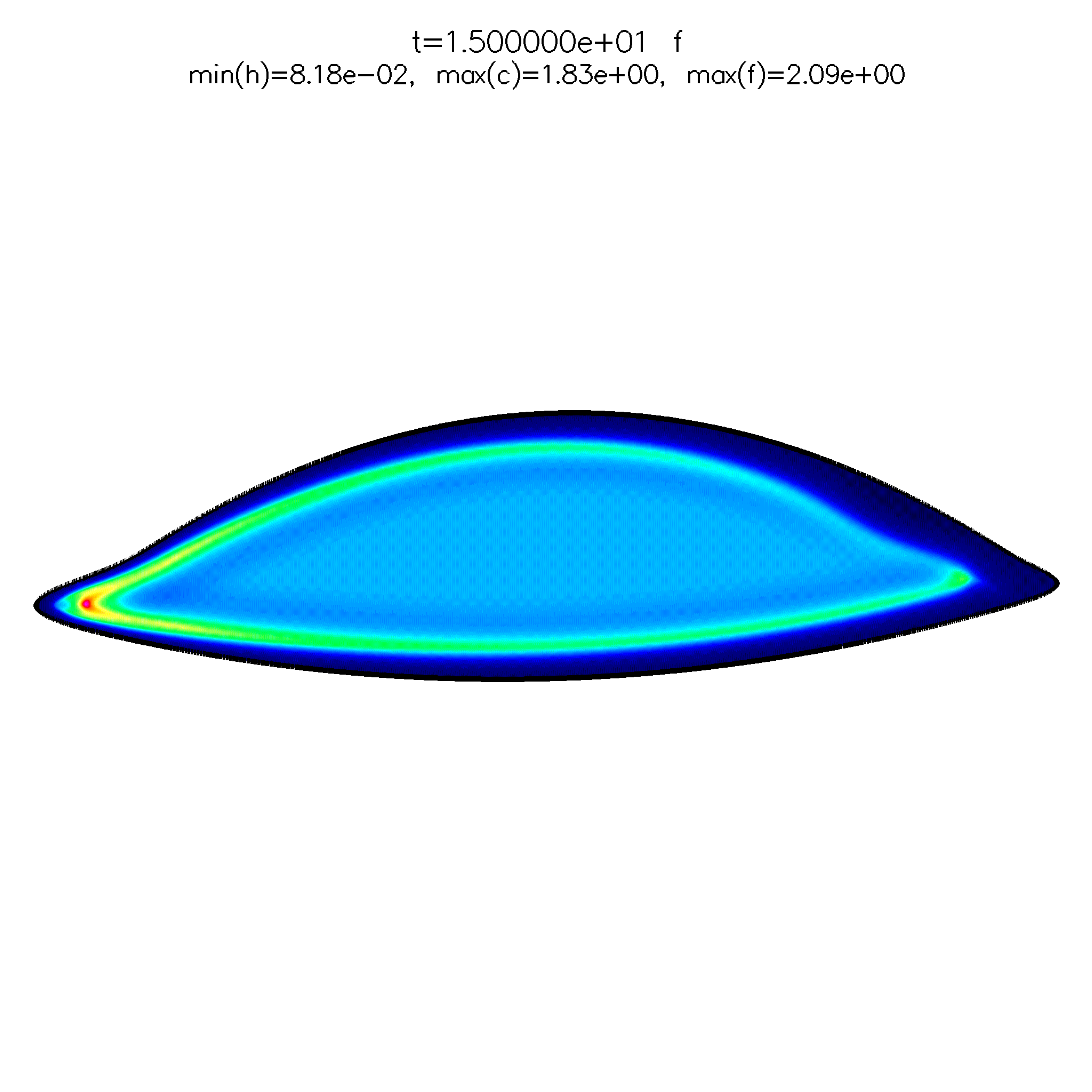}{\figWidth}};
\pgfmathparse{\xs+2.5*\figWidth} \let\xx\pgfmathresult
\draw(\xx,\row) node[anchor=north]{$\max(f)=$ 2.09};
\pgfmathparse{\row+\rowHeight} \let\row\pgfmathresult
\pgfmathparse{\row+1} \let\tP\pgfmathresult
\draw(0,\tP) node[anchor=west,xshift=-0.1cm] {$t=10$};
\pgfmathparse{0*\figWidth} \let\xf\pgfmathresult
\draw(\xf,\row) node[anchor=south west,xshift=\xs cm,yshift=\ys cm] {\trimfig{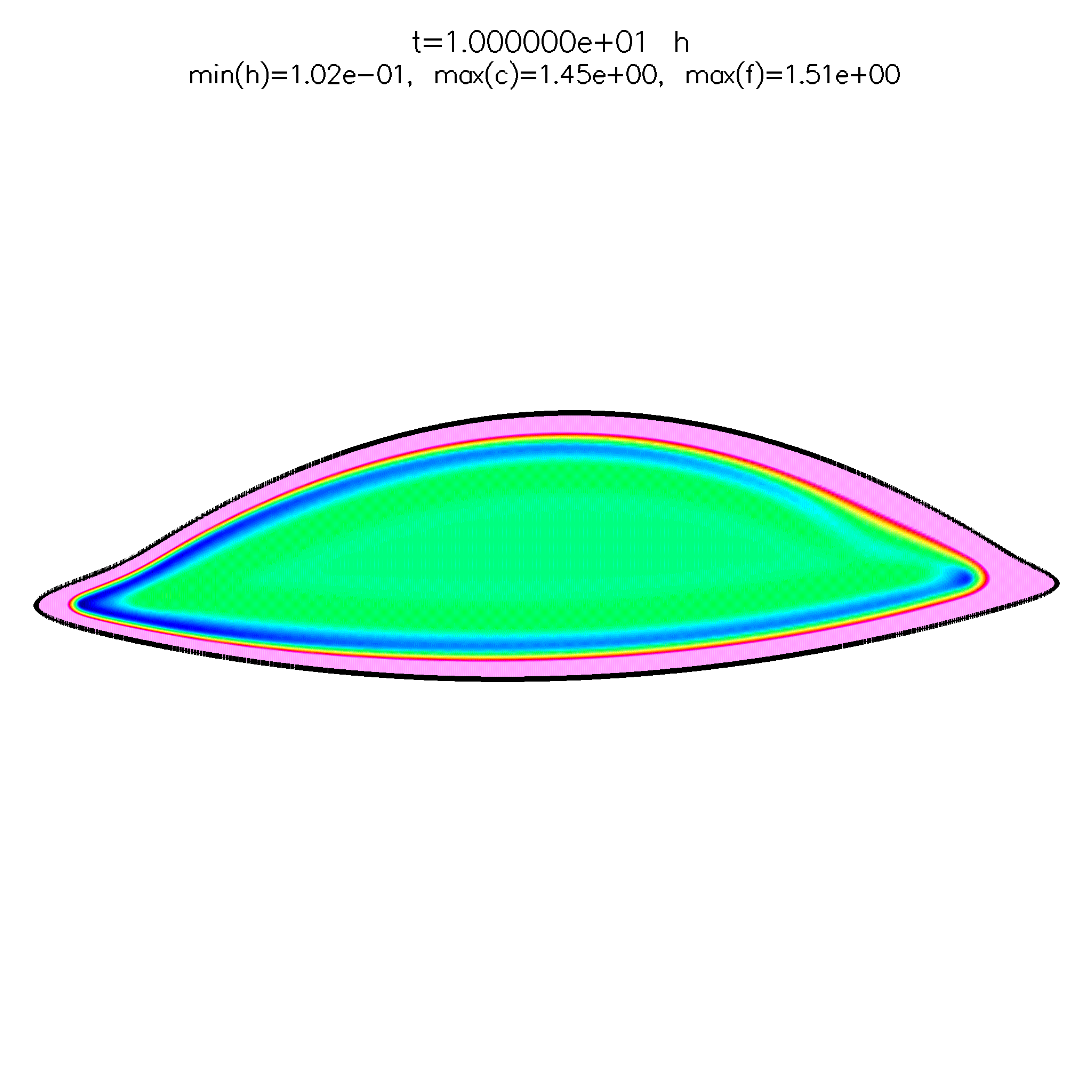}{\figWidth}};
\pgfmathparse{\xs+0.5*\figWidth} \let\xx\pgfmathresult
\draw(\xx,\row) node[anchor=north]{$\min(h)=$ 1.02e-1};
\pgfmathparse{1*\figWidth} \let\xf\pgfmathresult
\draw(\xf,\row) node[anchor=south west,xshift=\xs cm,yshift=\ys cm] {\trimfig{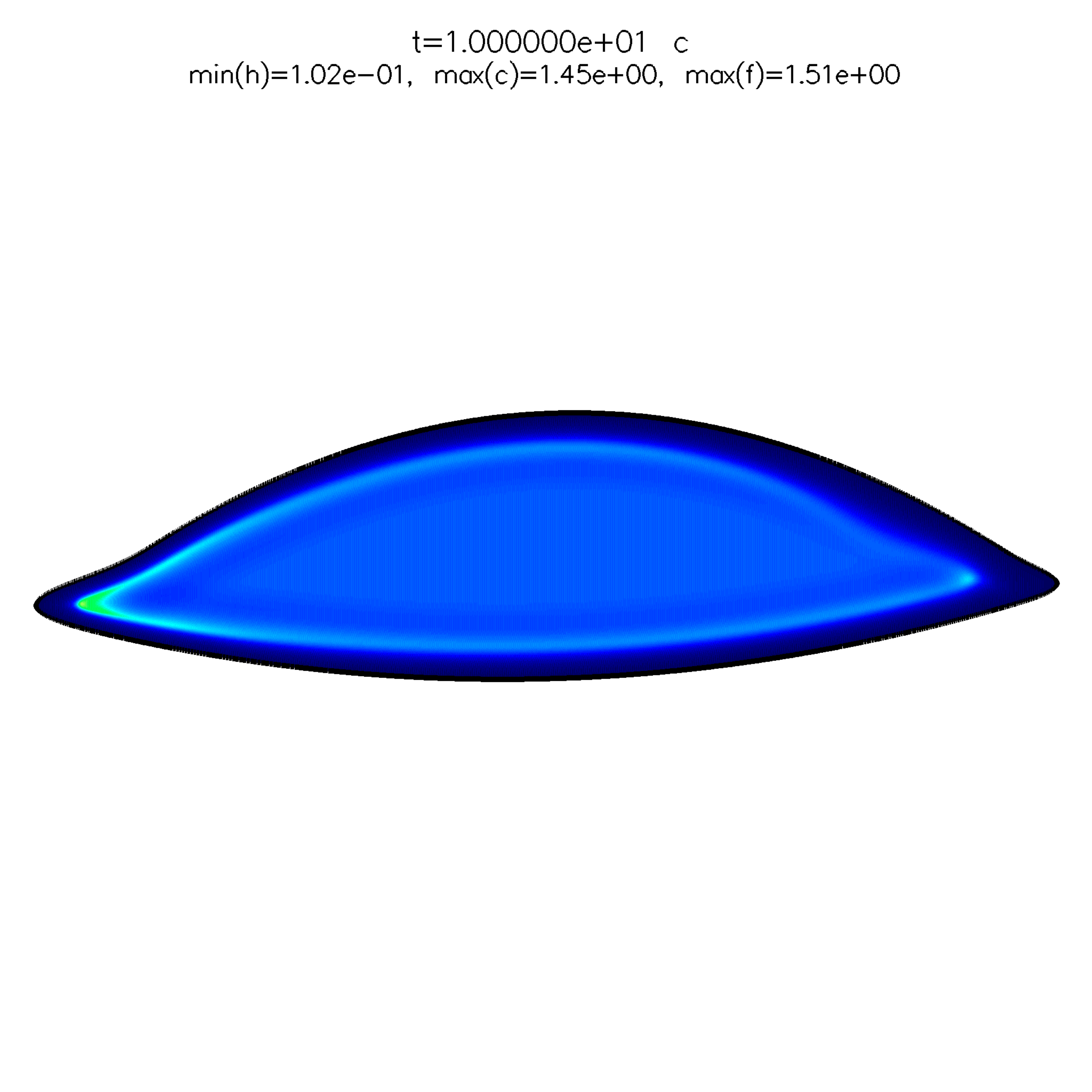}{\figWidth}};
\pgfmathparse{\xs+1.5*\figWidth} \let\xx\pgfmathresult
\draw(\xx,\row) node[anchor=north]{$\max(c)=$ 1.45};
\pgfmathparse{2*\figWidth} \let\xf\pgfmathresult
\draw(\xf,\row) node[anchor=south west,xshift=\xs cm,yshift=\ys cm] {\trimfig{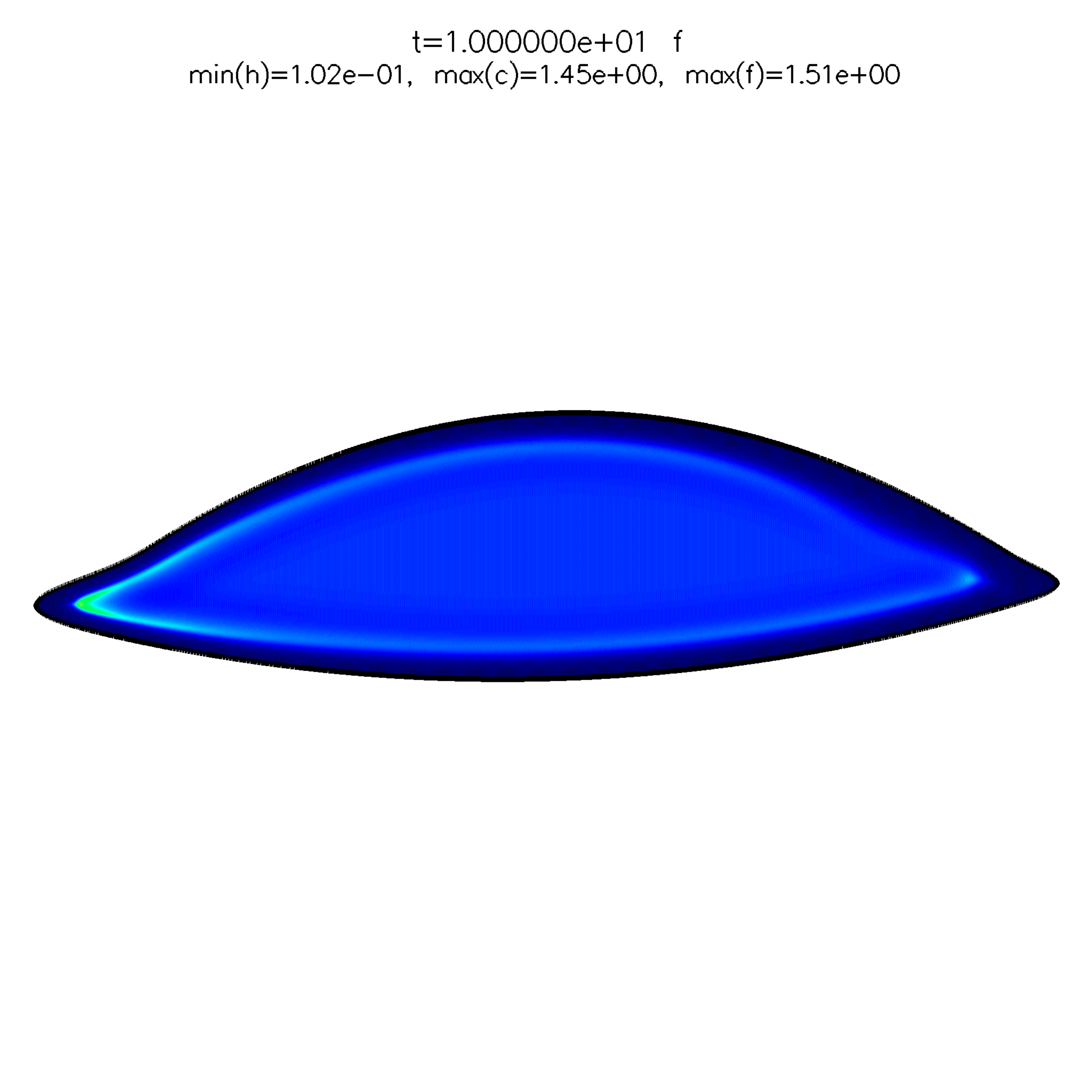}{\figWidth}};
\pgfmathparse{\xs+2.5*\figWidth} \let\xx\pgfmathresult
\draw(\xx,\row) node[anchor=north]{$\max(f)=$ 1.51};
\pgfmathparse{\row+\rowHeight} \let\row\pgfmathresult
\pgfmathparse{\row+1} \let\tP\pgfmathresult
\draw(0,\tP) node[anchor=west,xshift=-0.1cm] {$t=5$};
\pgfmathparse{0*\figWidth} \let\xf\pgfmathresult
\draw(\xf,\row) node[anchor=south west,xshift=\xs cm,yshift=\ys cm] {\trimfig{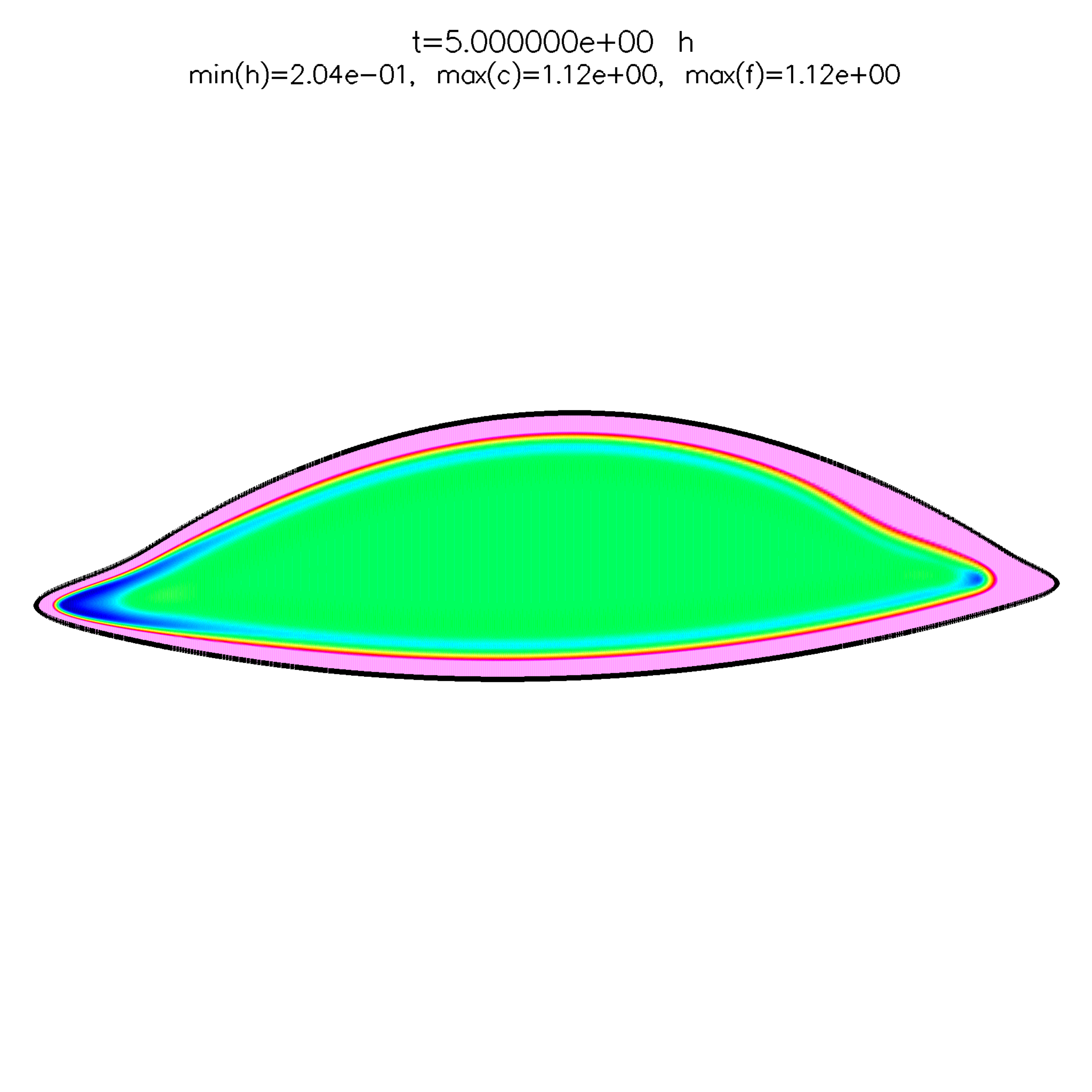}{\figWidth}};
\pgfmathparse{\xs+0.5*\figWidth} \let\xx\pgfmathresult
\draw(\xx,\row) node[anchor=north]{$\min(h)=$ 2.04e-1};
\pgfmathparse{1*\figWidth} \let\xf\pgfmathresult
\draw(\xf,\row) node[anchor=south west,xshift=\xs cm,yshift=\ys cm] {\trimfig{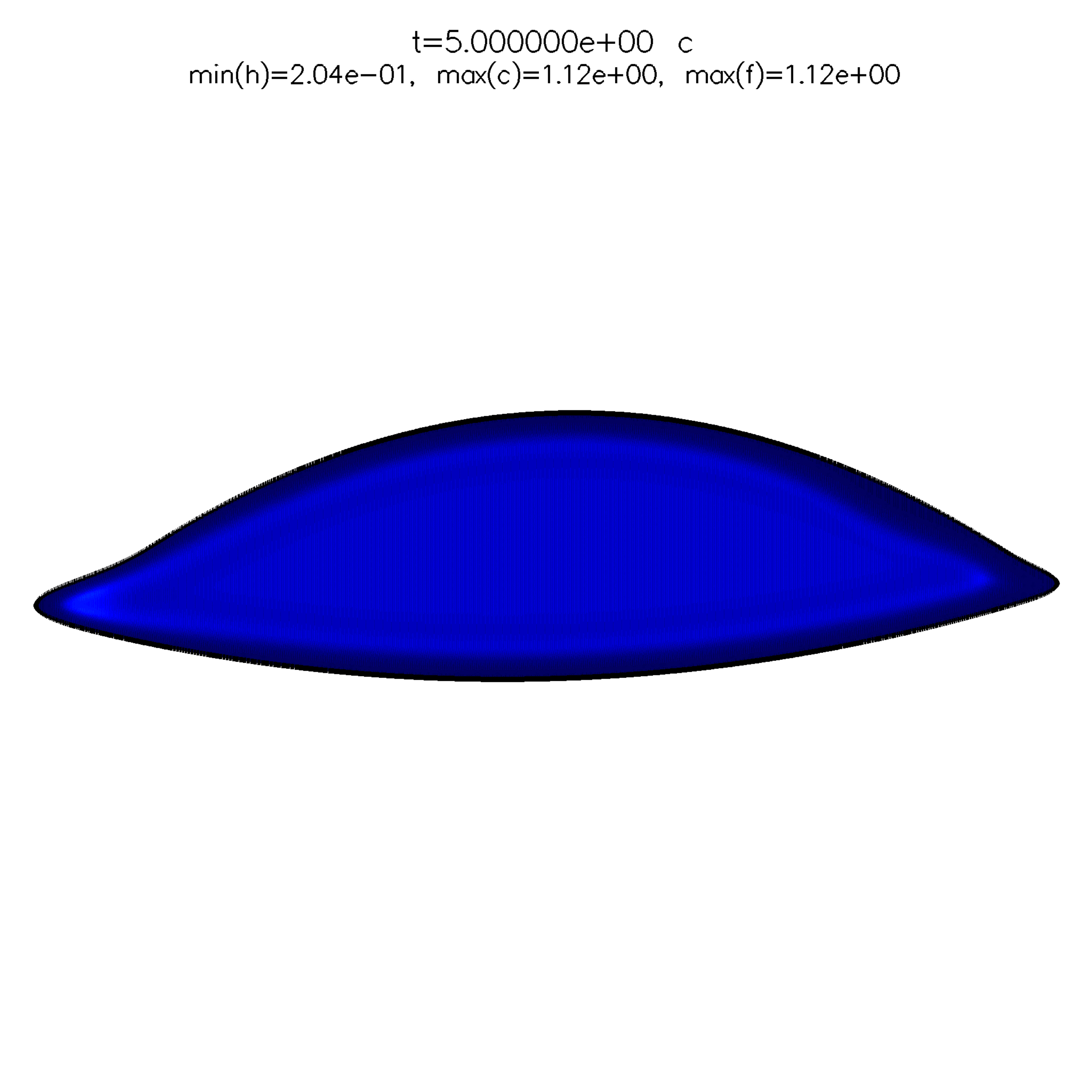}{\figWidth}};
\pgfmathparse{\xs+1.5*\figWidth} \let\xx\pgfmathresult
\draw(\xx,\row) node[anchor=north]{$\max(c)=$ 1.12};
\pgfmathparse{2*\figWidth} \let\xf\pgfmathresult
\draw(\xf,\row) node[anchor=south west,xshift=\xs cm,yshift=\ys cm] {\trimfig{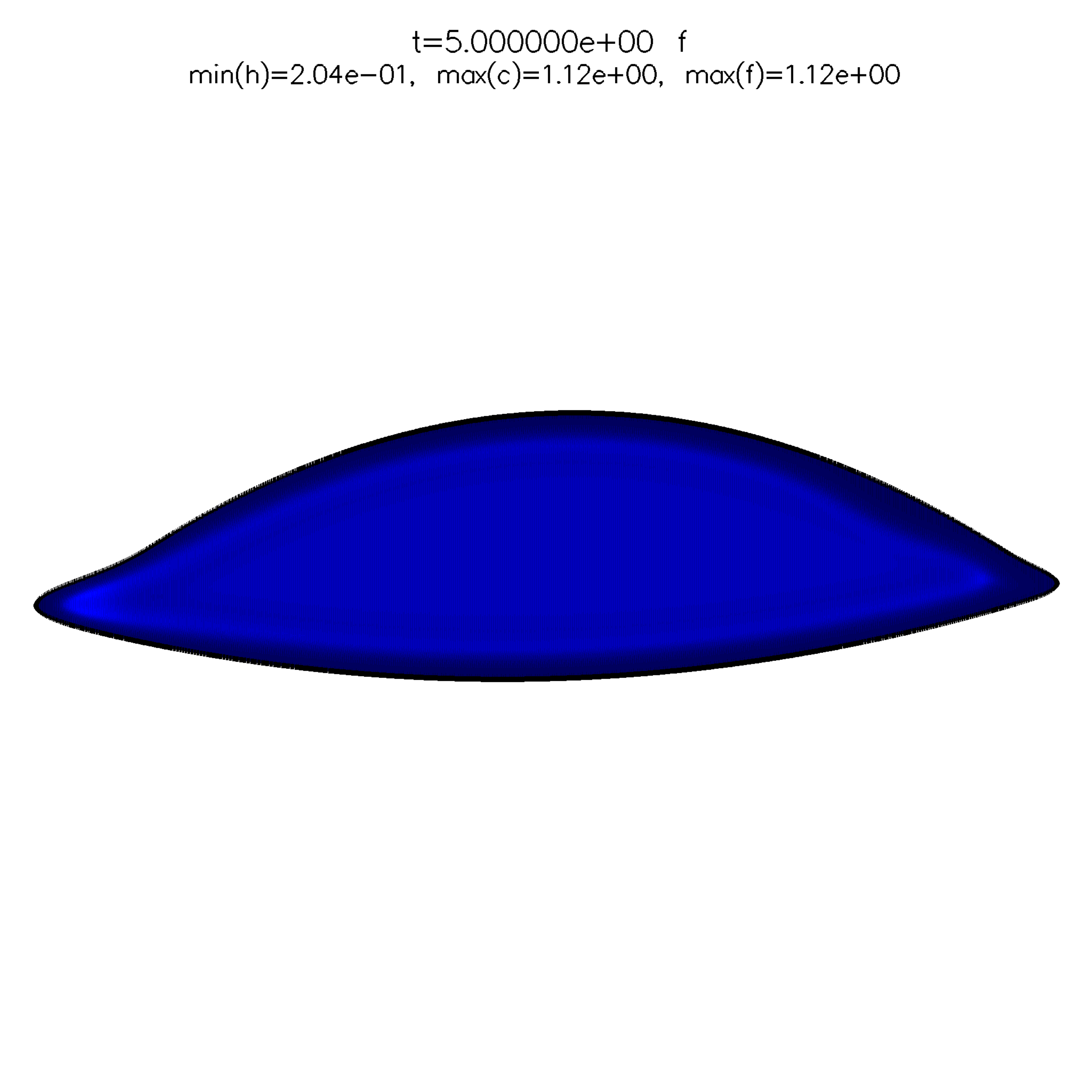}{\figWidth}};
\pgfmathparse{\xs+2.5*\figWidth} \let\xx\pgfmathresult
\draw(\xx,\row) node[anchor=north]{$\max(f)=$ 1.12};
\pgfmathparse{\row+\rowHeight} \let\row\pgfmathresult
%
\end{tikzpicture}
}
\end{center}
\caption{$h$, $c$ and $f$ for an impermeable ocular surface and a 4 $\mu$m/min nominal thinning rate.}
\label{fig:hcf_Pc=0_4mpm}
\end{sidewaysfigure}
}


With the computed distributions of $h$ and $f$, we can now use
(\ref{eqn:I-of-f}) to compute the intensity distribution from the fluorescein concentration.
Recall that the FL intensity may be the most widely used imaging technique but that the intensity observed depends on both $h$ and $f$.  This implies dependence on both the initial thickness and $f_0$, and the resulting intensity may then be more complicated than may be expected
from commonly used interpretations
\citep{NicholsEtal12,BraunEtal14}.
Figure~\ref{fig:Intensity_Pc=0_4mpm} shows the intensity for two different initial values of $f$.  The smaller value is in the so-called dilute regime, and the larger is the self-quenching regime \citep{Webber86,NicholsEtal12}.

The column on the left begins in the dilute regime
where $f<1$ \citep{Webber86,NicholsEtal12}.  The black line mentioned previously mentioned can be seen forming by $t=5$, and this kind of imaging is how it got its name.  The black line becomes that way because tangential flow into the menisci and the outer edge of the central region causing thinning and a subsequent reduction in intensity according to (\ref{eqn:I-of-f}). A slight increase in $I$ is seen on either side of the black line from the influx of fluid from the black line.  As time increases, water is lost due to evaporation, $f$ increases, but the intensity in the central region inside the black line changes relatively little.  This is because the product $fh$ that appears in the exponent is the most important term affecting the intensity in the dilute regime, and this product is nearly constant during evaporative thinning \citep{BraunEtal14}. A roughly unchanged $I$ for the central tear film continues throughout this case because any changes in thickness and concentration are driven primarily by evaporation in this model.

Turning to the self-quenching case in the right column of Figure~\ref{fig:Intensity_Pc=0_4mpm}, we again see the black line develop by $t=5$, with slight lightening on each side.  However, the central region becomes a bit darker as time progresses.  This is a feature of the self-quenching regime, where the intensity is dominated by the denominator of (\ref{eqn:I-of-f}), and in evaporative thinning, $f$ increases such that $f \propto 1/h$ and so $f$ will substantiantially affect the intensity observed \citep{BraunEtal14}.  For this case, the thickness does not change too much, but we will see more obvious differences in the subsequent cases.
{
\def\nRows{3}  
\def\nColumns{2}  
\begin{figure}[htb]
\begin{center}
\begin{tikzpicture}[scale=1]
\pgfmathparse{\columnWidth*\nColumns} \let\xb\pgfmathresult
\pgfmathparse{\rowHeight*\nRows+1} \let\yb\pgfmathresult 
\useasboundingbox (0.0,0.0) rectangle (\xb,\yb);  
\def\row{1}
\def\xs{1.1} 
\def\ys{0.3}
\pgfmathparse{\xs+0.5*\figWidth} \let\xcb\pgfmathresult 
\pgfmathparse{\xcb+0.*\cbWidth} \let\xx\pgfmathresult
\draw(\xx,\ys) node[anchor=south west] {\getCB{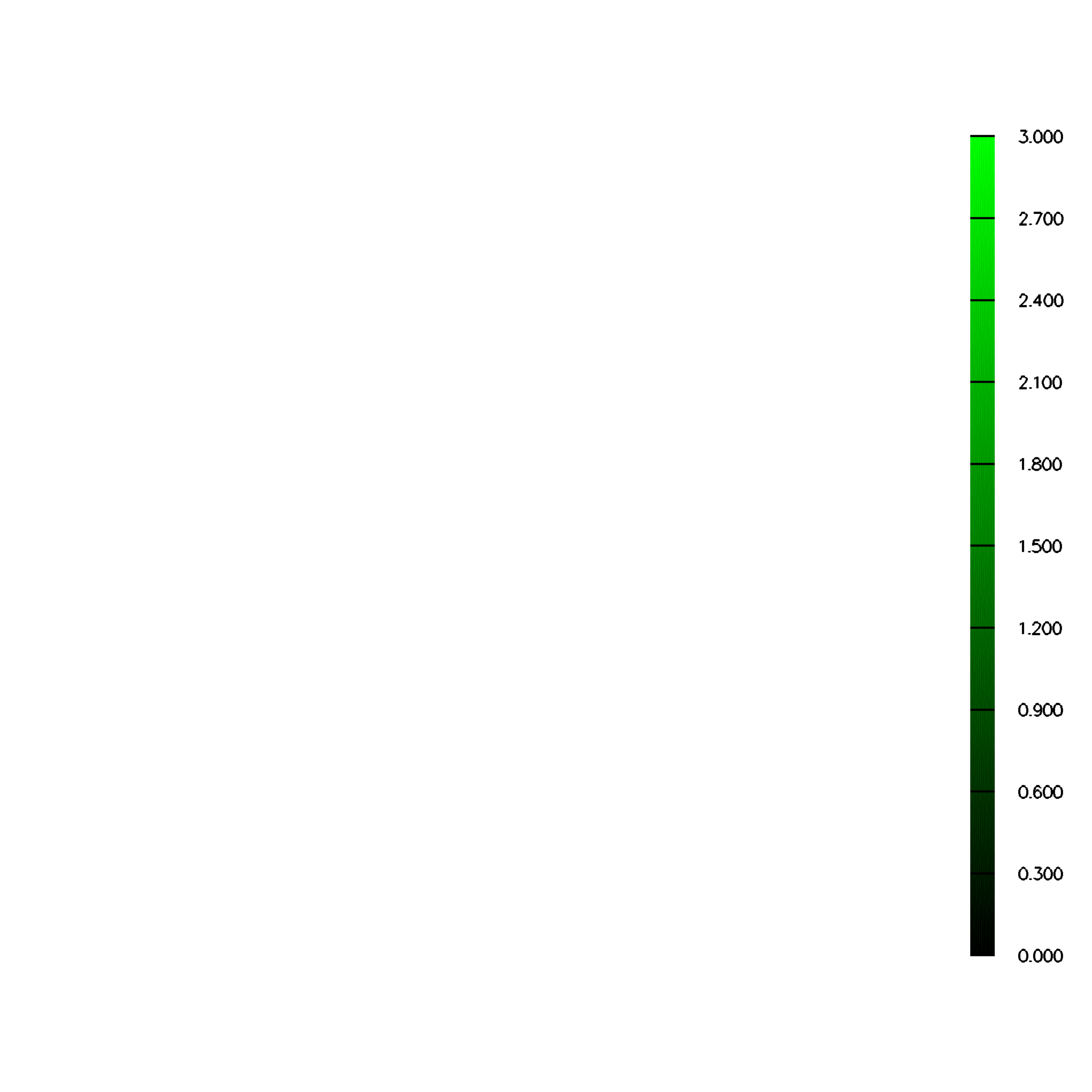}{\cbWidth}};
\draw(\xx,\ys) node[anchor= west]{$0$};
\pgfmathparse{\xcb+0.47*\cbWidth} \let\xx\pgfmathresult
\draw(\xx,\ys) node[anchor= west]{$50$};
\pgfmathparse{\xcb+0.97*\cbWidth} \let\xx\pgfmathresult
\draw(\xx,\ys) node[anchor= west]{$100$};
\def\xs{0.8}
\def\ys{0.}
\pgfmathparse{\xs+0.5*\figWidth} \let\xx\pgfmathresult
\pgfmathparse{\yb-1} \let\yy\pgfmathresult
\draw(\xx,\yy) node[anchor=south]{Intensity $I$ with $f_0=\frac{1}{3}$};
\pgfmathparse{\xs+1.5*\figWidth} \let\xx\pgfmathresult
\draw(\xx,\yy) node[anchor=south]{Intensity $I$ with $f_0=2$};
\pgfmathparse{\row+1} \let\tP\pgfmathresult
\draw(0,\tP) node[anchor=west,xshift=-0.1cm] {$t=15$};
\draw(0,\row) node[anchor=south west,xshift=\xs cm,yshift=\ys cm] {\trimfig{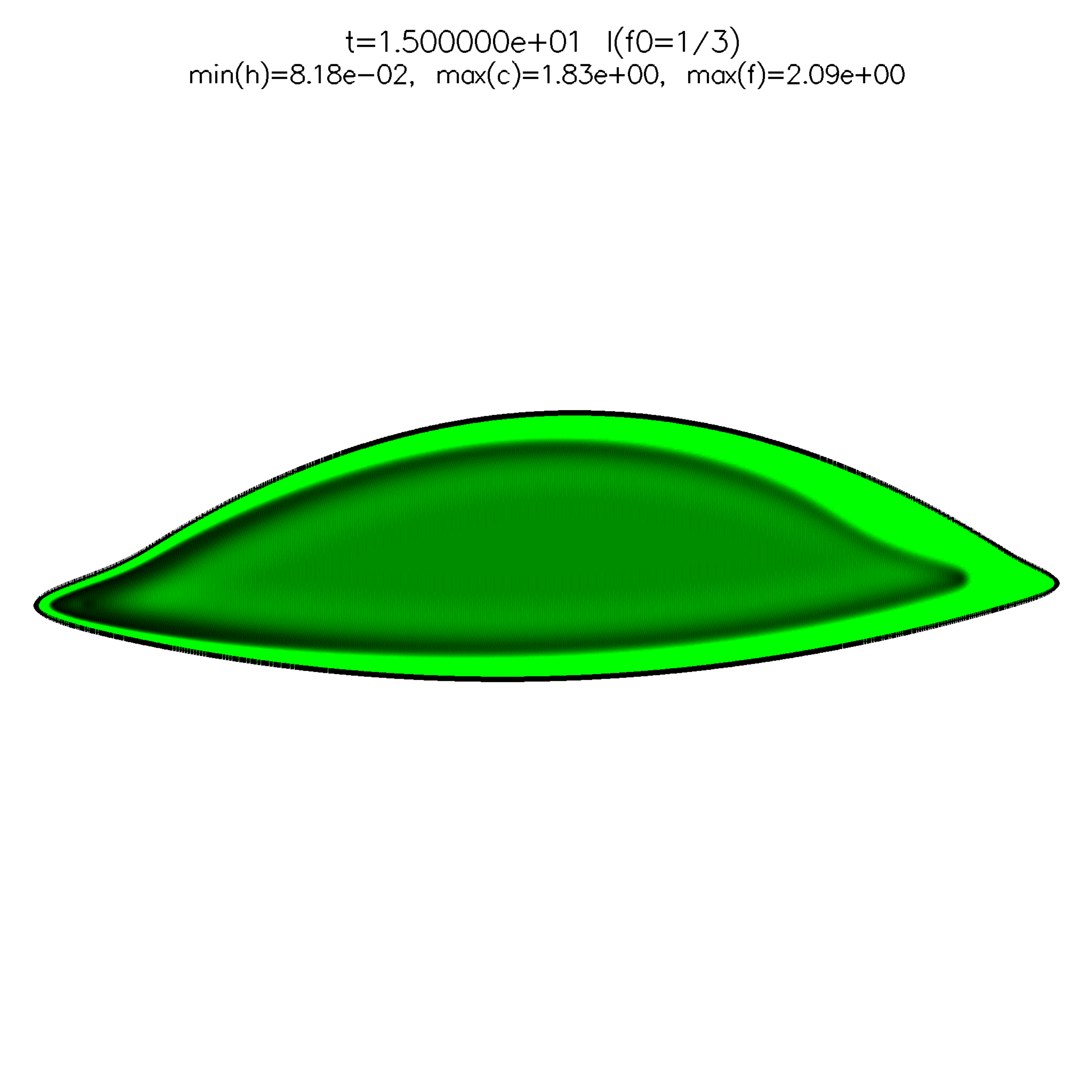}{\figWidth}};
\draw(\figWidth,\row) node[anchor=south west,xshift=\xs cm,yshift=\ys cm] {\trimfig{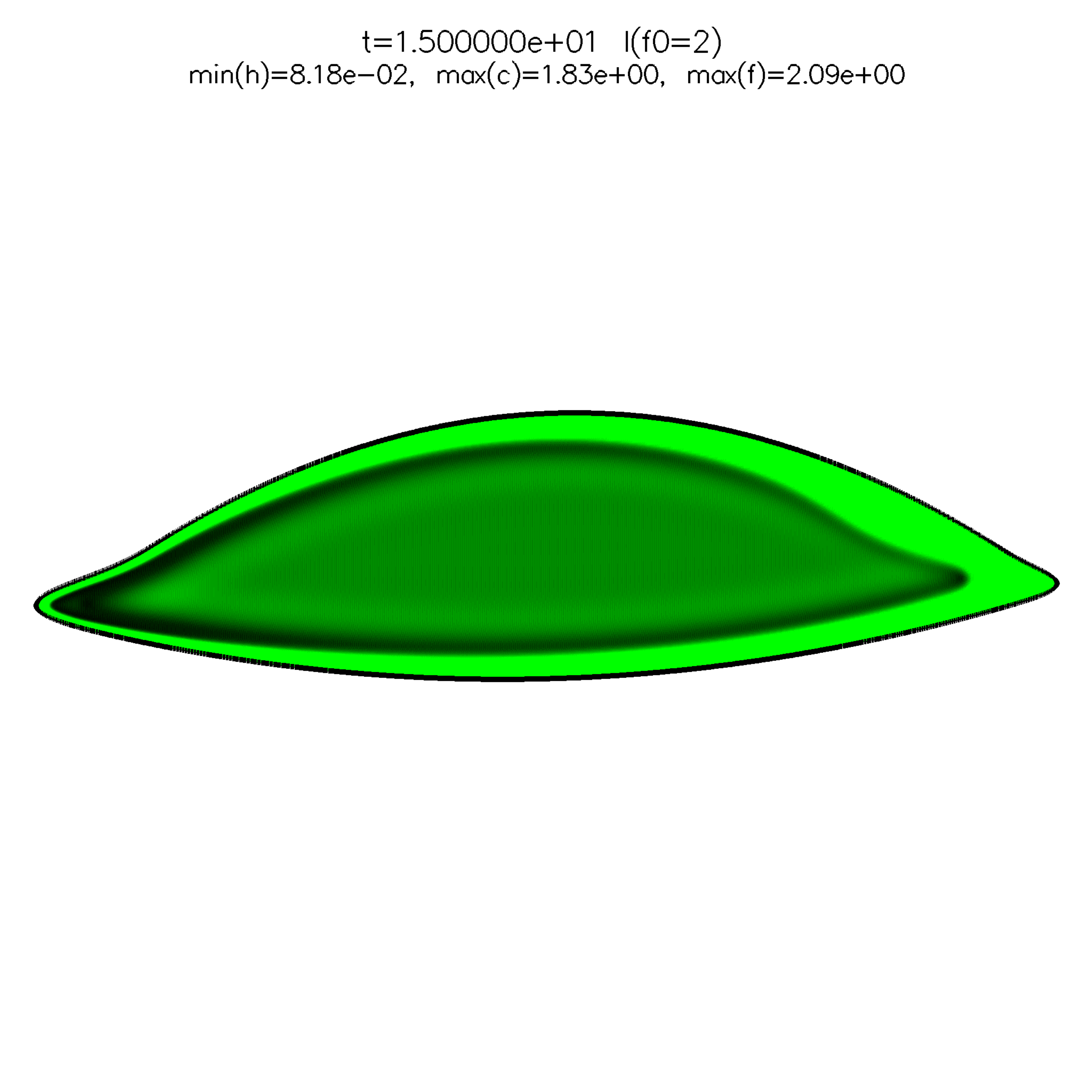}{\figWidth}};
\pgfmathparse{\row+\rowHeight} \let\row\pgfmathresult
\pgfmathparse{\row+1} \let\tP\pgfmathresult
\draw(0,\tP) node[anchor=west,xshift=-0.1cm] {$t=10$};
\draw(0,\row) node[anchor=south west,xshift=\xs cm,yshift=\ys cm] {\trimfig{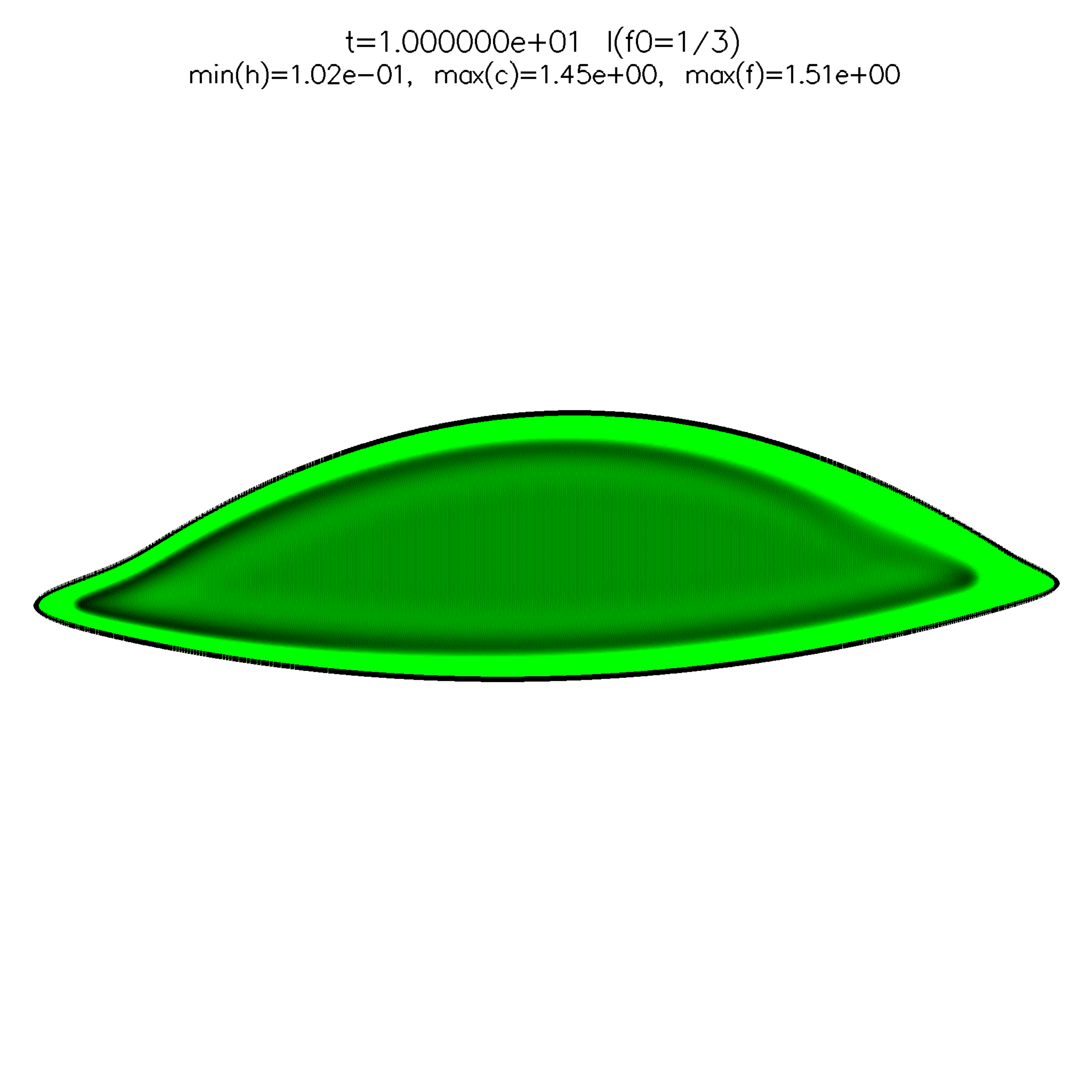}{\figWidth}};
\draw(\figWidth,\row) node[anchor=south west,xshift=\xs cm,yshift=\ys cm] {\trimfig{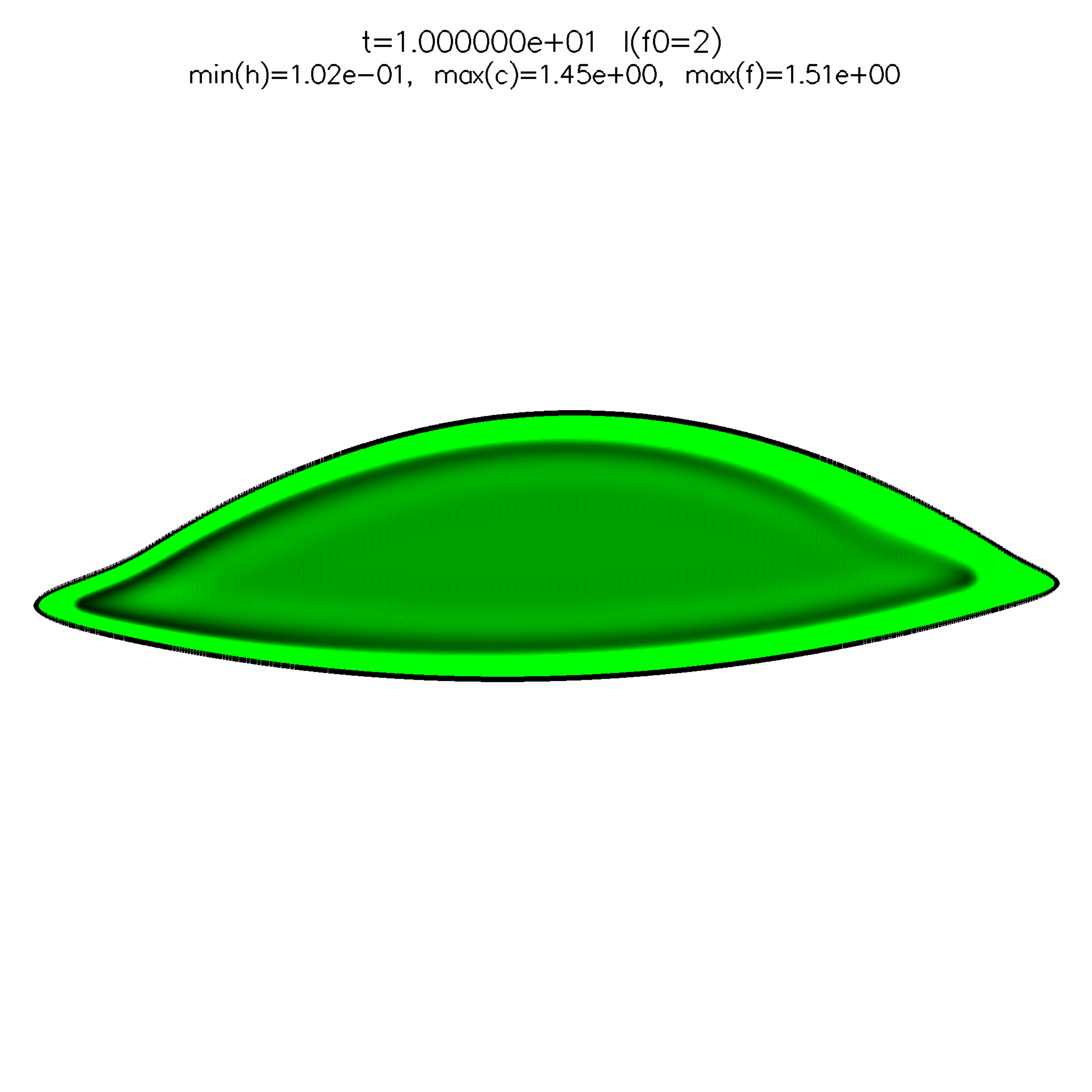}{\figWidth}};
\pgfmathparse{\row+\rowHeight} \let\row\pgfmathresult
\pgfmathparse{\row+1} \let\tP\pgfmathresult
\draw(0,\tP) node[anchor=west,xshift=-0.1cm] {$t=5$};
\draw(0,\row) node[anchor=south west,xshift=\xs cm,yshift=\ys cm] {\trimfig{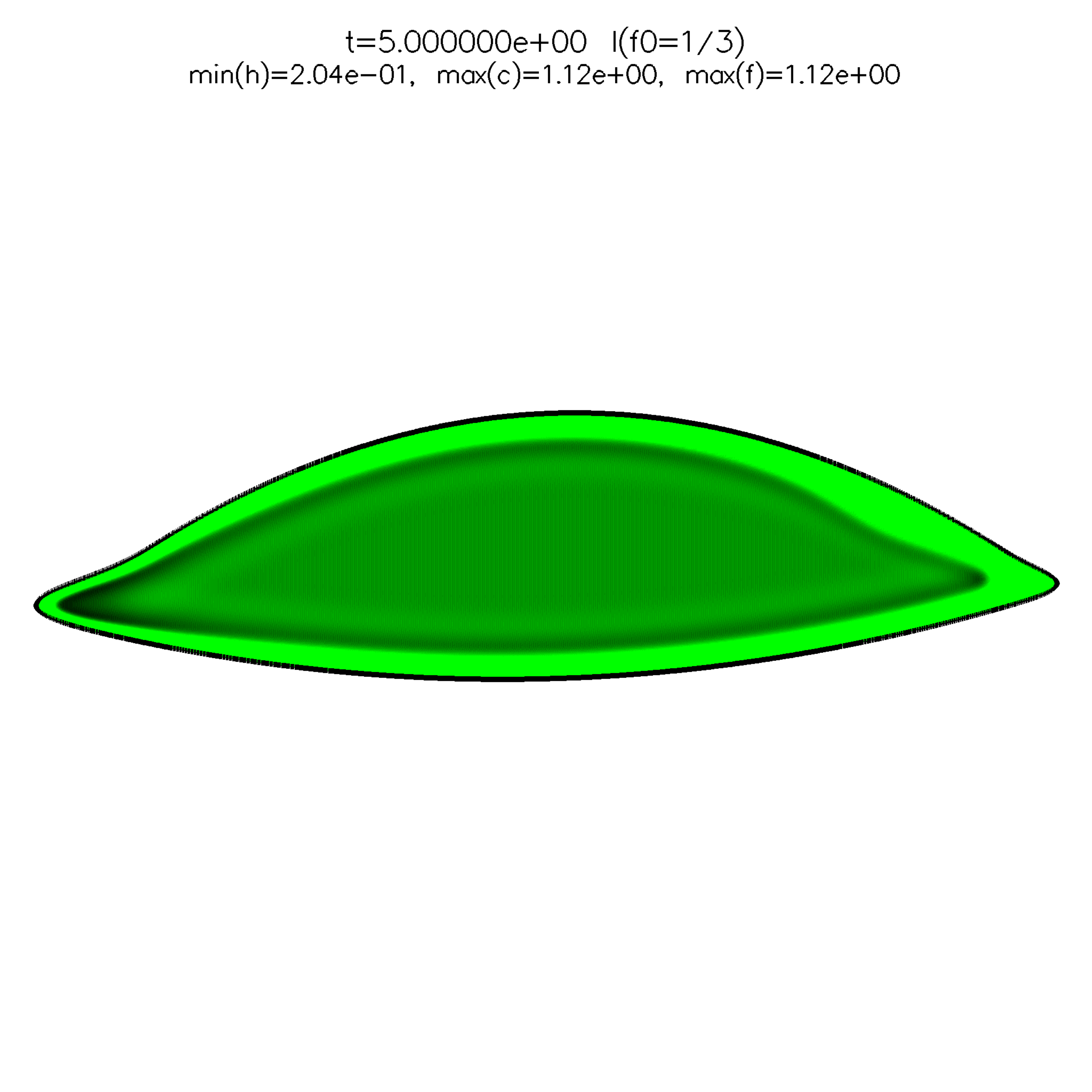}{\figWidth}};
\draw(\figWidth,\row) node[anchor=south west,xshift=\xs cm,yshift=\ys cm] {\trimfig{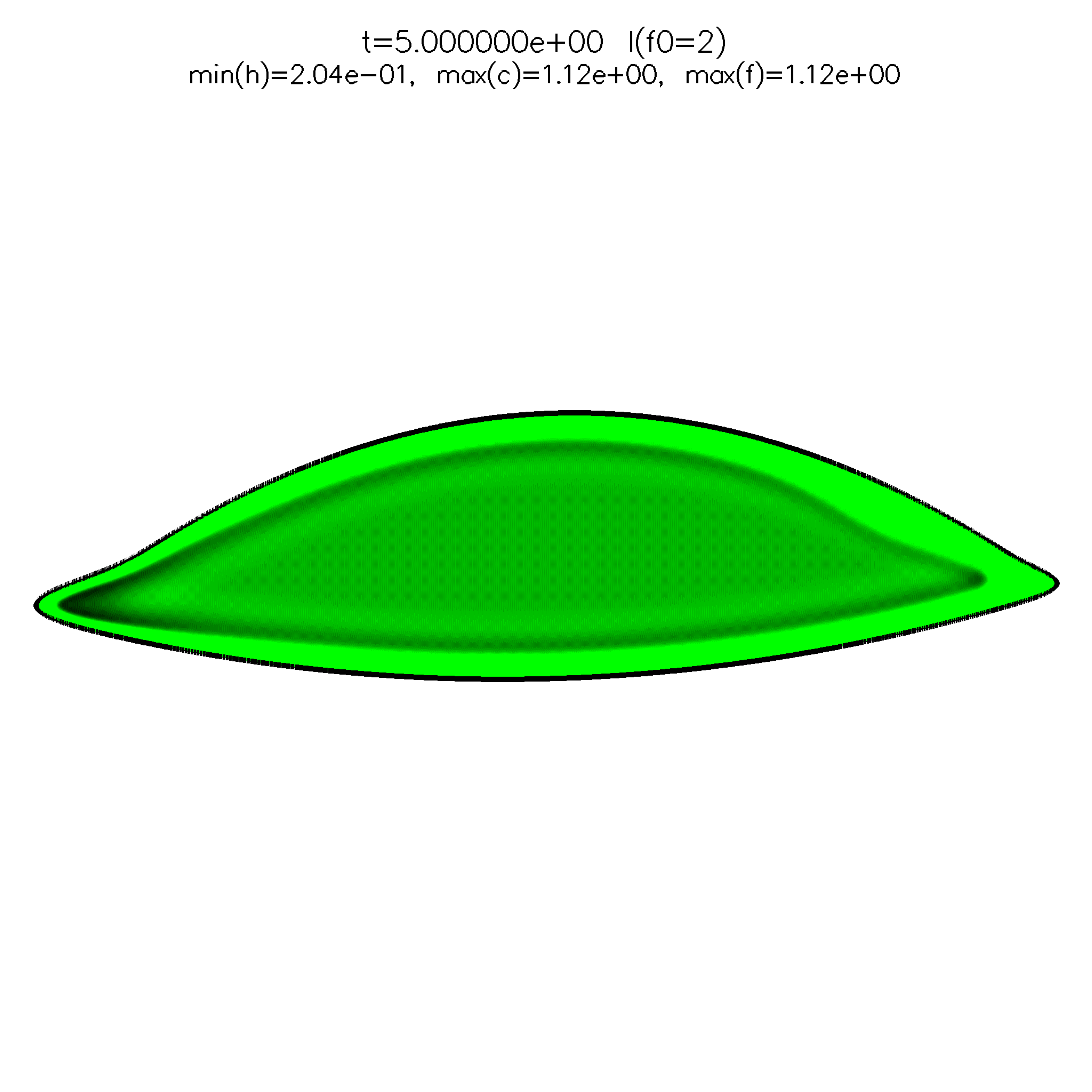}{\figWidth}};
\pgfmathparse{\row+\rowHeight} \let\row\pgfmathresult
%
\end{tikzpicture}
\caption{No permeability and evaporation is 4 micron/min, $I$ with $f_0=1/3$ and $f_0=2$}
\label{fig:Intensity_Pc=0_4mpm}
\end{center}
\end{figure}
}

There is an approximation of the intensity-thickness relationship that can be used fruitfully in the
self-quenching regime.  As suggested by \citet{NicholsEtal12}, (\ref{eqn:I-of-f}) can be approximated by
\begin{equation}
h = a \sqrt{I}
\end{equation}
where $a$ is a constant typically chosen to calibrate the observed intensity to a known thickness for a given optical system \cite{NicholsEtal12}.  In our case, we choose $a$ so that $a\sqrt{I}$ matches the initial
thickness of unity in the central tear film region.  Using our results for $I$ and $h$, we tested how well this approximation worked for this simulation for two different initial values of $f$.  The results are shown in Figure~\ref{fig:CompareMethods_Pc=0_4mpm}.  If the approximation is working well, the difference $|h-a\sqrt{I}|$ should be close to zero.  We see that for the dilute case on the left, each time has a large difference all over
the domain.  This is not unexpected for the dilute case, since the approximation is designed for the
self-quenching regime.  In that regime, the difference remains relatively small in the black line and
central regions.  The approximation did not work well in the menisci in either the dilute or self-quenching regimes, but this is to be expected because $a$ was chosen to work best in the central and black line regions.
{
\def\nRows{3}  
\def\nColumns{2}  
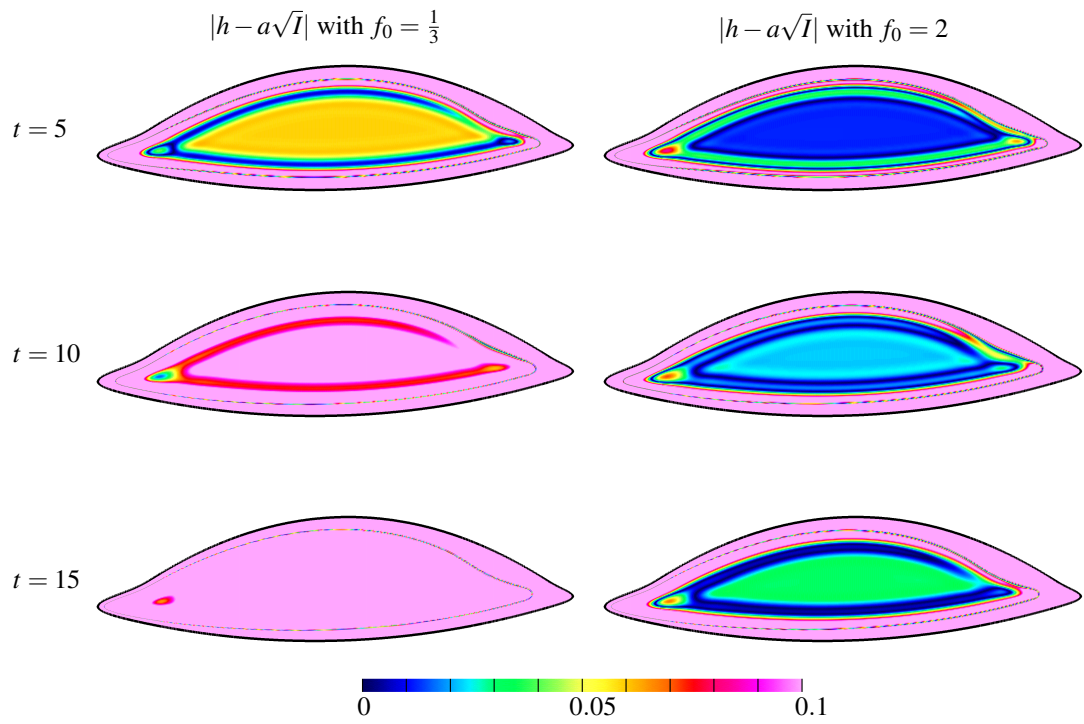
\begin{figure}[htb]
\begin{center}
\begin{tikzpicture}[scale=1]
\pgfmathparse{\columnWidth*\nColumns} \let\xb\pgfmathresult
\pgfmathparse{\rowHeight*\nRows+1} \let\yb\pgfmathresult 
\useasboundingbox (0.0,0.0) rectangle (\xb,\yb);  
\def\row{1}
\def\xs{1.1} 
\def\ys{0.3}
\pgfmathparse{\xs+0.5*\figWidth} \let\xcb\pgfmathresult 
\pgfmathparse{\xcb+0.*\cbWidth} \let\xx\pgfmathresult
\draw(\xx,\ys) node[anchor=south west] {\getCB{colorbar_rainbow.pdf}{\cbWidth}};
\draw(\xx,\ys) node[anchor= west]{$0$};
\pgfmathparse{\xcb+0.47*\cbWidth} \let\xx\pgfmathresult
\draw(\xx,\ys) node[anchor= west]{$0.05$};
\pgfmathparse{\xcb+0.97*\cbWidth} \let\xx\pgfmathresult
\draw(\xx,\ys) node[anchor= west]{$0.1$};
\def\xs{0.8}
\def\ys{0.}
\pgfmathparse{\xs+0.5*\figWidth} \let\xx\pgfmathresult
\pgfmathparse{\yb-1} \let\yy\pgfmathresult
\draw(\xx,\yy) node[anchor=south]{$|h-a\sqrt{I}|$ with $f_0=\frac{1}{3}$};
\pgfmathparse{\xs+1.5*\figWidth} \let\xx\pgfmathresult
\draw(\xx,\yy) node[anchor=south]{$|h-a\sqrt{I}|$ with $f_0=2$};
\pgfmathparse{\row+1} \let\tP\pgfmathresult
\draw(0,\tP) node[anchor=west,xshift=-0.1cm] {$t=15$};
\draw(0,\row) node[anchor=south west,xshift=\xs cm,yshift=\ys cm] {\trimfig{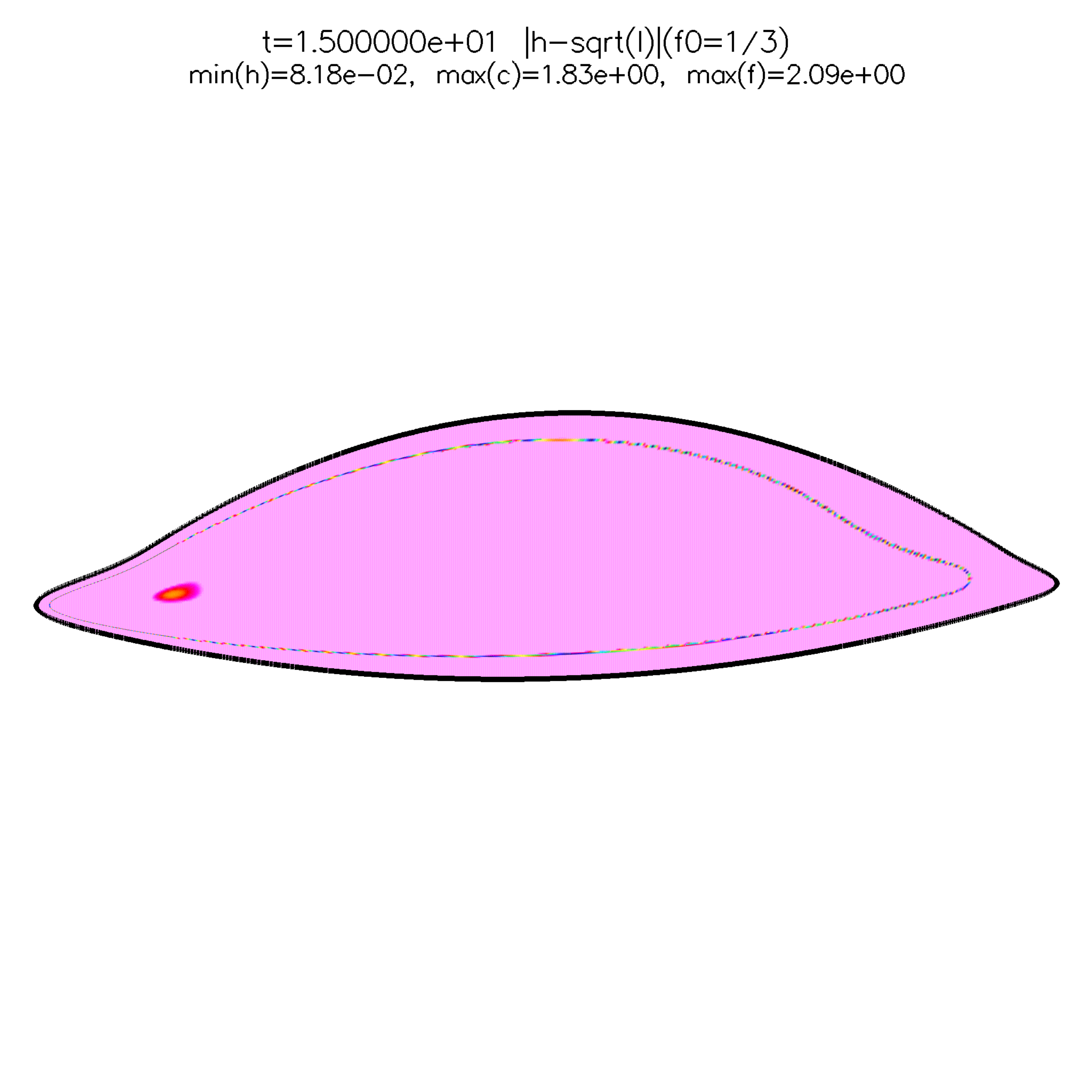}{\figWidth}};
\draw(\figWidth,\row) node[anchor=south west,xshift=\xs cm,yshift=\ys cm] {\trimfig{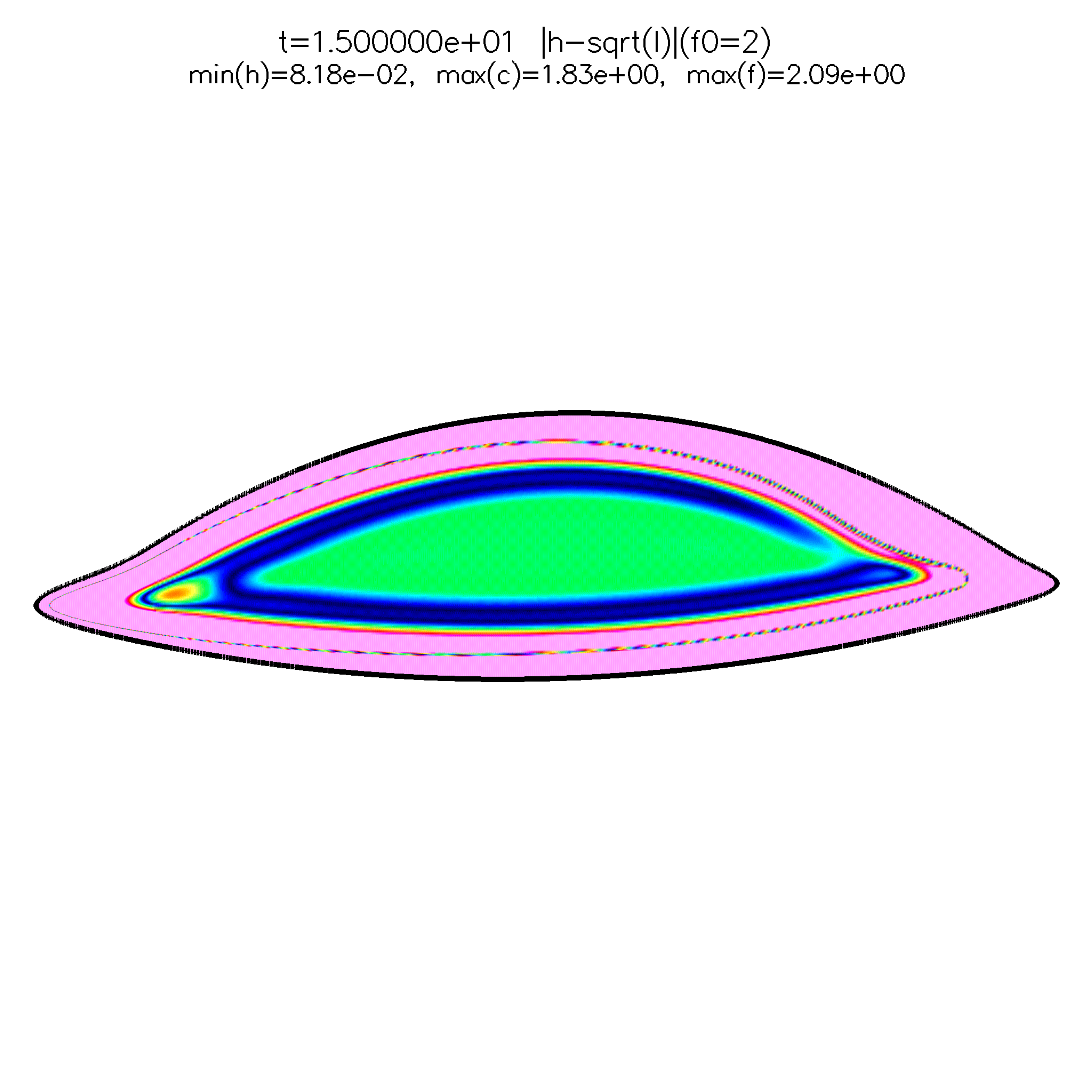}{\figWidth}};
\pgfmathparse{\row+\rowHeight} \let\row\pgfmathresult
\pgfmathparse{\row+1} \let\tP\pgfmathresult
\draw(0,\tP) node[anchor=west,xshift=-0.1cm] {$t=10$};
\draw(0,\row) node[anchor=south west,xshift=\xs cm,yshift=\ys cm] {\trimfig{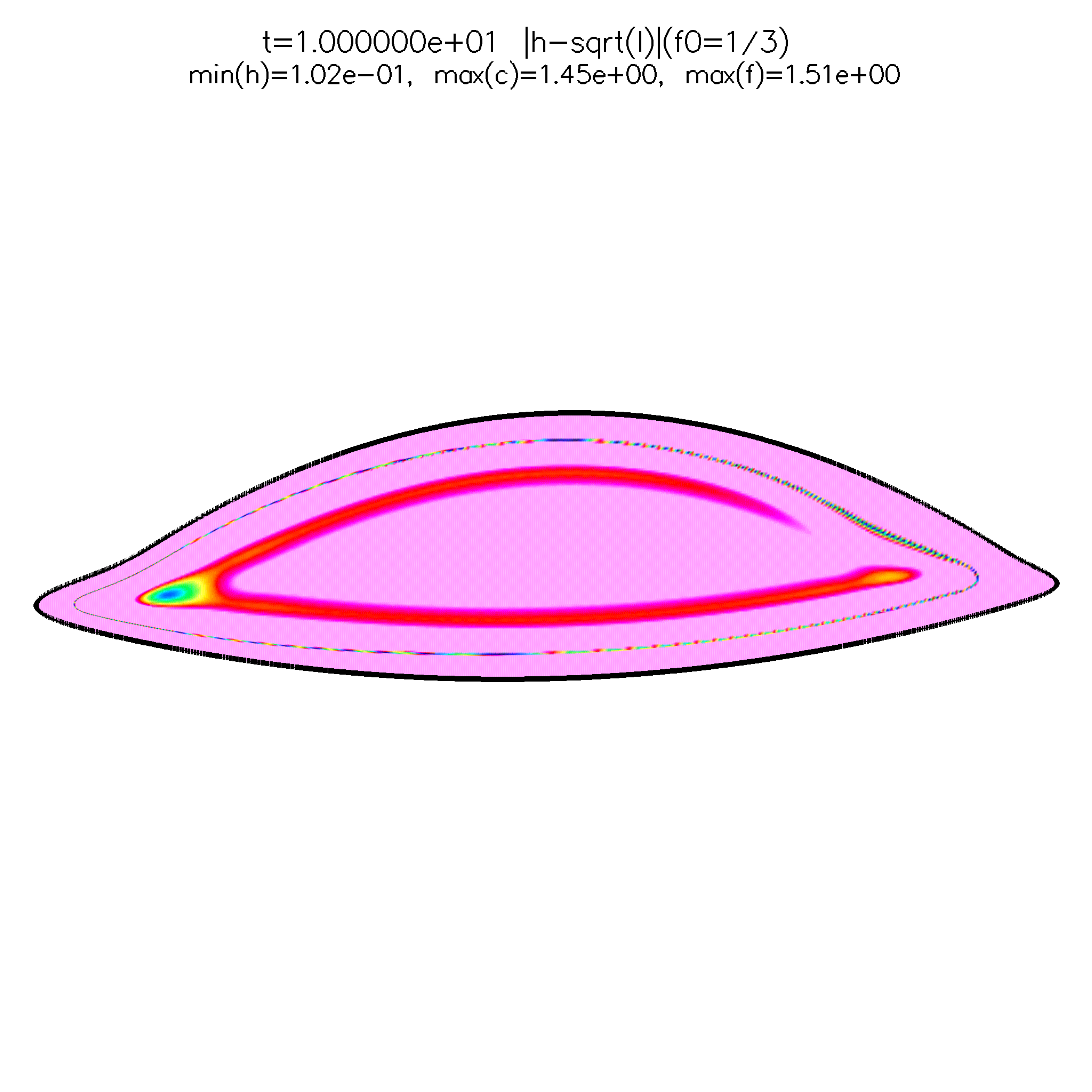}{\figWidth}};
\draw(\figWidth,\row) node[anchor=south west,xshift=\xs cm,yshift=\ys cm] {\trimfig{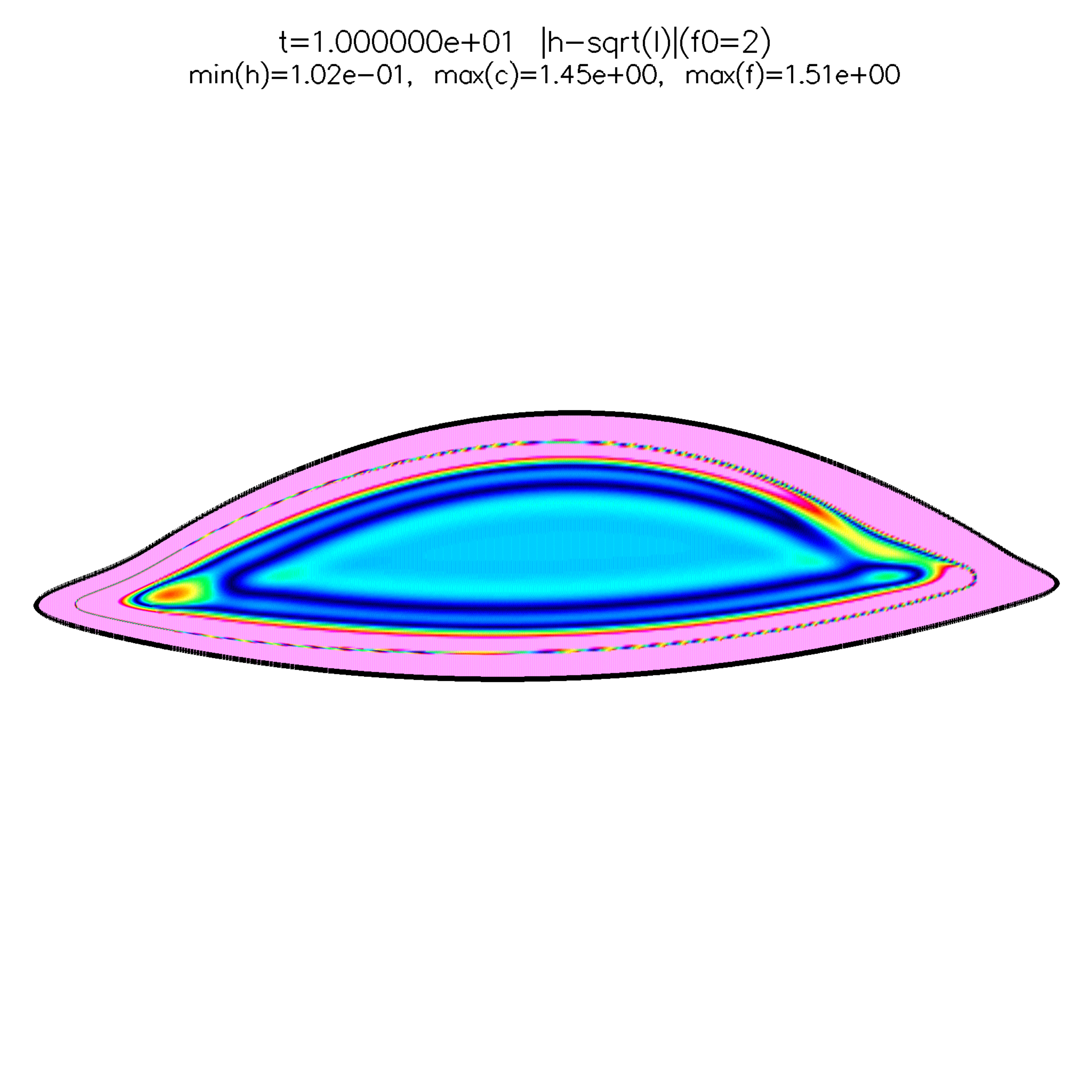}{\figWidth}};
\pgfmathparse{\row+\rowHeight} \let\row\pgfmathresult
\pgfmathparse{\row+1} \let\tP\pgfmathresult
\draw(0,\tP) node[anchor=west,xshift=-0.1cm] {$t=5$};
\draw(0,\row) node[anchor=south west,xshift=\xs cm,yshift=\ys cm] {\trimfig{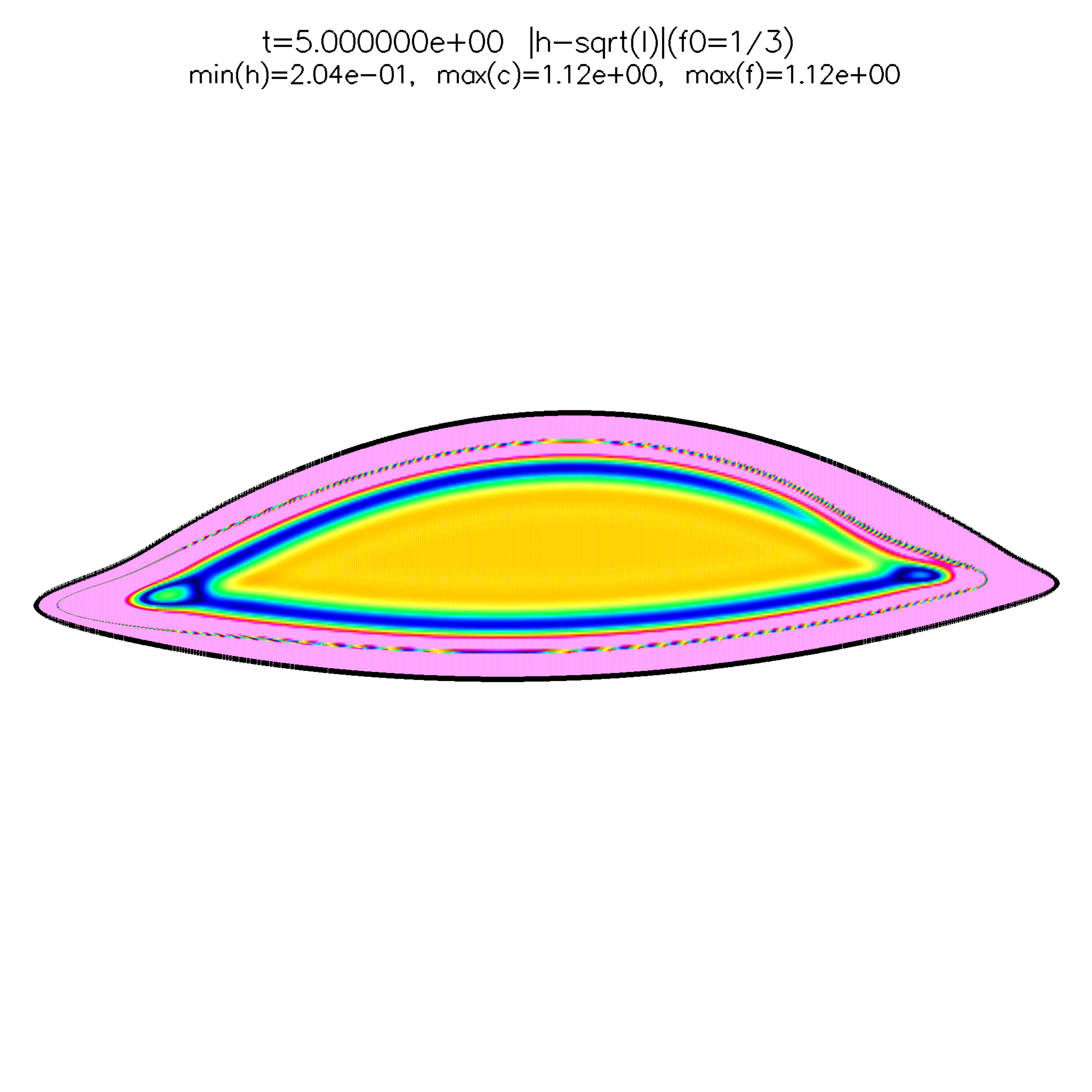}{\figWidth}};
\draw(\figWidth,\row) node[anchor=south west,xshift=\xs cm,yshift=\ys cm] {\trimfig{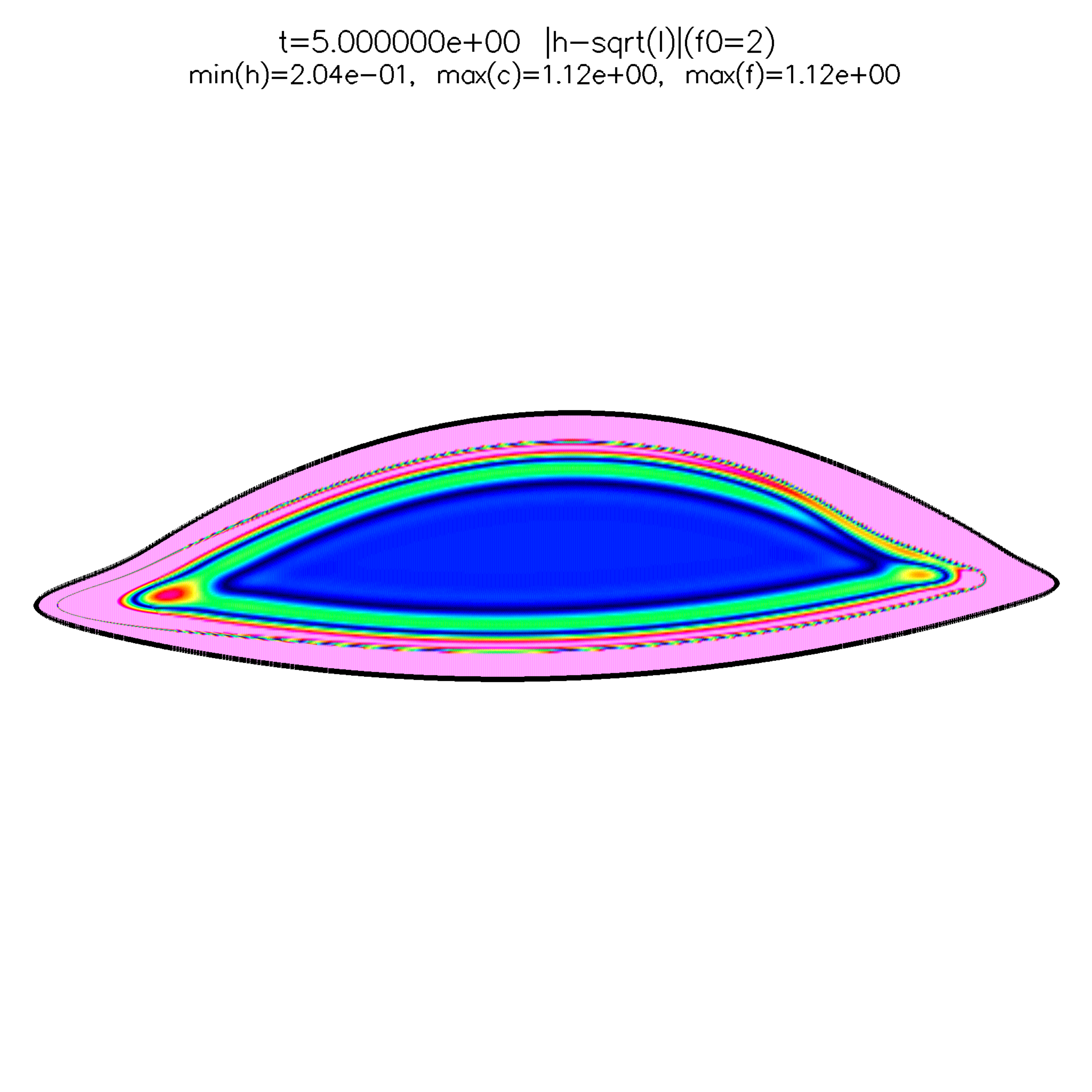}{\figWidth}};
\pgfmathparse{\row+\rowHeight} \let\row\pgfmathresult
%
\end{tikzpicture}
\caption{No permeability and evaporation is 4 micron/min, $|h-a\sqrt{I}|$ with $f_0=1/3$ and $f_0=2$}
\label{fig:CompareMethods_Pc=0_4mpm}
\end{center}
\end{figure}
}

\clearpage
\subsection{Variable Permeability $(P_c=P_c(x,y))$}

We now proceed to the case where the ocular surface is permeable and the cornea is about
four times less permeable than the surrounding conjunctival area \cite{LiBraun15}.  We consider two different
evaporation rates: 4 $\mu$m/min and 20 $\mu$m/min.  The former corresponds to the high end of the normal evaporation rates, while the latter is near the highest end of measured thinning rates \cite{KSHinNic10}.

\subsubsection{Evaporation rate 4 $\mu$m/min}

The dynamics for $h$, $c$ and $f$ are shown in Fig.\ref{fig:hcf_Pc(x,y)_4mpm}.
In the first column, $h$ is seen to thin overall as time increases,
and this is more noticeable away from the menisci. The black lines are still seen.
The tear film becomes
thinner over the cornea due to its lower permeability, and this
appears as the circular central disk region as seen in our previous work without
fluorescein \citep{LiBraun15}.

In the second and third columns,
$c$ and $f$ increase as the tear film thins due to mass
conservation of solutes. The osmolarity increases more
over the cornea than the conjunctiva, and the osmolarity is
highest in black lines over the cornea.  The largest value appears over the cornea
because of the lower permeability there \citep{LiBraun15}.
The distribution of $f$ is very similar in its pattern to $c$,
despite the difference in diffusivity.  In this simulation lasting
15 s, the maximum value is about 1.3 times the initial value,
corresponding to about 390 mOsM peak concentration for the osmolarity
or $1.3f_0$ for $f$.
At this evaporation rate and with the supply of water via osmosis, the
solute concentrations don't build up as much as the impermeable case,
and because of this the diffusivities are insufficiently different to separate
the solute distributions for this evaporation rate and end time.
{
\def\nRows{3}  
\def\nColumns{3}  
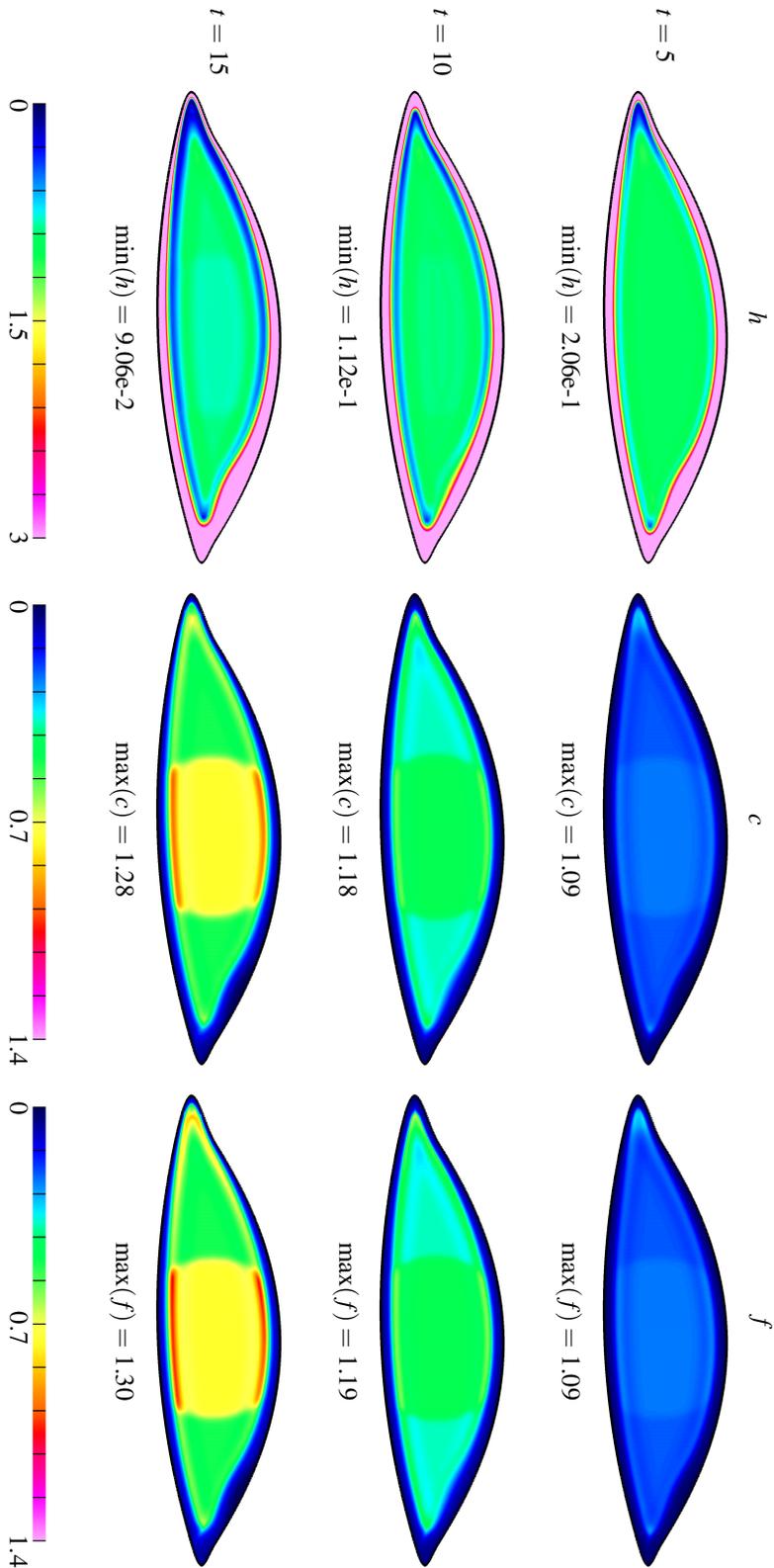
\begin{sidewaysfigure}[htb]
\begin{center}
\begin{tikzpicture}[scale=1]
\pgfmathparse{\columnWidth*\nColumns} \let\xb\pgfmathresult
\pgfmathparse{\rowHeight*\nRows+2} \let\yb\pgfmathresult 
\useasboundingbox (0.0,0.0) rectangle (\xb,\yb);  
\def\row{2}
\def\xs{1.1} 
\def\ys{0.3}
\pgfmathparse{\xs+0.*\figWidth} \let\xcb\pgfmathresult 
\pgfmathparse{\xcb+0.*\cbWidth} \let\xx\pgfmathresult
\draw(\xx,\ys) node[anchor=south west] {\getCB{colorbar_rainbow.pdf}{\cbWidth}};
\draw(\xx,\ys) node[anchor= west]{$0$};
\pgfmathparse{\xcb+0.47*\cbWidth} \let\xx\pgfmathresult
\draw(\xx,\ys) node[anchor= west]{$1.5$};
\pgfmathparse{\xcb+0.97*\cbWidth} \let\xx\pgfmathresult
\draw(\xx,\ys) node[anchor= west]{$3$};
\pgfmathparse{\xs+1.*\figWidth} \let\xcb\pgfmathresult 
\pgfmathparse{\xcb+0.*\cbWidth} \let\xx\pgfmathresult
\draw(\xx,\ys) node[anchor=south west] {\getCB{colorbar_rainbow.pdf}{\cbWidth}};
\draw(\xx,\ys) node[anchor= west]{$0$};
\pgfmathparse{\xcb+0.47*\cbWidth} \let\xx\pgfmathresult
\draw(\xx,\ys) node[anchor= west]{$0.7$};
\pgfmathparse{\xcb+0.97*\cbWidth} \let\xx\pgfmathresult
\draw(\xx,\ys) node[anchor= west]{$1.4$};
\pgfmathparse{\xs+2.*\figWidth} \let\xcb\pgfmathresult 
\pgfmathparse{\xcb+0.*\cbWidth} \let\xx\pgfmathresult
\draw(\xx,\ys) node[anchor=south west] {\getCB{colorbar_rainbow.pdf}{\cbWidth}};
\draw(\xx,\ys) node[anchor= west]{$0$};
\pgfmathparse{\xcb+0.47*\cbWidth} \let\xx\pgfmathresult
\draw(\xx,\ys) node[anchor= west]{$0.7$};
\pgfmathparse{\xcb+0.97*\cbWidth} \let\xx\pgfmathresult
\draw(\xx,\ys) node[anchor= west]{$1.4$};
\def\xs{0.8}
\def\ys{0.}
\pgfmathparse{\yb-1} \let\yy\pgfmathresult
\pgfmathparse{\xs+0.5*\figWidth} \let\xx\pgfmathresult
\draw(\xx,\yy) node[anchor=south]{$h$};
\pgfmathparse{\xs+1.5*\figWidth} \let\xx\pgfmathresult
\draw(\xx,\yy) node[anchor=south]{$c$};
\pgfmathparse{\xs+2.5*\figWidth} \let\xx\pgfmathresult
\draw(\xx,\yy) node[anchor=south]{$f$};
\pgfmathparse{\row+1} \let\tP\pgfmathresult
\draw(0,\tP) node[anchor=west,xshift=-0.1cm] {$t=15$};
\pgfmathparse{0.*\figWidth} \let\xf\pgfmathresult
\draw(\xf,\row) node[anchor=south west,xshift=\xs cm,yshift=\ys cm] {\trimfig{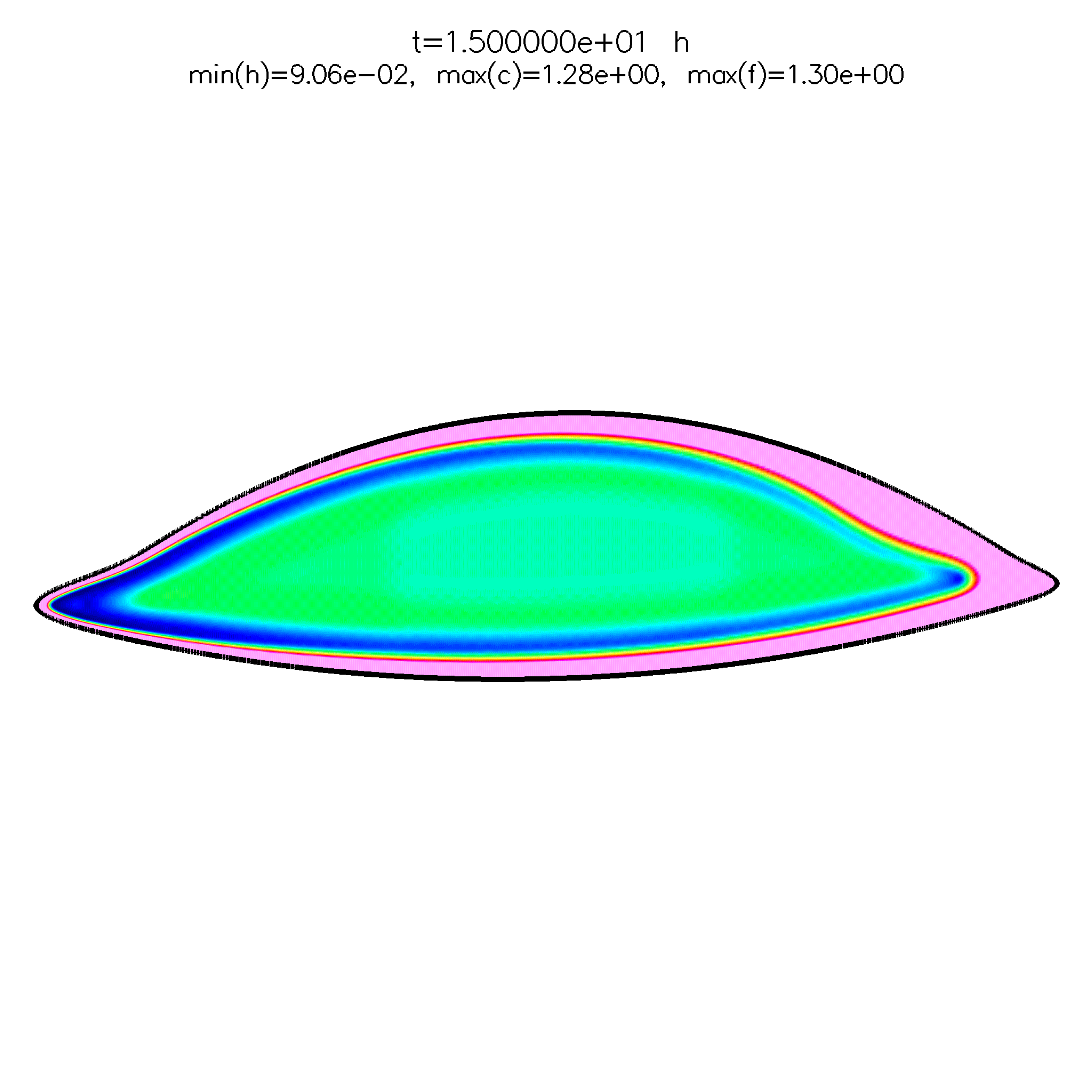}{\figWidth}};
\pgfmathparse{\xs+0.5*\figWidth} \let\xx\pgfmathresult
\draw(\xx,\row) node[anchor=north]{$\min(h)=$ 9.06e-2};
\pgfmathparse{1.*\figWidth} \let\xf\pgfmathresult
\draw(\xf,\row) node[anchor=south west,xshift=\xs cm,yshift=\ys cm] {\trimfig{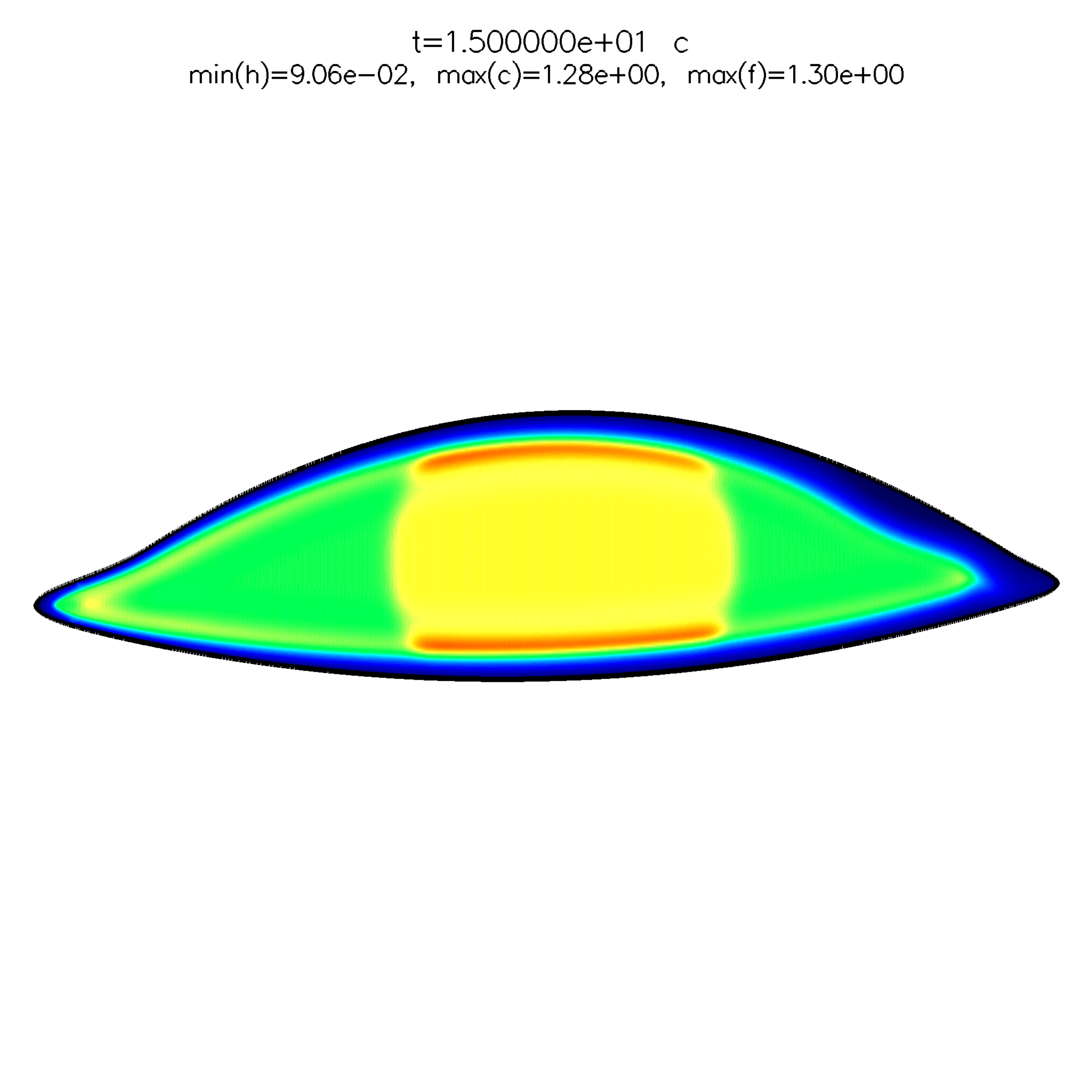}{\figWidth}};
\pgfmathparse{\xs+1.5*\figWidth} \let\xx\pgfmathresult
\draw(\xx,\row) node[anchor=north]{$\max(c)=$ 1.28};
\pgfmathparse{2.*\figWidth} \let\xf\pgfmathresult
\draw(\xf,\row) node[anchor=south west,xshift=\xs cm,yshift=\ys cm] {\trimfig{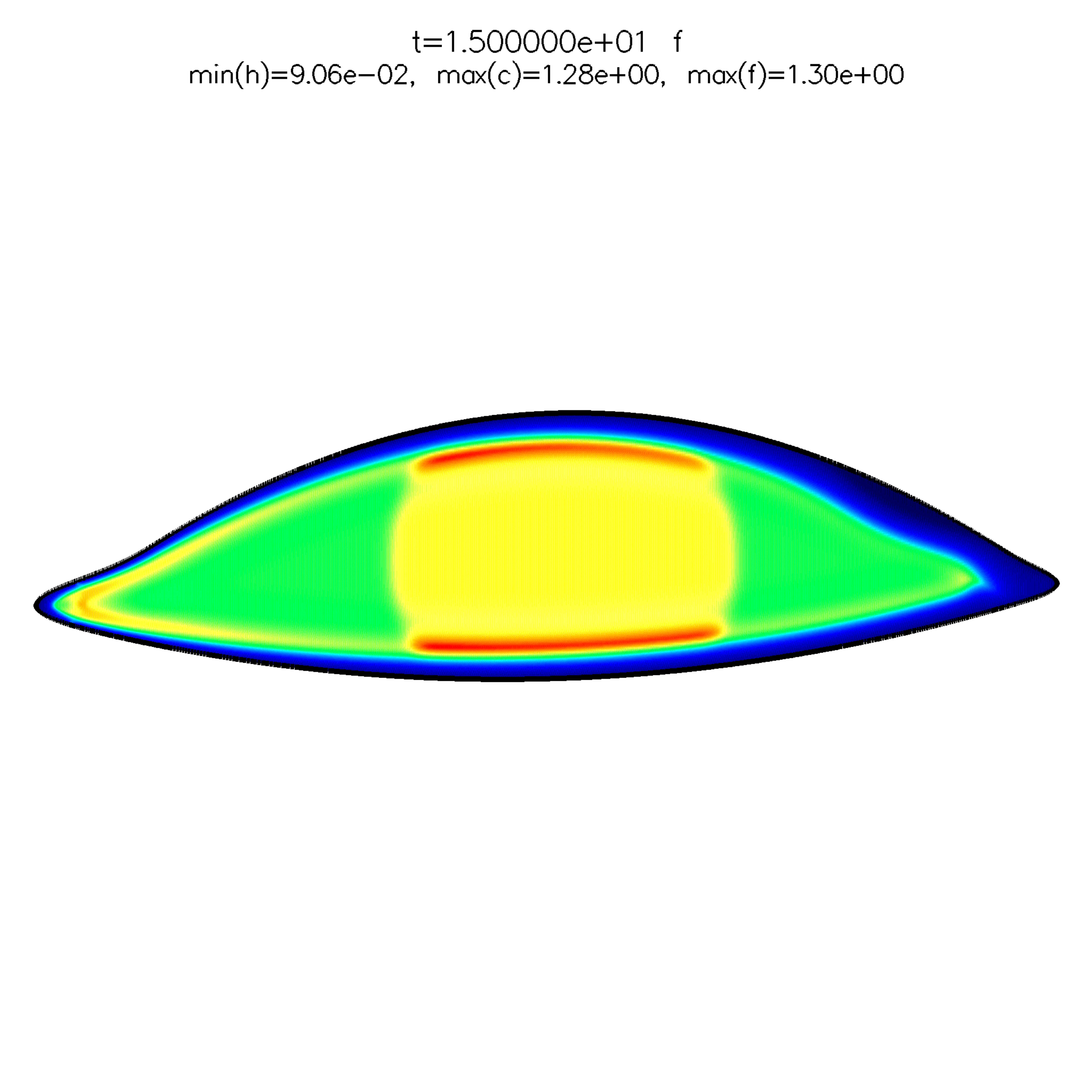}{\figWidth}};
\pgfmathparse{\xs+2.5*\figWidth} \let\xx\pgfmathresult
\draw(\xx,\row) node[anchor=north]{$\max(f)=$ 1.30};
\pgfmathparse{\row+\rowHeight} \let\row\pgfmathresult
\pgfmathparse{\row+1} \let\tP\pgfmathresult
\draw(0,\tP) node[anchor=west,xshift=-0.1cm] {$t=10$};
\pgfmathparse{0.*\figWidth} \let\xf\pgfmathresult
\draw(\xf,\row) node[anchor=south west,xshift=\xs cm,yshift=\ys cm] {\trimfig{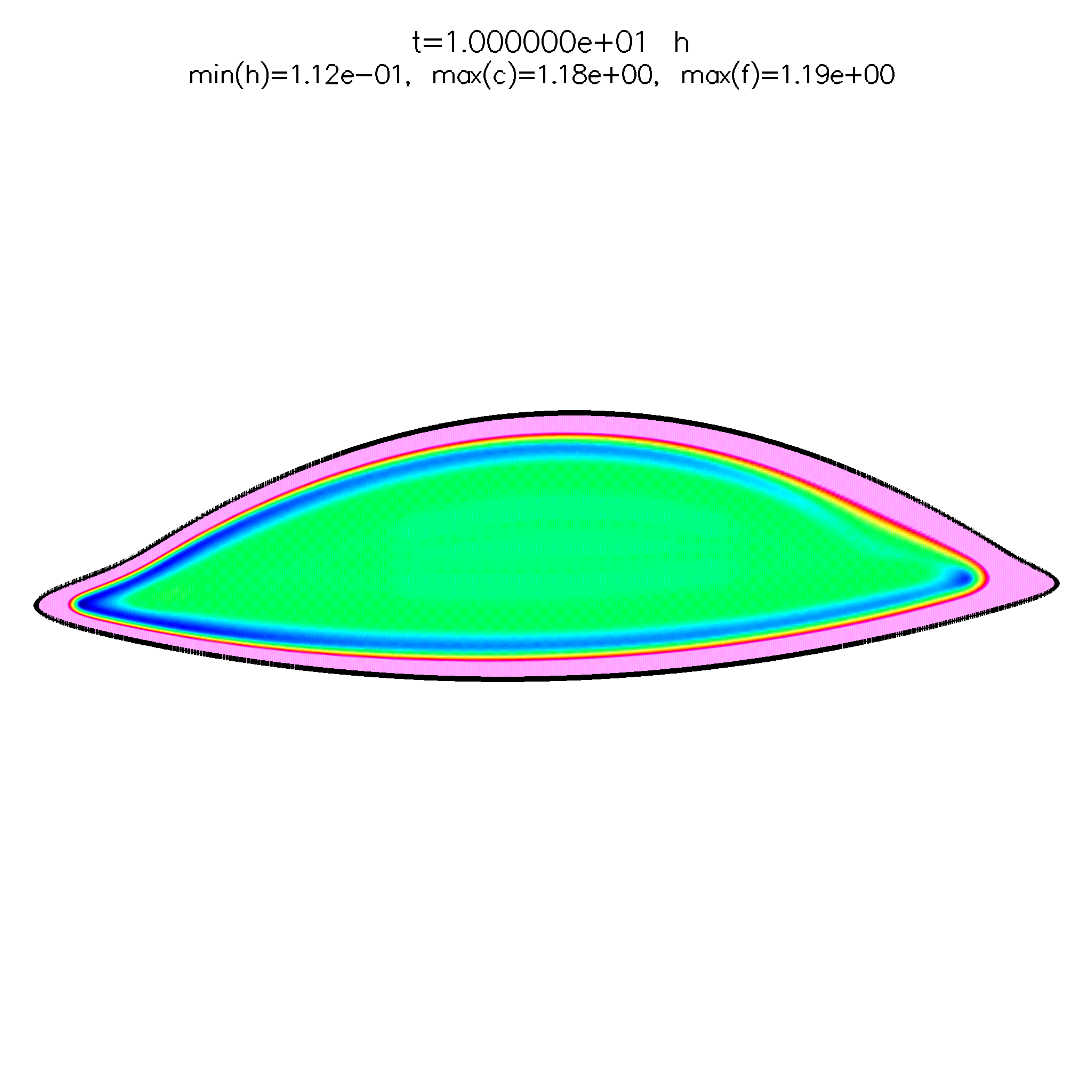}{\figWidth}};
\pgfmathparse{\xs+0.5*\figWidth} \let\xx\pgfmathresult
\draw(\xx,\row) node[anchor=north]{$\min(h)=$ 1.12e-1};
\pgfmathparse{1.*\figWidth} \let\xf\pgfmathresult
\draw(\xf,\row) node[anchor=south west,xshift=\xs cm,yshift=\ys cm] {\trimfig{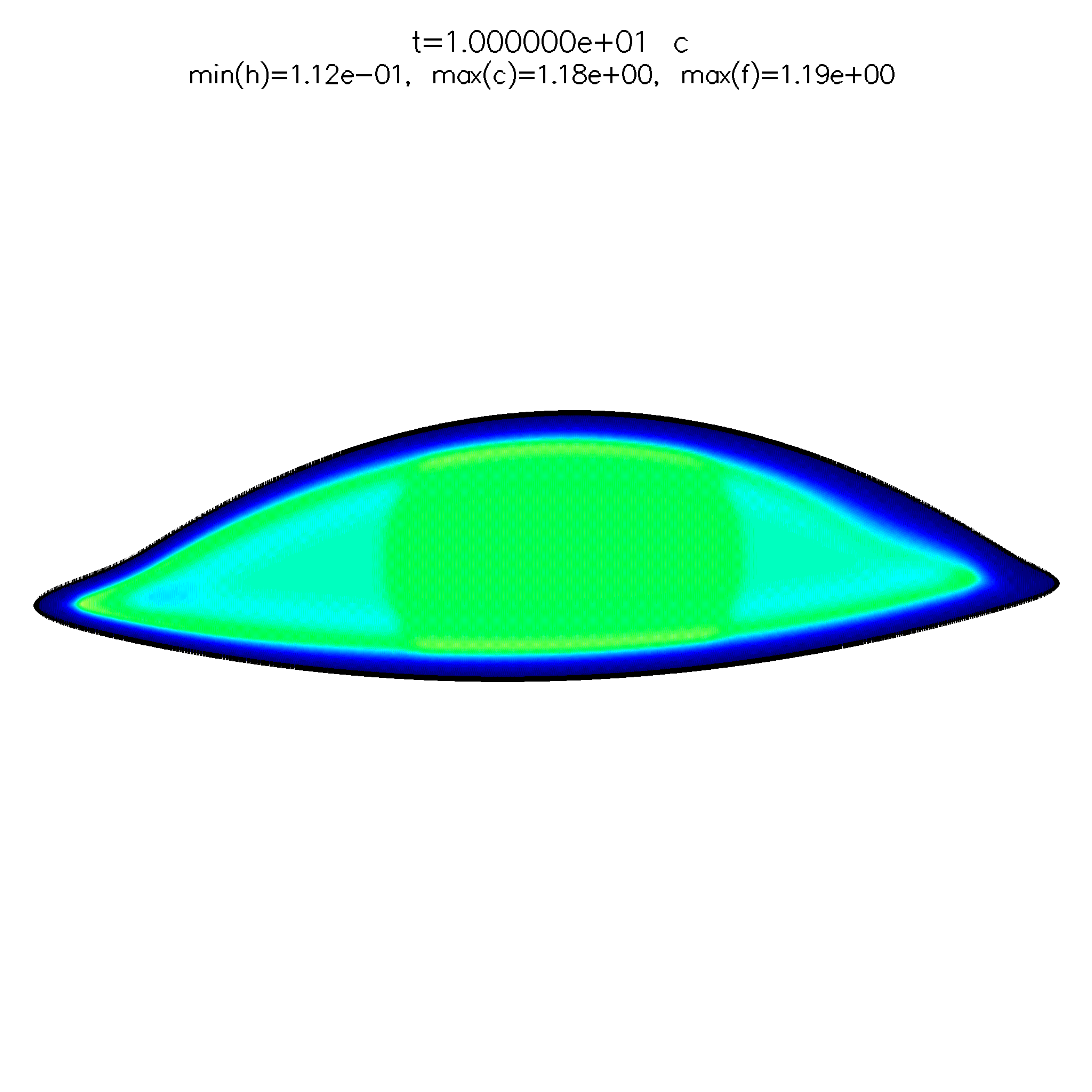}{\figWidth}};
\pgfmathparse{\xs+1.5*\figWidth} \let\xx\pgfmathresult
\draw(\xx,\row) node[anchor=north]{$\max(c)=$ 1.18};
\pgfmathparse{2.*\figWidth} \let\xf\pgfmathresult
\draw(\xf,\row) node[anchor=south west,xshift=\xs cm,yshift=\ys cm] {\trimfig{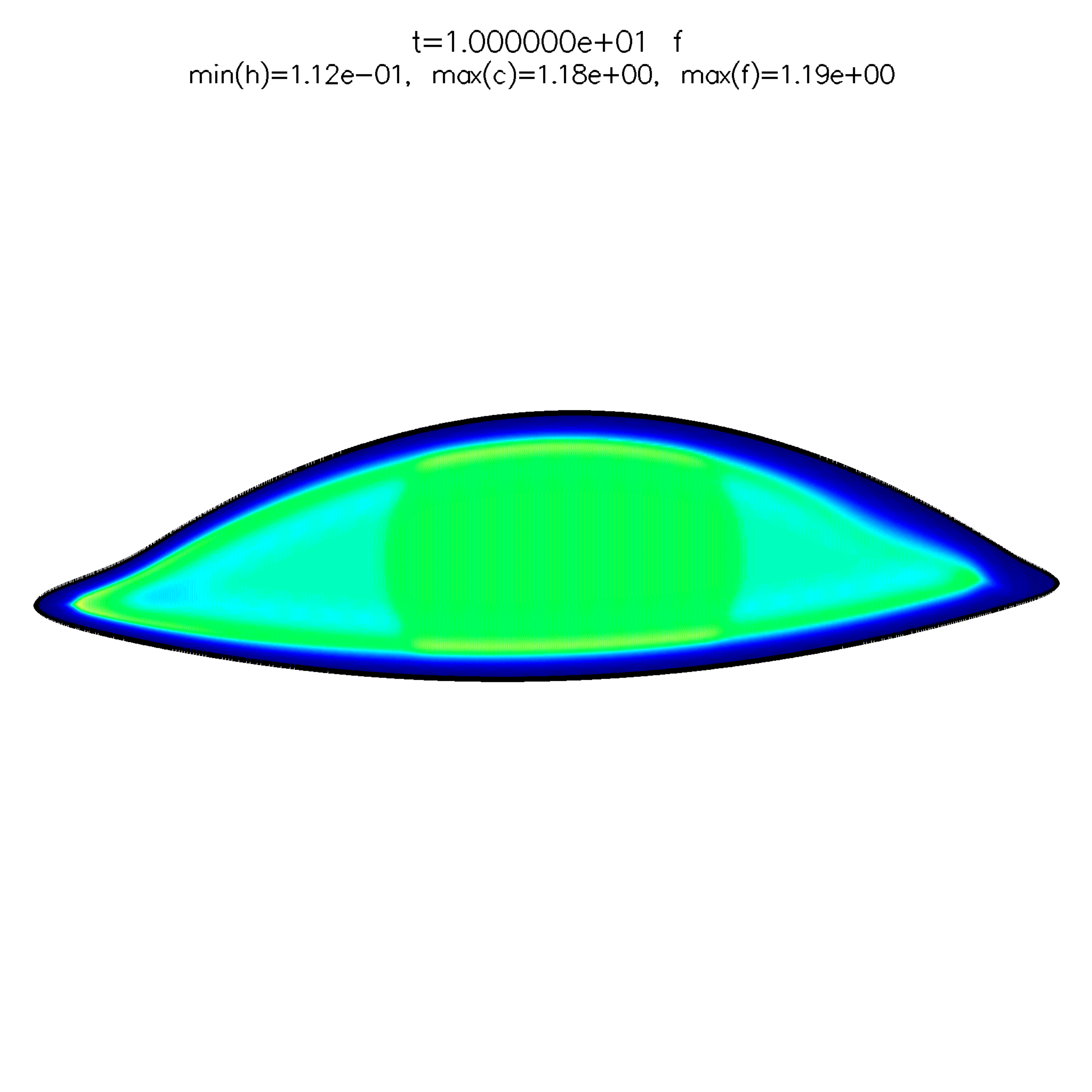}{\figWidth}};
\pgfmathparse{\xs+2.5*\figWidth} \let\xx\pgfmathresult
\draw(\xx,\row) node[anchor=north]{$\max(f)=$ 1.19};
\pgfmathparse{\row+\rowHeight} \let\row\pgfmathresult
\pgfmathparse{\row+1} \let\tP\pgfmathresult
\draw(0,\tP) node[anchor=west,xshift=-0.1cm] {$t=5$};
\pgfmathparse{0.*\figWidth} \let\xf\pgfmathresult
\draw(\xf,\row) node[anchor=south west,xshift=\xs cm,yshift=\ys cm] {\trimfig{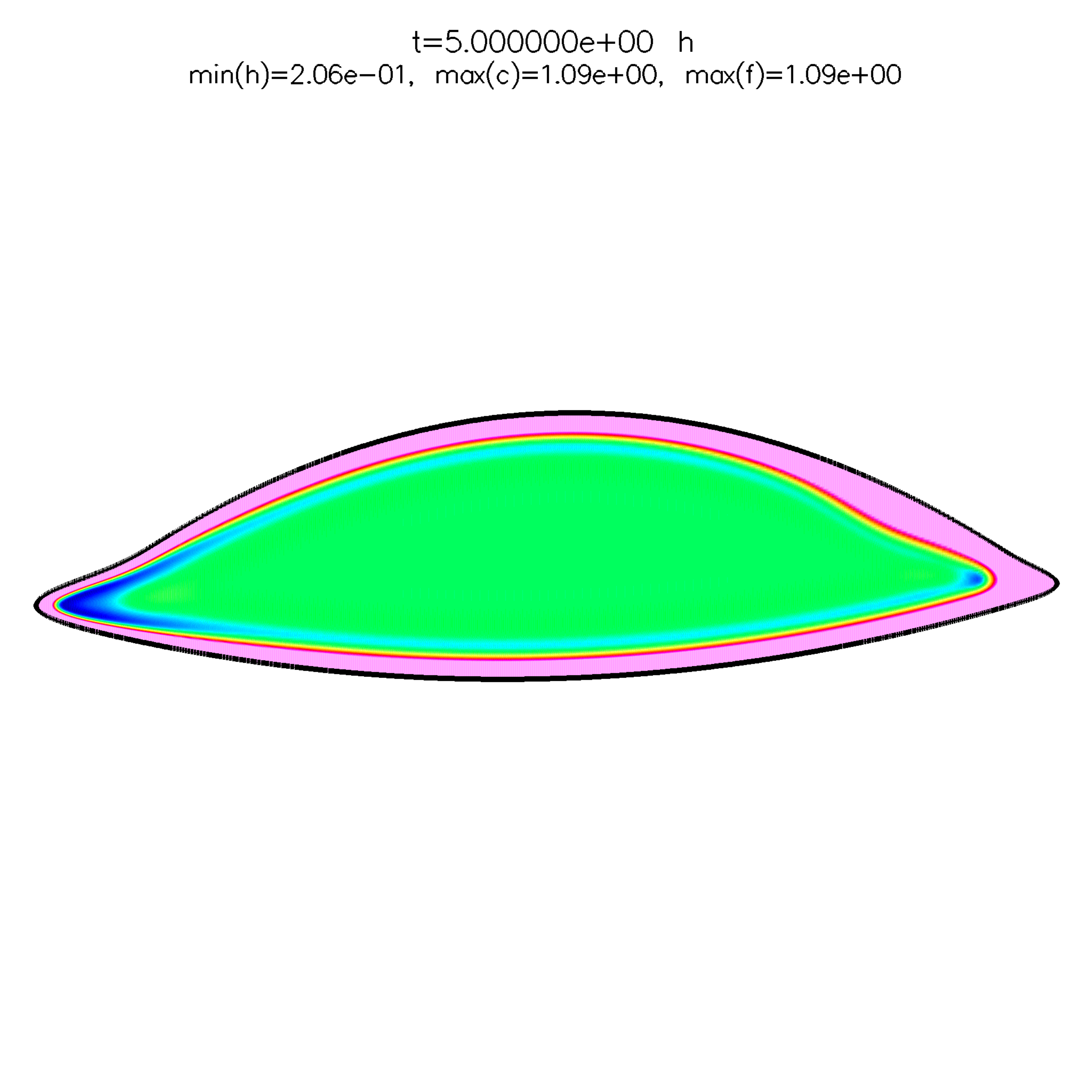}{\figWidth}};
\pgfmathparse{\xs+0.5*\figWidth} \let\xx\pgfmathresult
\draw(\xx,\row) node[anchor=north]{$\min(h)=$ 2.06e-1};
\pgfmathparse{1.*\figWidth} \let\xf\pgfmathresult
\draw(\xf,\row) node[anchor=south west,xshift=\xs cm,yshift=\ys cm] {\trimfig{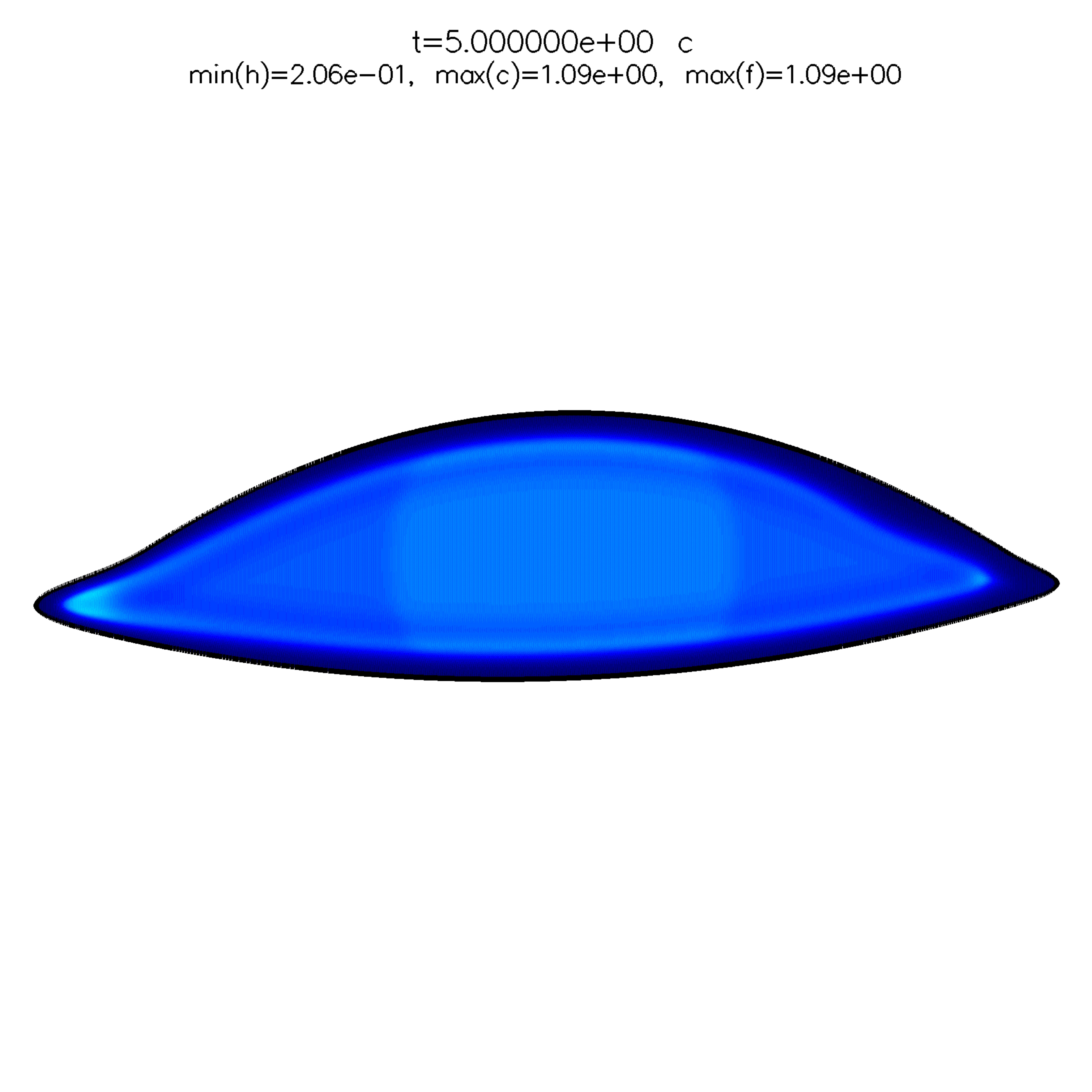}{\figWidth}};
\pgfmathparse{\xs+1.5*\figWidth} \let\xx\pgfmathresult
\draw(\xx,\row) node[anchor=north]{$\max(c)=$ 1.09};
\pgfmathparse{2.*\figWidth} \let\xf\pgfmathresult
\draw(\xf,\row) node[anchor=south west,xshift=\xs cm,yshift=\ys cm] {\trimfig{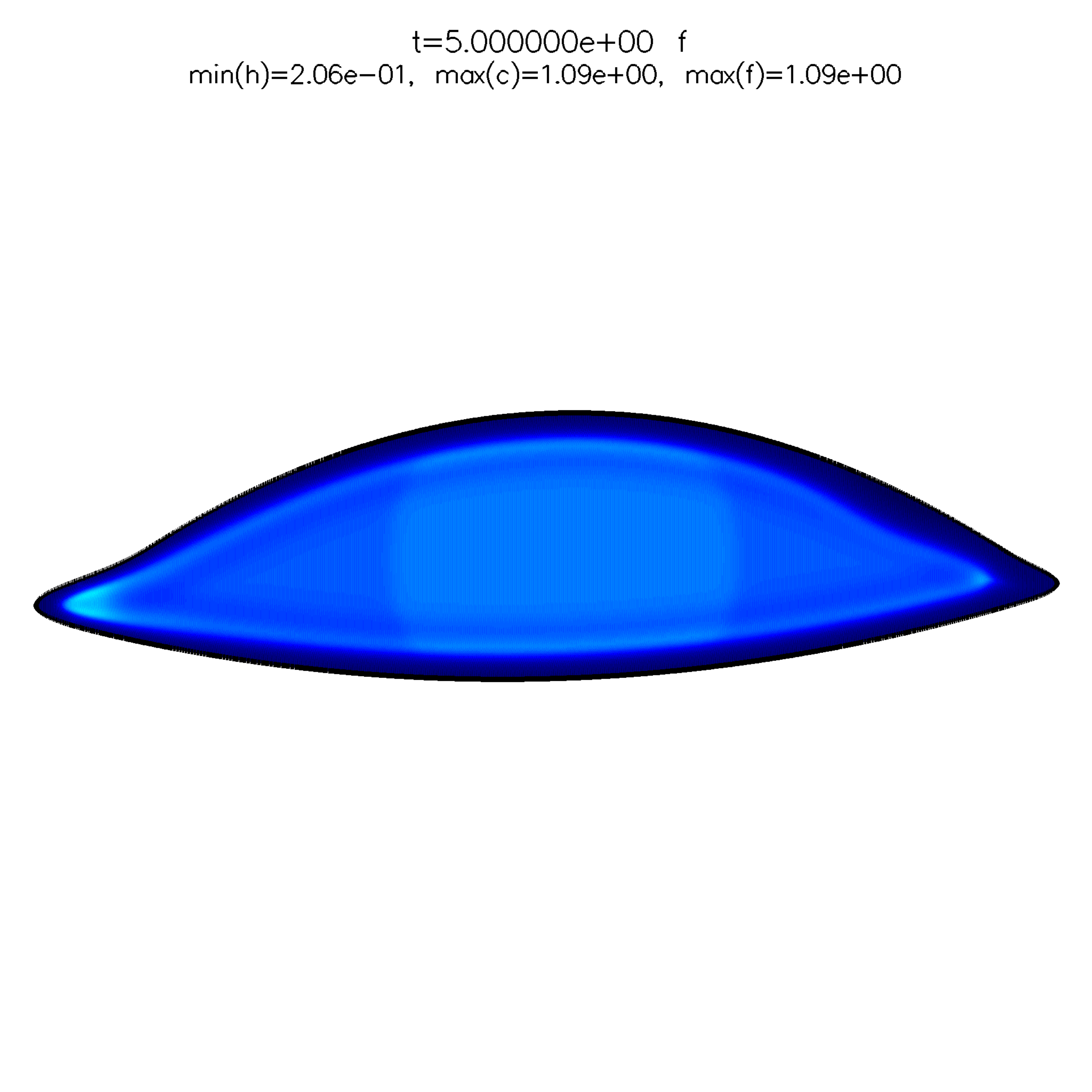}{\figWidth}};
\pgfmathparse{\xs+2.5*\figWidth} \let\xx\pgfmathresult
\draw(\xx,\row) node[anchor=north]{$\max(f)=$ 1.09};
\pgfmathparse{\row+\rowHeight} \let\row\pgfmathresult
%
\end{tikzpicture}
\end{center}
\caption{Variable permeability and evaporation is 4 micron/min, $h$, $c$ and $f$}
\label{fig:hcf_Pc(x,y)_4mpm}
\end{sidewaysfigure}
}

Fig.~\ref{fig:Intensity_Pc(x,y)_4mpm} shows the distribution of fluorescent intensity over the
exposed ocular surface after using (\ref{eqn:I-of-f}) together with the computed $f$ and
$h$. In either case, the meniscus stays at essentially constant intensity because $h$ and $f$ change little.
The dilute case on the left has nearly no change in the central region inside the black line, but previously seen changes in the width of the meniscus is made visible by the fluorescence \citep{MakiBraun10b,LiBraun12}.
The central region is similar to the impermeable case, and even though osmosis occurs,
there is little tangential flow.
As a result, the product $hf$ remains nearly constant from mass conservation, so that
$I$ remains nearly constant \cite{BraunEtal14}.
For the self-quenching case (right),
besides the black line, some overall dimming in the central region can be seen.  Enhanced dimming is seen
over the cornea, where the permeability is lower and the thinning is greater
(see previous figure).  Experimentally, we would expect the difference in
dimming over the cornea compared to that over the conjunctiva to be unobservable due to the refective
nature of the stroma beneath the conjunctiva.  The drop in intensity over the central region
is because $f$ is larger than unity and
the dominance of the denominator in Eq.~\ref{eqn:I-of-f} causes the intensity to drop with inverse square
power of $f$ \citep{NicholsEtal12,BraunEtal14}.
{
\def\nRows{3}  
\def\nColumns{2}  
\begin{figure}[htb]
\begin{center}
\begin{tikzpicture}[scale=1]
\pgfmathparse{\columnWidth*\nColumns} \let\xb\pgfmathresult
\pgfmathparse{\rowHeight*\nRows+1} \let\yb\pgfmathresult 
\useasboundingbox (0.0,0.0) rectangle (\xb,\yb);  
\def\row{1}
\def\xs{1.1} 
\def\ys{0.3}
\pgfmathparse{\xs+0.5*\figWidth} \let\xcb\pgfmathresult 
\pgfmathparse{\xcb+0.*\cbWidth} \let\xx\pgfmathresult
\draw(\xx,\ys) node[anchor=south west] {\getCB{colorbar_green.pdf}{\cbWidth}};
\draw(\xx,\ys) node[anchor= west]{$0$};
\pgfmathparse{\xcb+0.47*\cbWidth} \let\xx\pgfmathresult
\draw(\xx,\ys) node[anchor= west]{$50$};
\pgfmathparse{\xcb+0.97*\cbWidth} \let\xx\pgfmathresult
\draw(\xx,\ys) node[anchor= west]{$100$};
\def\xs{0.8}
\def\ys{0.}
\pgfmathparse{\xs+0.5*\figWidth} \let\xx\pgfmathresult
\pgfmathparse{\yb-1} \let\yy\pgfmathresult
\draw(\xx,\yy) node[anchor=south]{Intensity $I$ with $f_0=\frac{1}{3}$};
\pgfmathparse{\xs+1.5*\figWidth} \let\xx\pgfmathresult
\draw(\xx,\yy) node[anchor=south]{Intensity $I$ with $f_0=2$};
\pgfmathparse{\row+1} \let\tP\pgfmathresult
\draw(0,\tP) node[anchor=west,xshift=-0.1cm] {$t=15$};
\draw(0,\row) node[anchor=south west,xshift=\xs cm,yshift=\ys cm] {\trimfig{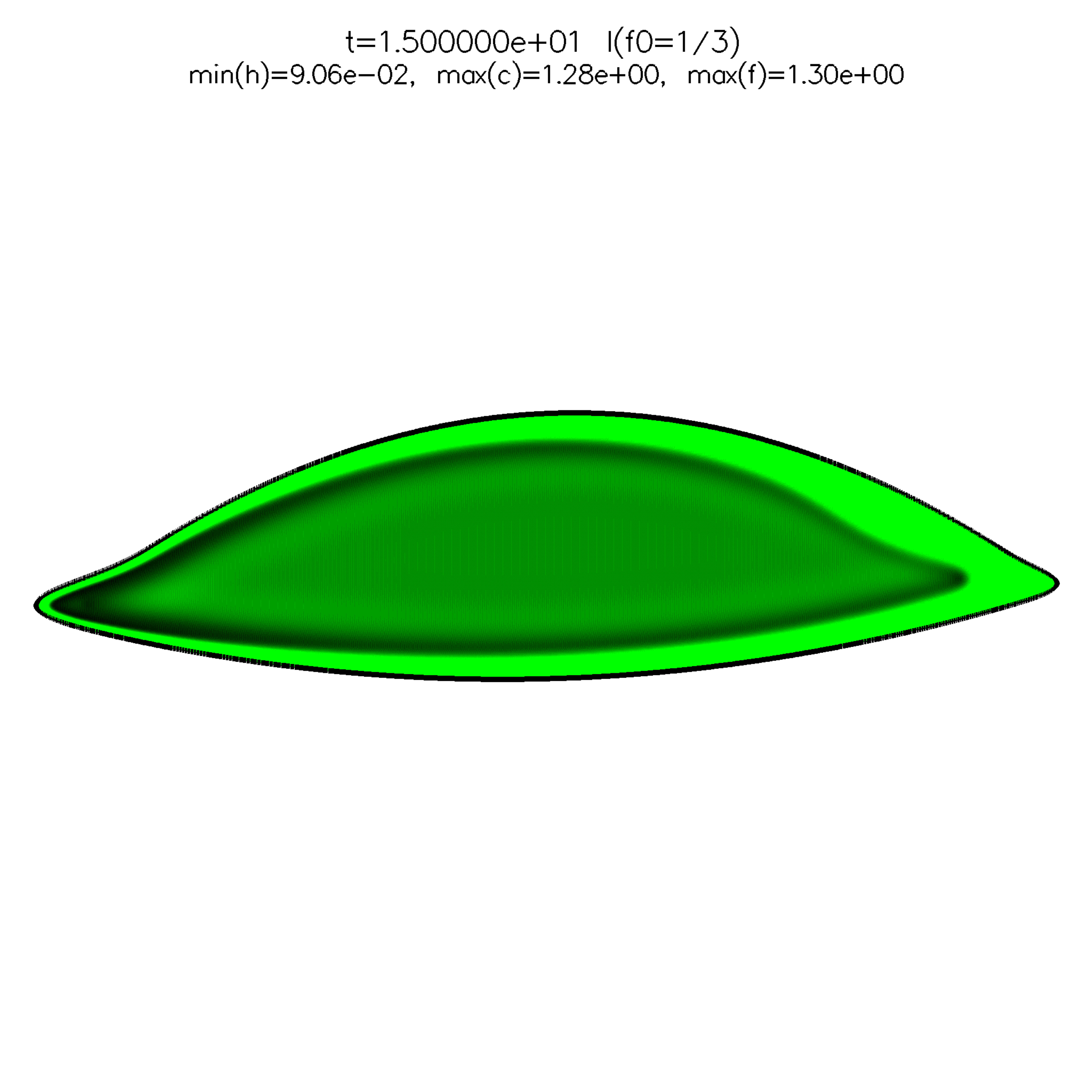}{\figWidth}};
\draw(\figWidth,\row) node[anchor=south west,xshift=\xs cm,yshift=\ys cm] {\trimfig{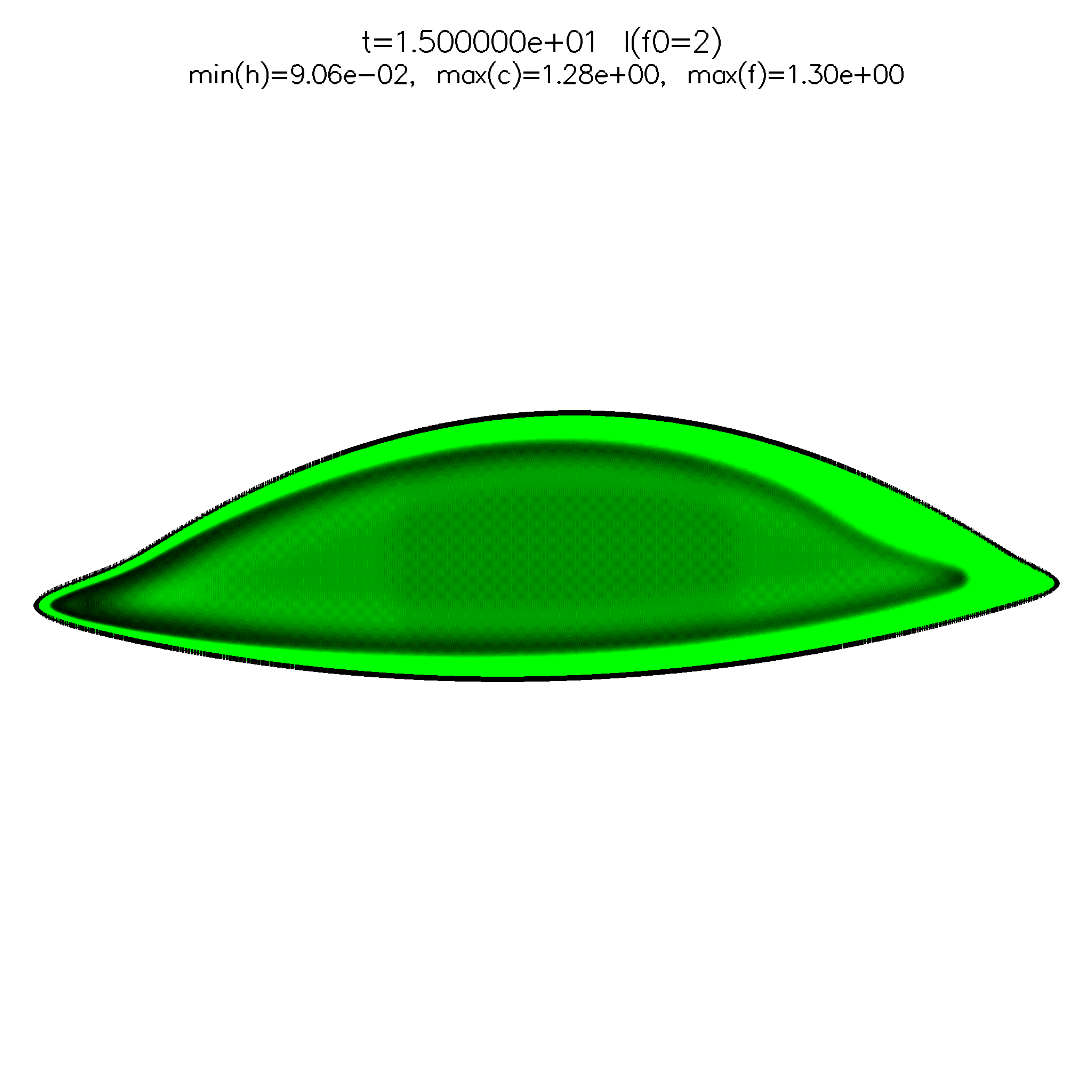}{\figWidth}};
\pgfmathparse{\row+\rowHeight} \let\row\pgfmathresult
\pgfmathparse{\row+1} \let\tP\pgfmathresult
\draw(0,\tP) node[anchor=west,xshift=-0.1cm] {$t=10$};
\draw(0,\row) node[anchor=south west,xshift=\xs cm,yshift=\ys cm] {\trimfig{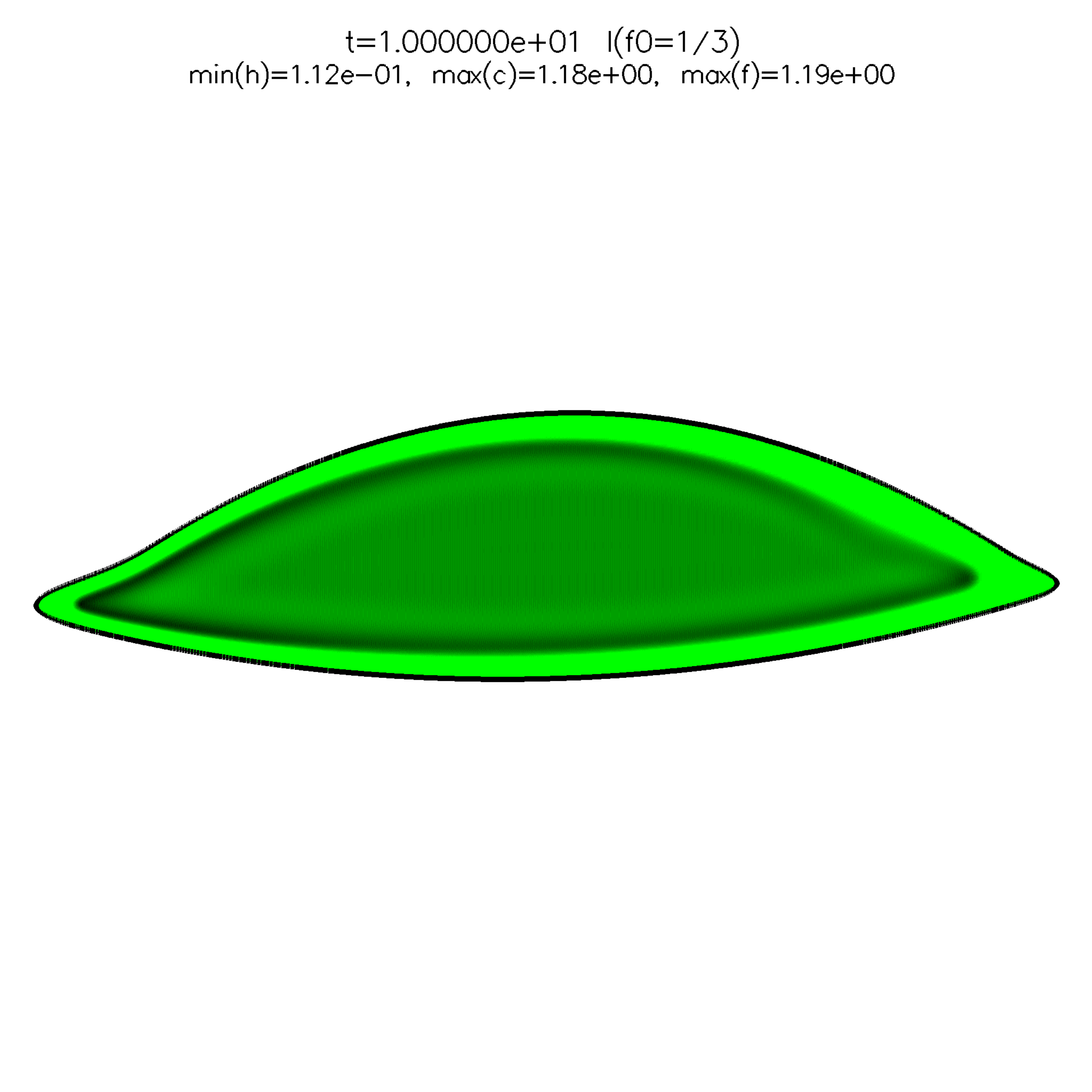}{\figWidth}};
\draw(\figWidth,\row) node[anchor=south west,xshift=\xs cm,yshift=\ys cm] {\trimfig{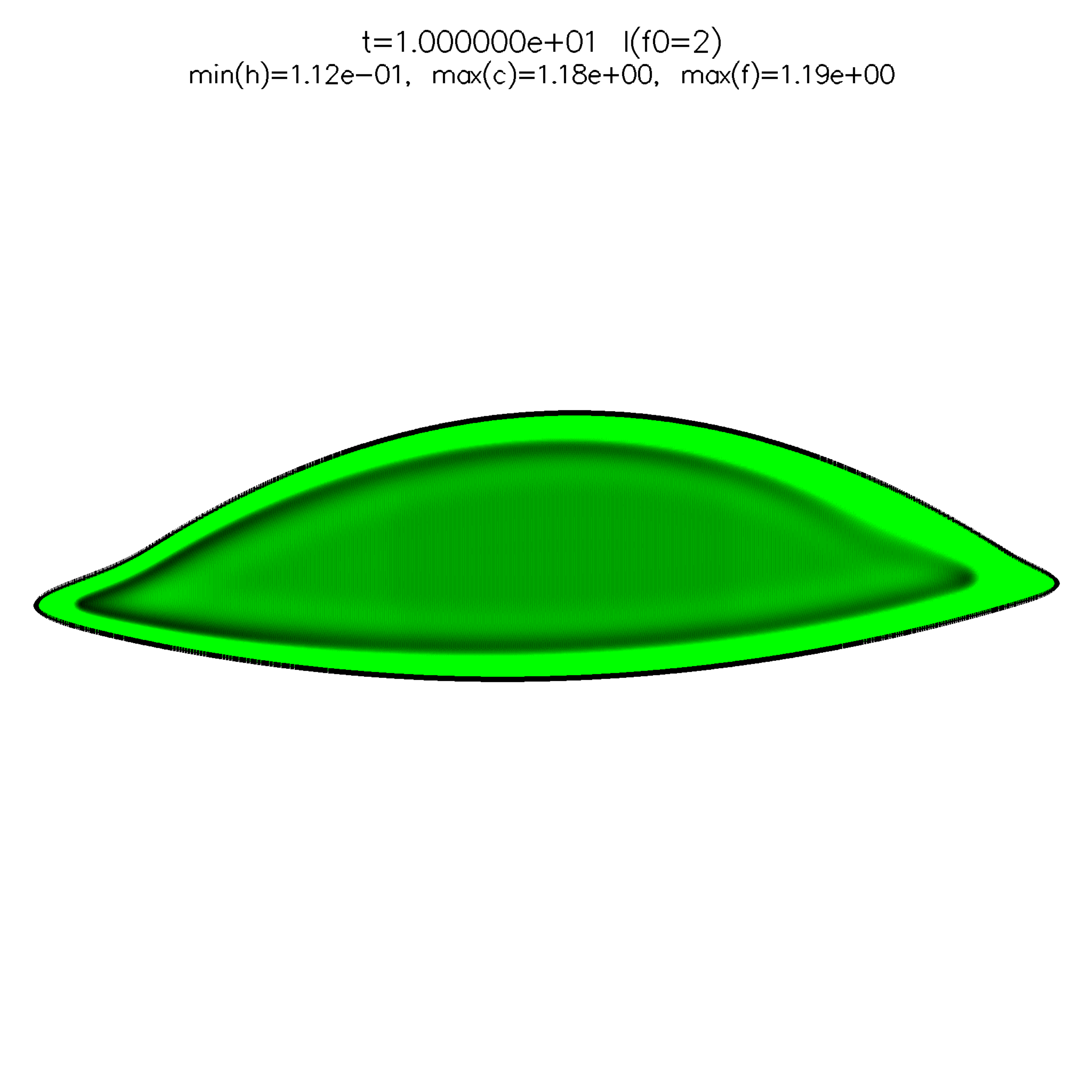}{\figWidth}};
\pgfmathparse{\row+\rowHeight} \let\row\pgfmathresult
\pgfmathparse{\row+1} \let\tP\pgfmathresult
\draw(0,\tP) node[anchor=west,xshift=-0.1cm] {$t=5$};
\draw(0,\row) node[anchor=south west,xshift=\xs cm,yshift=\ys cm] {\trimfig{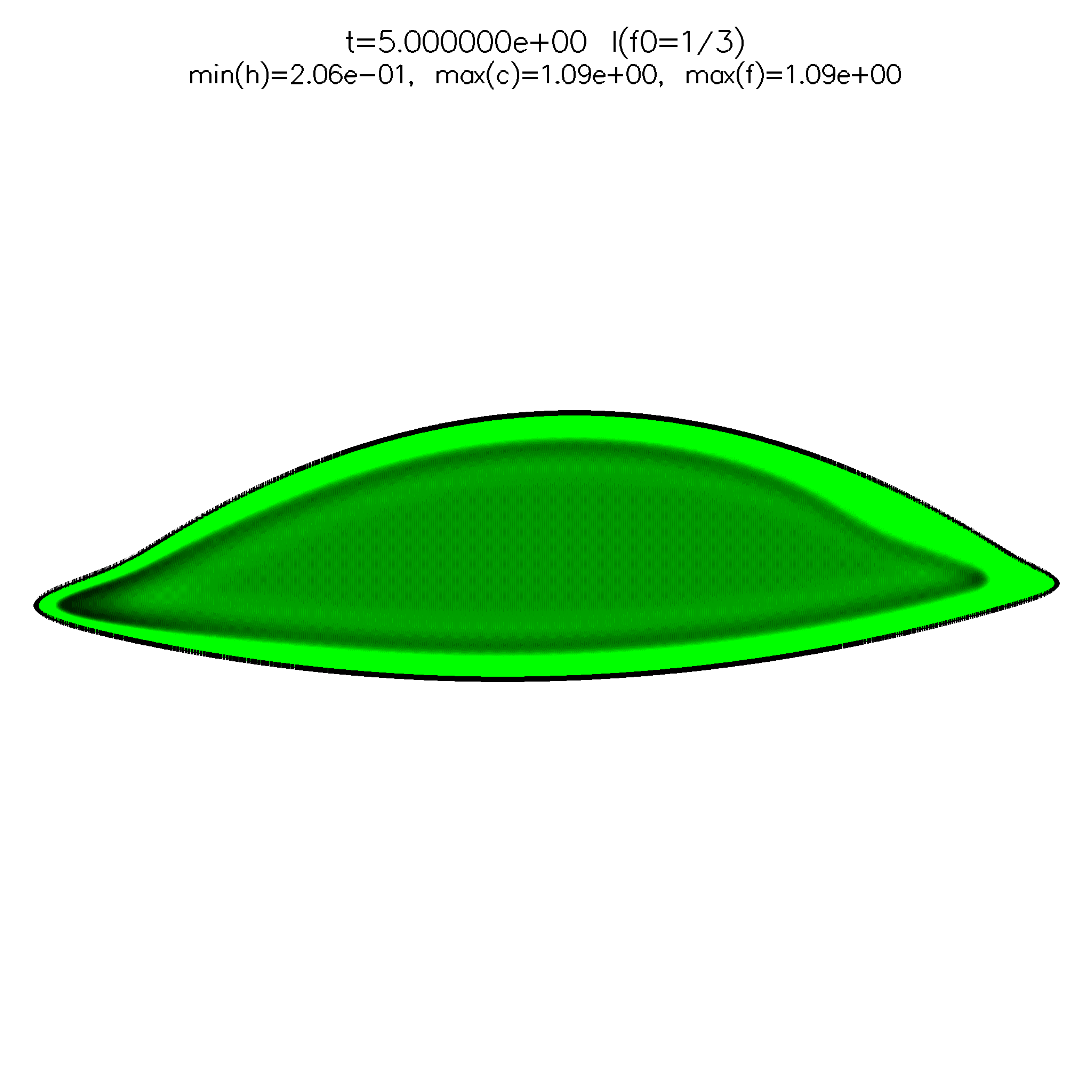}{\figWidth}};
\draw(\figWidth,\row) node[anchor=south west,xshift=\xs cm,yshift=\ys cm] {\trimfig{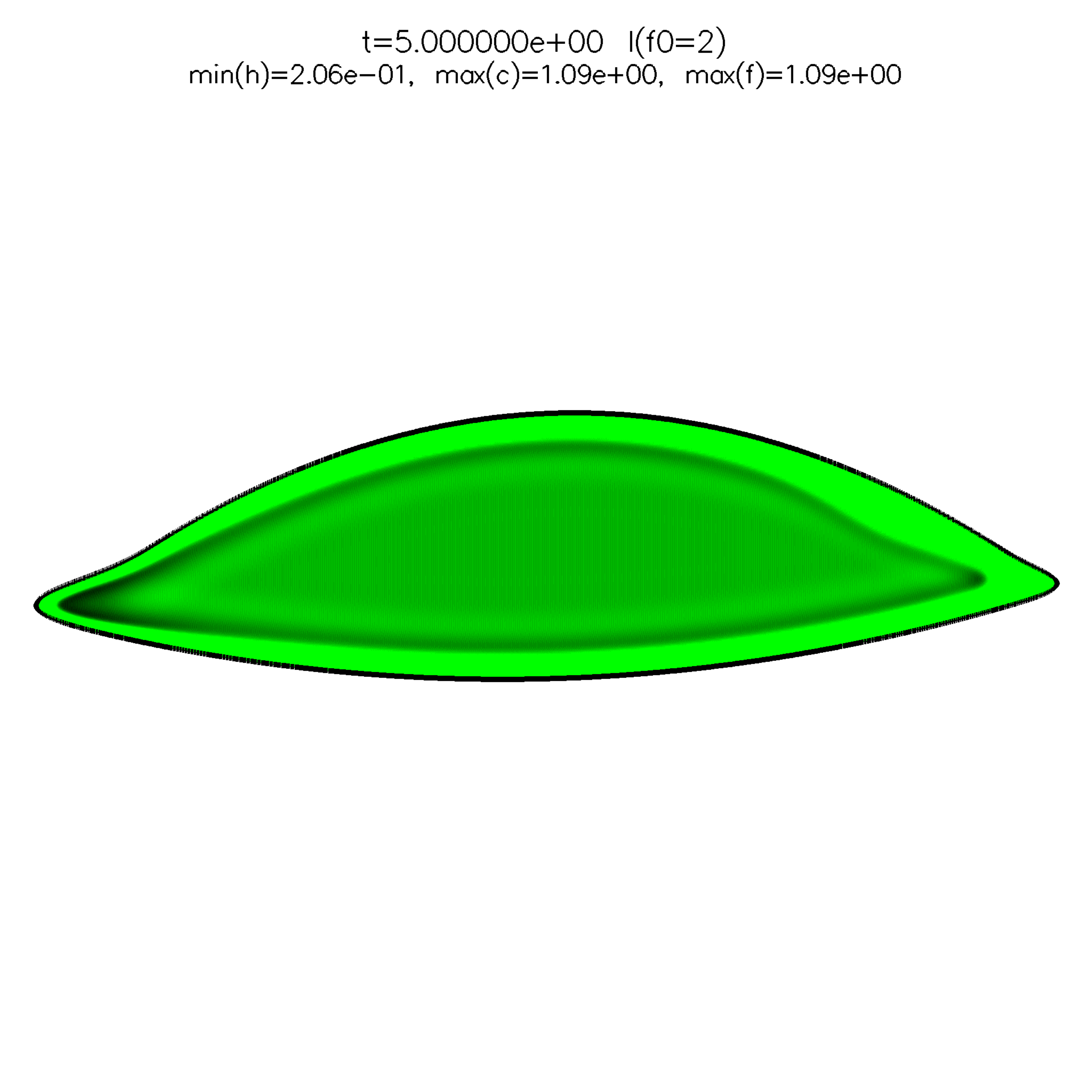}{\figWidth}};
\pgfmathparse{\row+\rowHeight} \let\row\pgfmathresult
%
\end{tikzpicture}
\end{center}
\caption{Variable permeability and evaporation is 4 micron/min, $I$ with $f_0=1/3$ and $f_0=2$}
\label{fig:Intensity_Pc(x,y)_4mpm}
\end{figure}
}

Despite the similarities of the solute plots, the intensity plots had some differences.  These differences
are highlighted by the difference between the computed thickness $h$ and the approximation to the thickness from
$a\sqrt{I}$.  Results for the two different initial fluorescein concentrations are shown in
Figure~\ref{fig:CompareMethods_Pc(x,y)_4mpm}.  In either the dilute case with $f_0=1/3$ (left) or the quenching regime with $f_0=2$, the menisci have a large difference $|h-a\sqrt{I}|$ because $a$ was chosen to work in the initially flat region outside the mensici.  In the dilute case (left), the difference between computed and estimated values is large as well.  In the quenching regime, the difference remains small in almost all of the black line and central regions.  Near the nasal canthus where the black lines meet is the exception where
the error becomes larger; this is not an area of measurement interest, to our knowledge.
Thus the approximation to the thickness based on the intensity appears to be a good estimate of the thickness
where the flow in the central region is small \cite{NicholsEtal12}.
{
\def\nRows{3}  
\def\nColumns{2}  
\begin{figure}[htb]
\begin{center}
\begin{tikzpicture}[scale=1]
\pgfmathparse{\columnWidth*\nColumns} \let\xb\pgfmathresult
\pgfmathparse{\rowHeight*\nRows+1} \let\yb\pgfmathresult 
\useasboundingbox (0.0,0.0) rectangle (\xb,\yb);  
\def\row{1}
\def\xs{1.1} 
\def\ys{0.3}
\pgfmathparse{\xs+0.5*\figWidth} \let\xcb\pgfmathresult 
\pgfmathparse{\xcb+0.*\cbWidth} \let\xx\pgfmathresult
\draw(\xx,\ys) node[anchor=south west] {\getCB{colorbar_rainbow.pdf}{\cbWidth}};
\draw(\xx,\ys) node[anchor= west]{$0$};
\pgfmathparse{\xcb+0.47*\cbWidth} \let\xx\pgfmathresult
\draw(\xx,\ys) node[anchor= west]{$0.05$};
\pgfmathparse{\xcb+0.97*\cbWidth} \let\xx\pgfmathresult
\draw(\xx,\ys) node[anchor= west]{$0.1$};
\def\xs{0.8}
\def\ys{0.}
\pgfmathparse{\xs+0.5*\figWidth} \let\xx\pgfmathresult
\pgfmathparse{\yb-1} \let\yy\pgfmathresult
\draw(\xx,\yy) node[anchor=south]{$|h-a\sqrt{I}|$ with $f_0=\frac{1}{3}$};
\pgfmathparse{\xs+1.5*\figWidth} \let\xx\pgfmathresult
\draw(\xx,\yy) node[anchor=south]{$|h-a\sqrt{I}|$ with $f_0=2$};
\pgfmathparse{\row+1} \let\tP\pgfmathresult
\draw(0,\tP) node[anchor=west,xshift=-0.1cm] {$t=15$};
\draw(0,\row) node[anchor=south west,xshift=\xs cm,yshift=\ys cm] {\trimfig{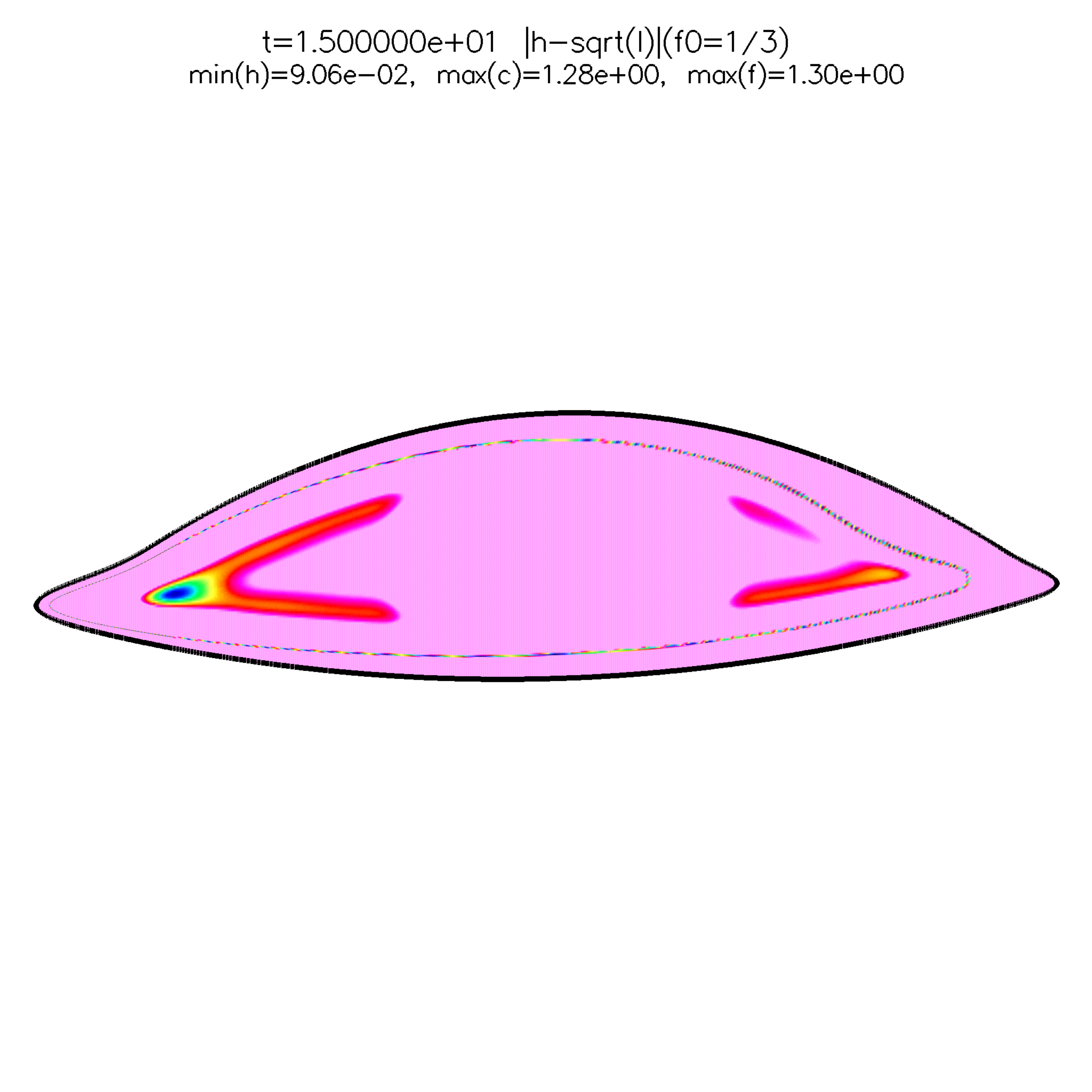}{\figWidth}};
\draw(\figWidth,\row) node[anchor=south west,xshift=\xs cm,yshift=\ys cm] {\trimfig{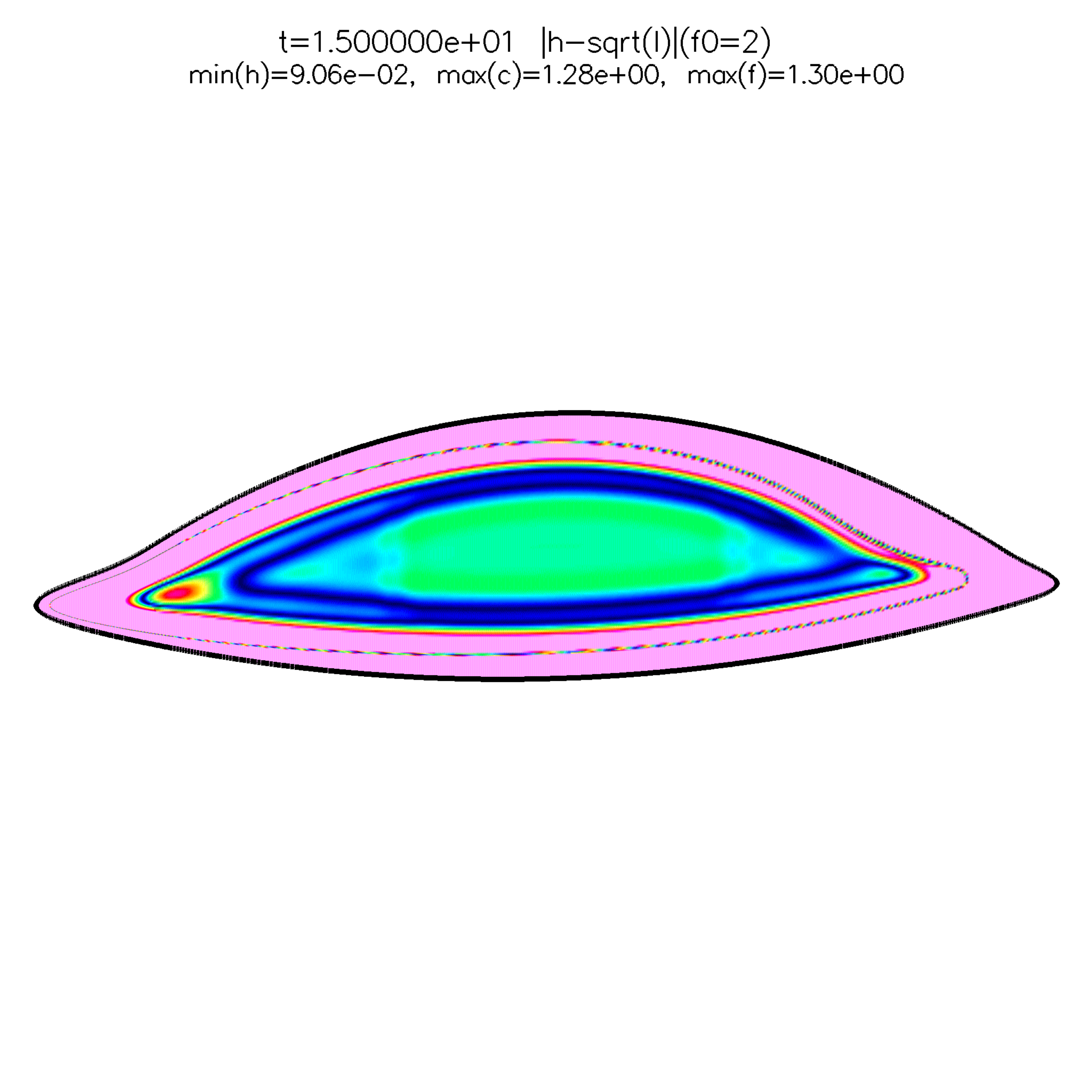}{\figWidth}};
\pgfmathparse{\row+\rowHeight} \let\row\pgfmathresult
\pgfmathparse{\row+1} \let\tP\pgfmathresult
\draw(0,\tP) node[anchor=west,xshift=-0.1cm] {$t=10$};
\draw(0,\row) node[anchor=south west,xshift=\xs cm,yshift=\ys cm] {\trimfig{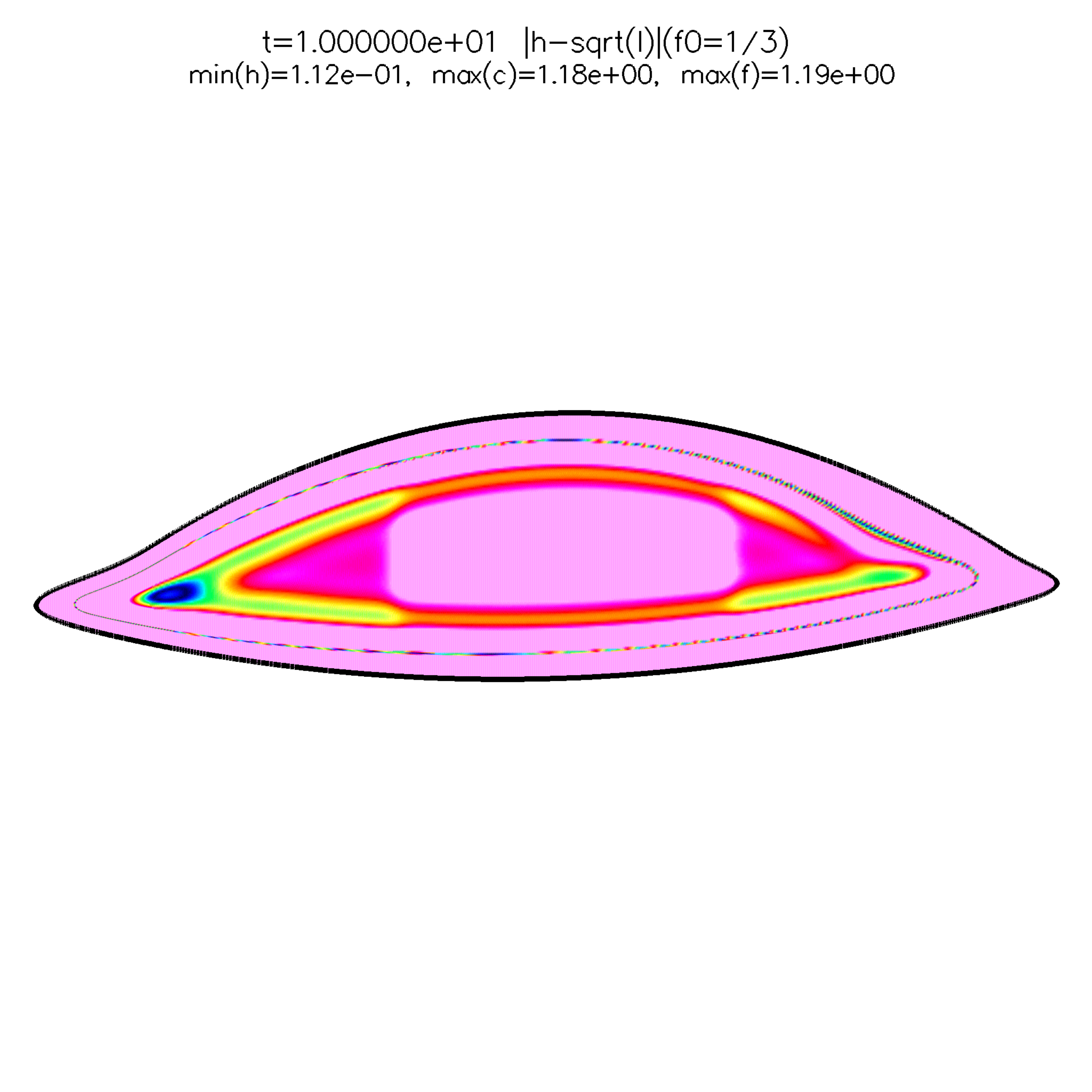}{\figWidth}};
\draw(\figWidth,\row) node[anchor=south west,xshift=\xs cm,yshift=\ys cm] {\trimfig{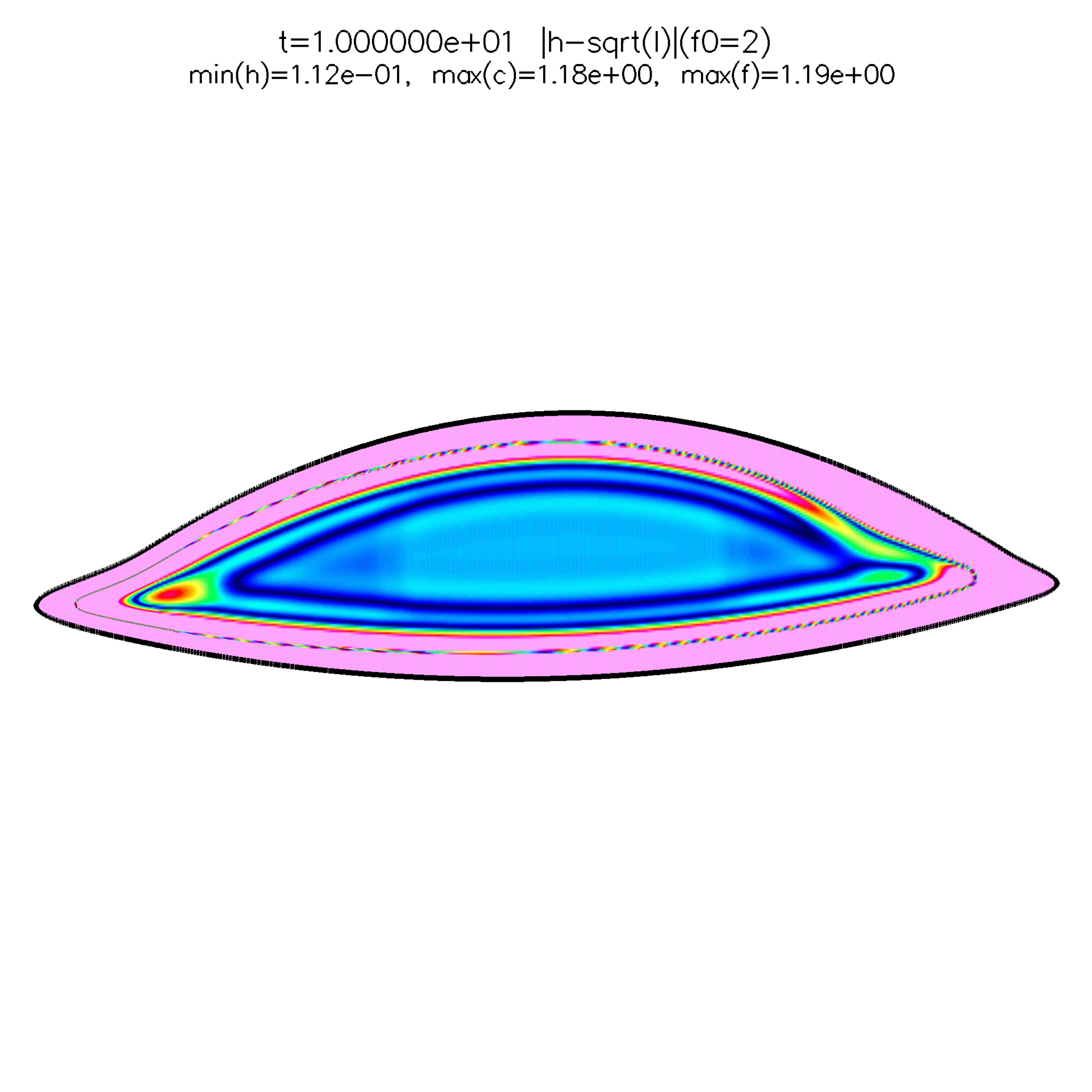}{\figWidth}};
\pgfmathparse{\row+\rowHeight} \let\row\pgfmathresult
\pgfmathparse{\row+1} \let\tP\pgfmathresult
\draw(0,\tP) node[anchor=west,xshift=-0.1cm] {$t=5$};
\draw(0,\row) node[anchor=south west,xshift=\xs cm,yshift=\ys cm] {\trimfig{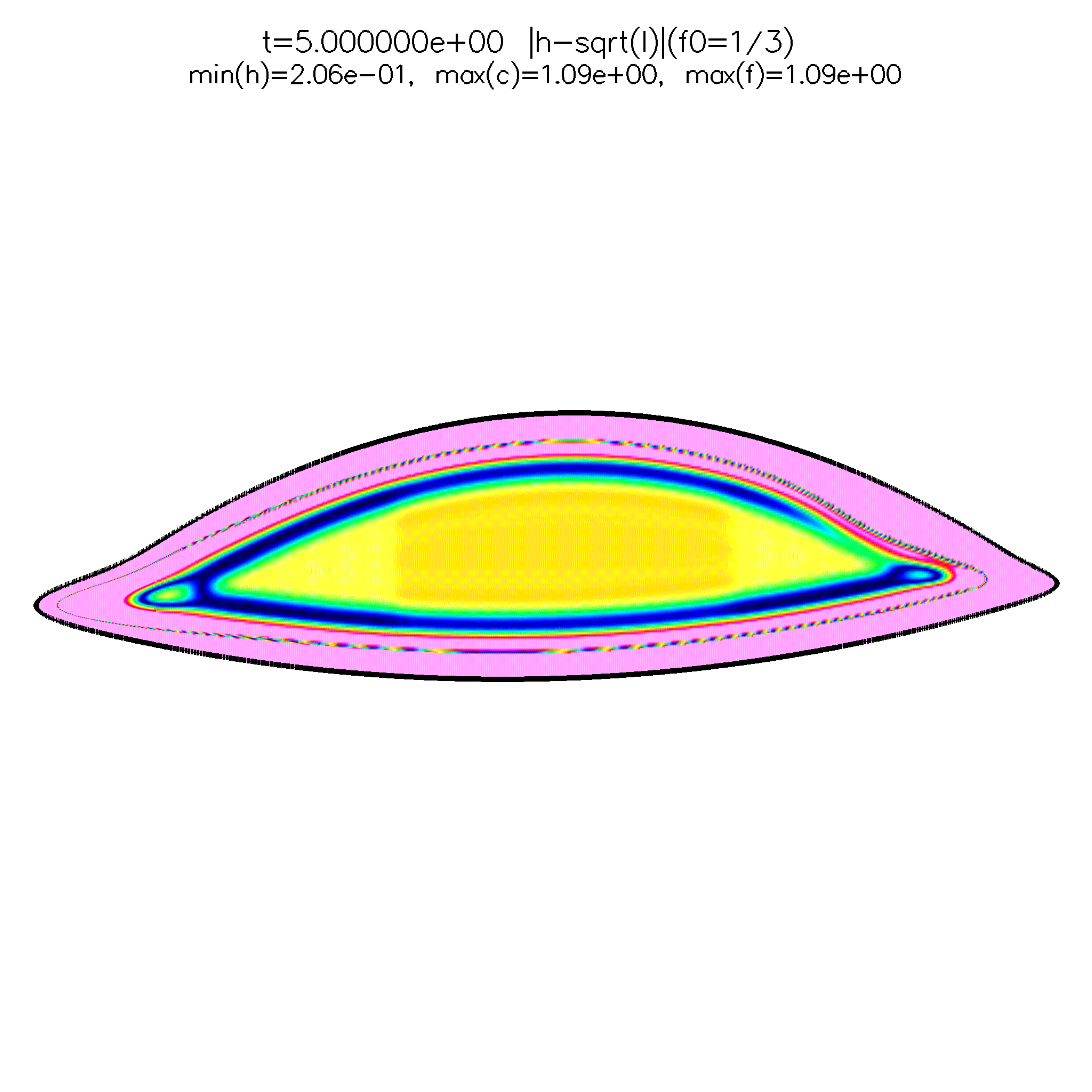}{\figWidth}};
\draw(\figWidth,\row) node[anchor=south west,xshift=\xs cm,yshift=\ys cm] {\trimfig{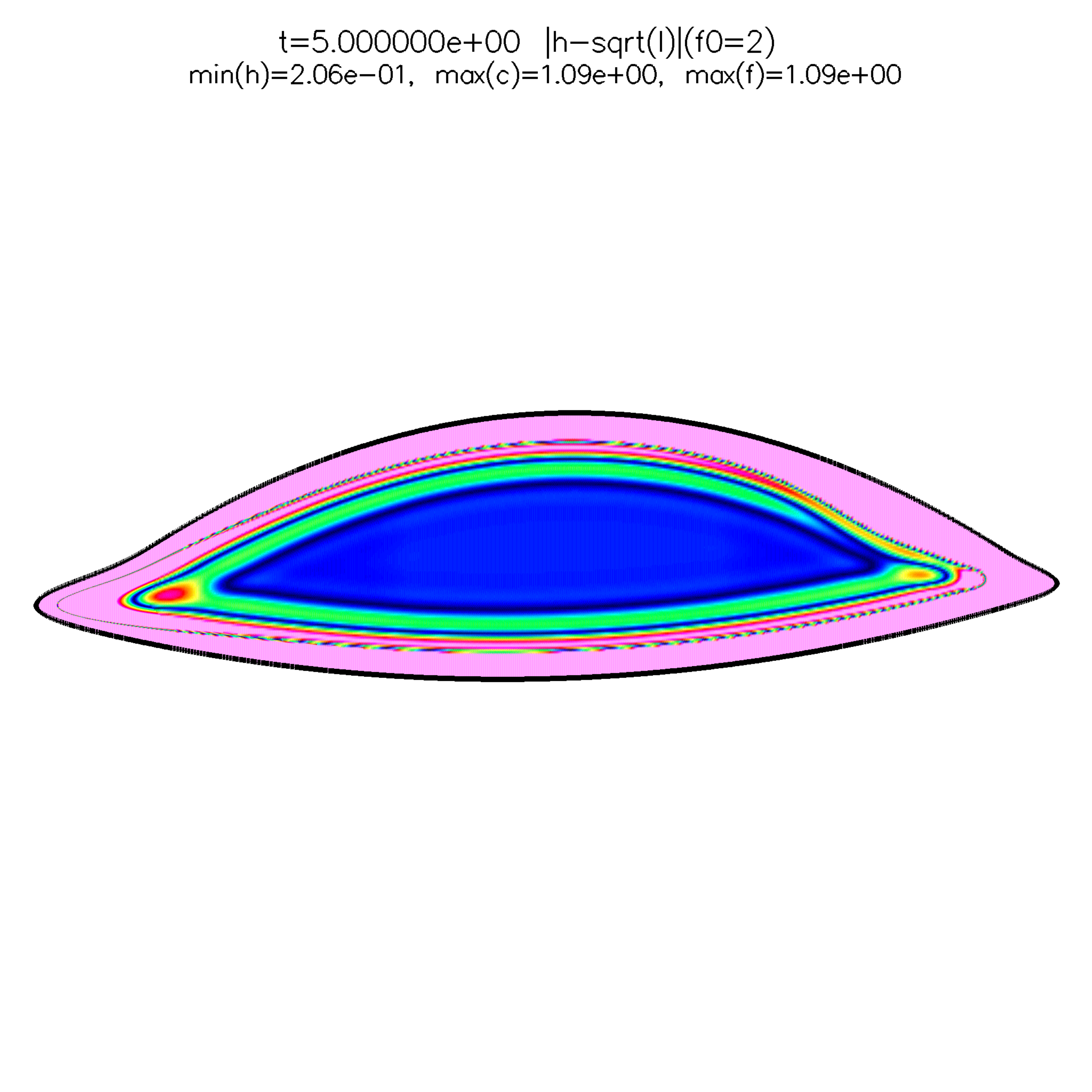}{\figWidth}};
\pgfmathparse{\row+\rowHeight} \let\row\pgfmathresult
%
\end{tikzpicture}
\end{center}
\caption{Variable permeability and evaporation rate of 4 $\mu$m/min, $|h-a\sqrt{I}|$ with $f_0=1/3$ and $f_0=2$}
\label{fig:CompareMethods_Pc(x,y)_4mpm}
\end{figure}
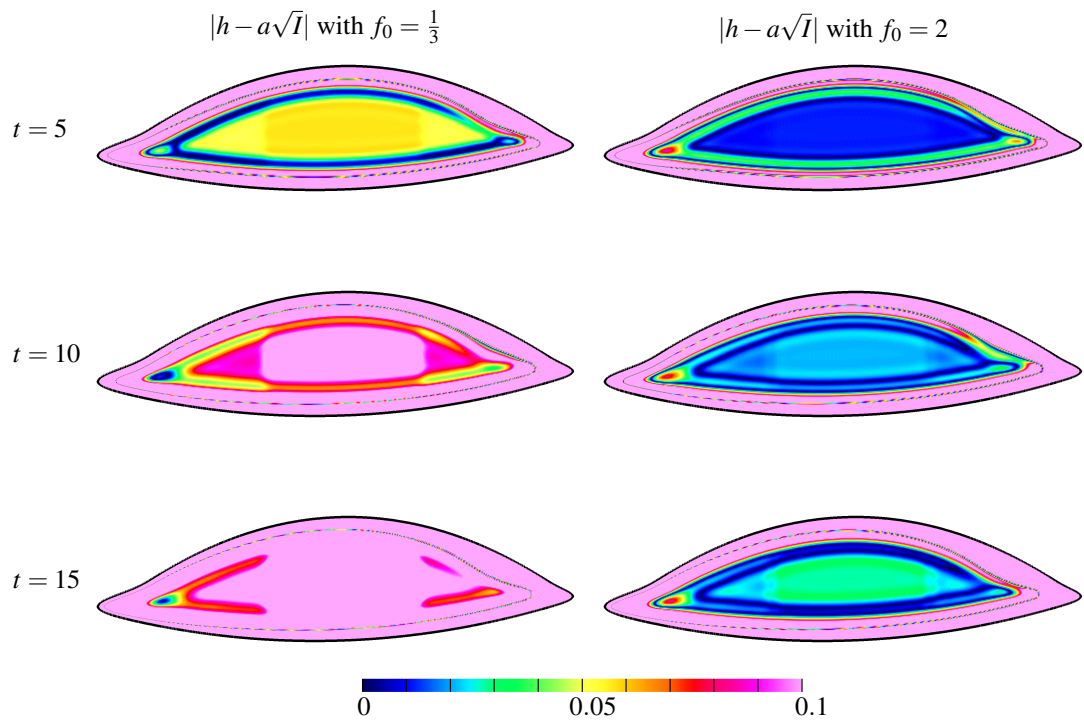
}

\clearpage
\subsubsection{Evaporation rate 20 $\mu$m/min}

We now consider the variable permeability ocular surface with a nominal 20 $\mu$m/min thinning rate for the evaporation.
This is a high evaporation rate, near the upper end of observed thinning rates in vivo \citep{KSHinNic10}.
The dynamics for $h$, $c$ and $f$ are shown in Fig.\ref{fig:hcf_Pc(x,y)_20mpm}.
In the first column, $h$ is seen to thin rapidly outside of the menisci as time increases.
The black lines are seen, but the central region steadily becomes thin as well.
The tear film becomes even thinner over the cornea due to its lower permeability, appearing as
the darker circular central region.  The thinnest tear film occurs in the inferior meniscus
over the cornea.

In the second and third columns,
both $c$ and $f$ increase outside of the menisci as the tear film thins
due to mass conservation of solutes.  The osmolarity increases significantly more
over the cornea than the conjunctiva due to the lower corneal permeability to water,
and the osmolarity is
highest in the black lines over the cornea as in the lower evaporation rate
case.  However, the max value at $t=15$ of 5.43, or 1630 mOsM, is much larger in this case.
The distribution of $f$ is similar in its pattern to $c$,
with steadily increasing $f$ outside the menisci and the highest values
occurring over the cornea.  However, the features in the $f$ distribution are narrower
than those in the corresponding $c$ distribution because of the smaller diffusivity for $f$.
The maximum $f$ is higher as well, reaching 6.59 times the initial value $f_0$.
At this high evaporation rate, the simulation time is sufficiently long to
show  the distinction that develops between the solutes due to differing
diffusion rates.
{
\def\nRows{3}  
\def\nColumns{3}  
\begin{sidewaysfigure}[htb]
\begin{center}
\begin{tikzpicture}[scale=1]
\pgfmathparse{\columnWidth*\nColumns} \let\xb\pgfmathresult
\pgfmathparse{\rowHeight*\nRows+2} \let\yb\pgfmathresult 
\useasboundingbox (0.0,0.0) rectangle (\xb,\yb);  
\def\row{2}
\def\xs{1.1} 
\def\ys{0.3}
\pgfmathparse{\xs+0.*\figWidth} \let\xcb\pgfmathresult 
\pgfmathparse{\xcb+0.*\cbWidth} \let\xx\pgfmathresult
\draw(\xx,\ys) node[anchor=south west] {\getCB{colorbar_rainbow.pdf}{\cbWidth}};
\draw(\xx,\ys) node[anchor= west]{$0$};
\pgfmathparse{\xcb+0.47*\cbWidth} \let\xx\pgfmathresult
\draw(\xx,\ys) node[anchor= west]{$1.5$};
\pgfmathparse{\xcb+0.97*\cbWidth} \let\xx\pgfmathresult
\draw(\xx,\ys) node[anchor= west]{$3$};
\pgfmathparse{\xs+1.*\figWidth} \let\xcb\pgfmathresult 
\pgfmathparse{\xcb+0.*\cbWidth} \let\xx\pgfmathresult
\draw(\xx,\ys) node[anchor=south west] {\getCB{colorbar_rainbow.pdf}{\cbWidth}};
\draw(\xx,\ys) node[anchor= west]{$0$};
\pgfmathparse{\xcb+0.47*\cbWidth} \let\xx\pgfmathresult
\draw(\xx,\ys) node[anchor= west]{$2.8$};
\pgfmathparse{\xcb+0.97*\cbWidth} \let\xx\pgfmathresult
\draw(\xx,\ys) node[anchor= west]{$5.6$};
\pgfmathparse{\xs+2.*\figWidth} \let\xcb\pgfmathresult 
\pgfmathparse{\xcb+0.*\cbWidth} \let\xx\pgfmathresult
\draw(\xx,\ys) node[anchor=south west] {\getCB{colorbar_rainbow.pdf}{\cbWidth}};
\draw(\xx,\ys) node[anchor= west]{$0$};
\pgfmathparse{\xcb+0.47*\cbWidth} \let\xx\pgfmathresult
\draw(\xx,\ys) node[anchor= west]{$3.3$};
\pgfmathparse{\xcb+0.97*\cbWidth} \let\xx\pgfmathresult
\draw(\xx,\ys) node[anchor= west]{$6.6$};
\def\xs{0.8}
\def\ys{0.}
\pgfmathparse{\xs+0.5*\figWidth} \let\xx\pgfmathresult
\pgfmathparse{\yb-1} \let\yy\pgfmathresult
\draw(\xx,\yy) node[anchor=south]{$h$};
\pgfmathparse{\xs+1.5*\figWidth} \let\xx\pgfmathresult
\pgfmathparse{\yb-1} \let\yy\pgfmathresult
\draw(\xx,\yy) node[anchor=south]{$c$};
\pgfmathparse{\xs+2.5*\figWidth} \let\xx\pgfmathresult
\draw(\xx,\yy) node[anchor=south]{$f$};
\pgfmathparse{\row+1} \let\tP\pgfmathresult
\draw(0,\tP) node[anchor=west,xshift=-0.1cm] {$t=15$};
\pgfmathparse{0.*\figWidth} \let\xf\pgfmathresult
\draw(\xf,\row) node[anchor=south west,xshift=\xs cm,yshift=\ys cm] {\trimfig{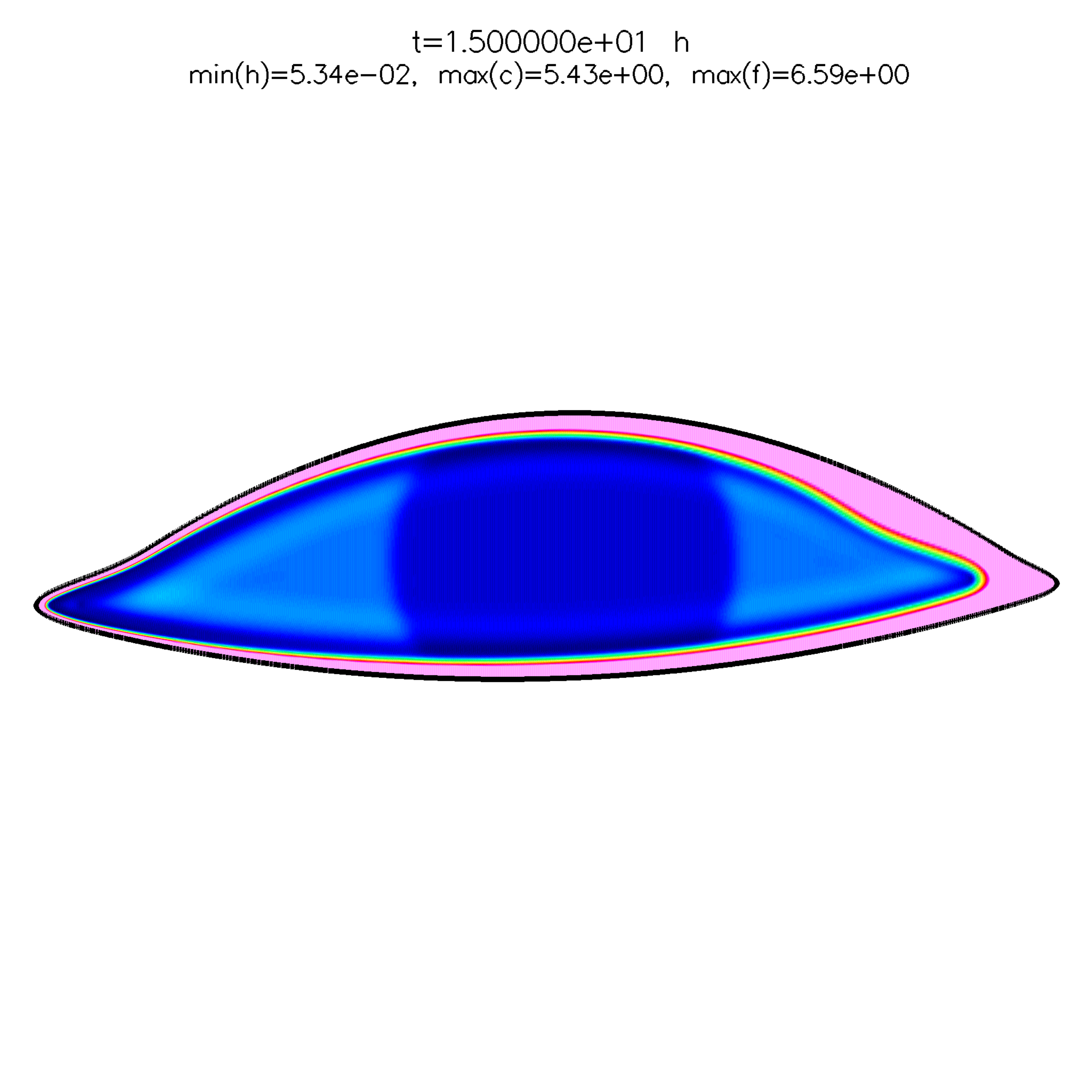}{\figWidth}};
\pgfmathparse{\xs+0.5*\figWidth} \let\xx\pgfmathresult
\draw(\xx,\row) node[anchor=north]{$\min(h)=$ 5.34e-2};
\pgfmathparse{1.*\figWidth} \let\xf\pgfmathresult
\draw(\xf,\row) node[anchor=south west,xshift=\xs cm,yshift=\ys cm] {\trimfig{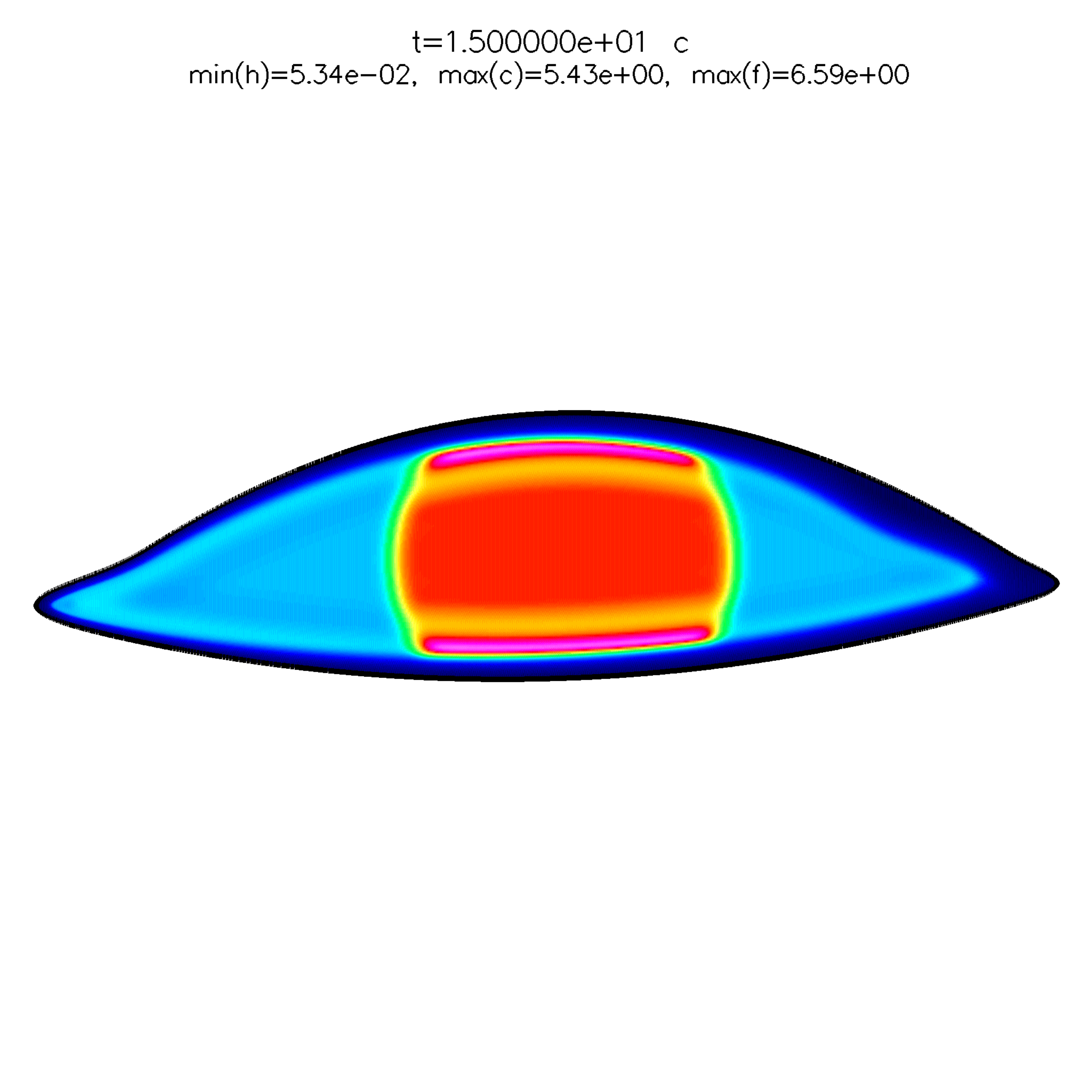}{\figWidth}};
\pgfmathparse{\xs+1.5*\figWidth} \let\xx\pgfmathresult
\draw(\xx,\row) node[anchor=north]{$\max(c)=$ 5.43};
\pgfmathparse{2.*\figWidth} \let\xf\pgfmathresult
\draw(\xf,\row) node[anchor=south west,xshift=\xs cm,yshift=\ys cm] {\trimfig{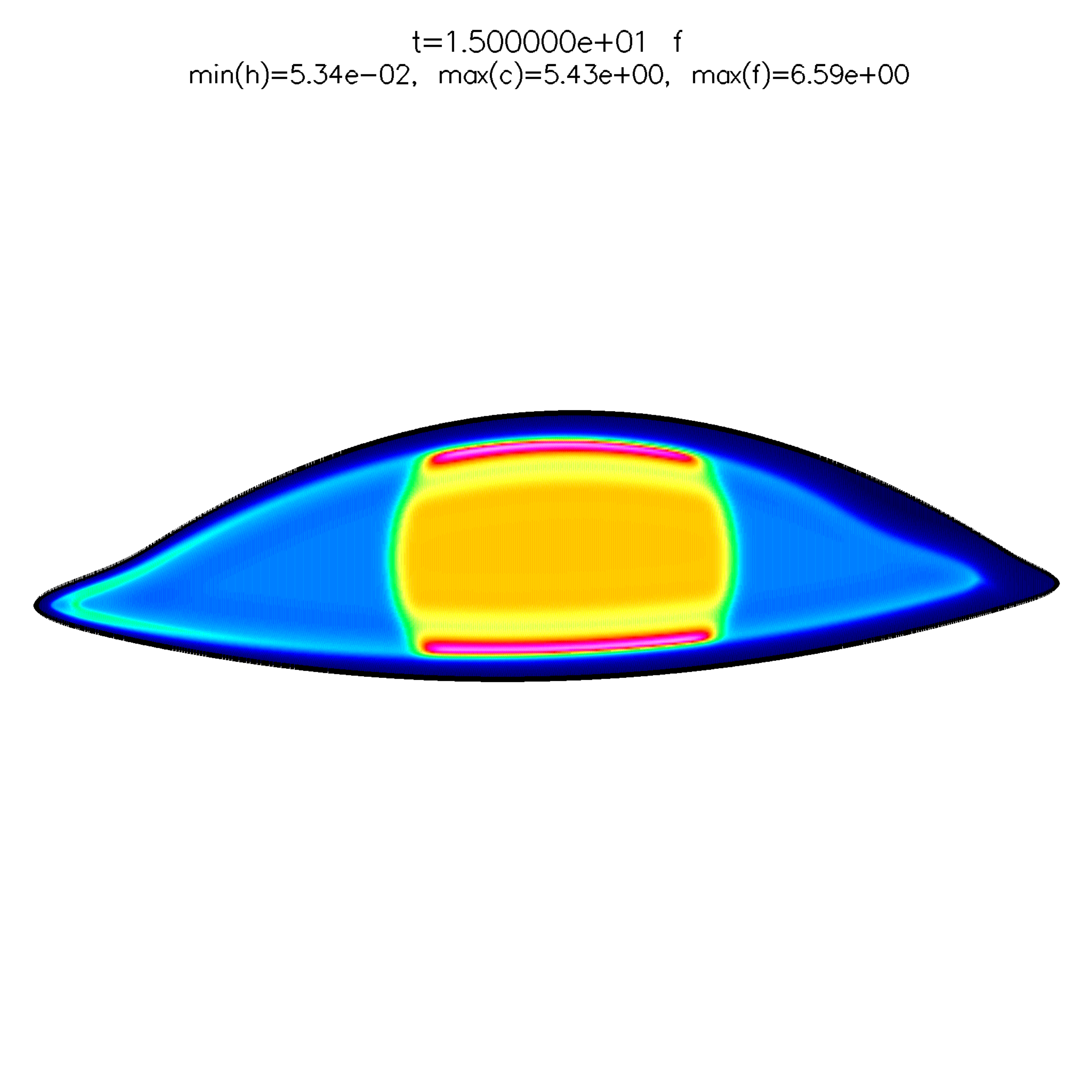}{\figWidth}};
\pgfmathparse{\xs+2.5*\figWidth} \let\xx\pgfmathresult
\draw(\xx,\row) node[anchor=north]{$\max(f)=$ 6.59};
\pgfmathparse{\row+\rowHeight} \let\row\pgfmathresult
\pgfmathparse{\row+1} \let\tP\pgfmathresult
\draw(0,\tP) node[anchor=west,xshift=-0.1cm] {$t=10$};
\pgfmathparse{0.*\figWidth} \let\xf\pgfmathresult
\draw(\xf,\row) node[anchor=south west,xshift=\xs cm,yshift=\ys cm] {\trimfig{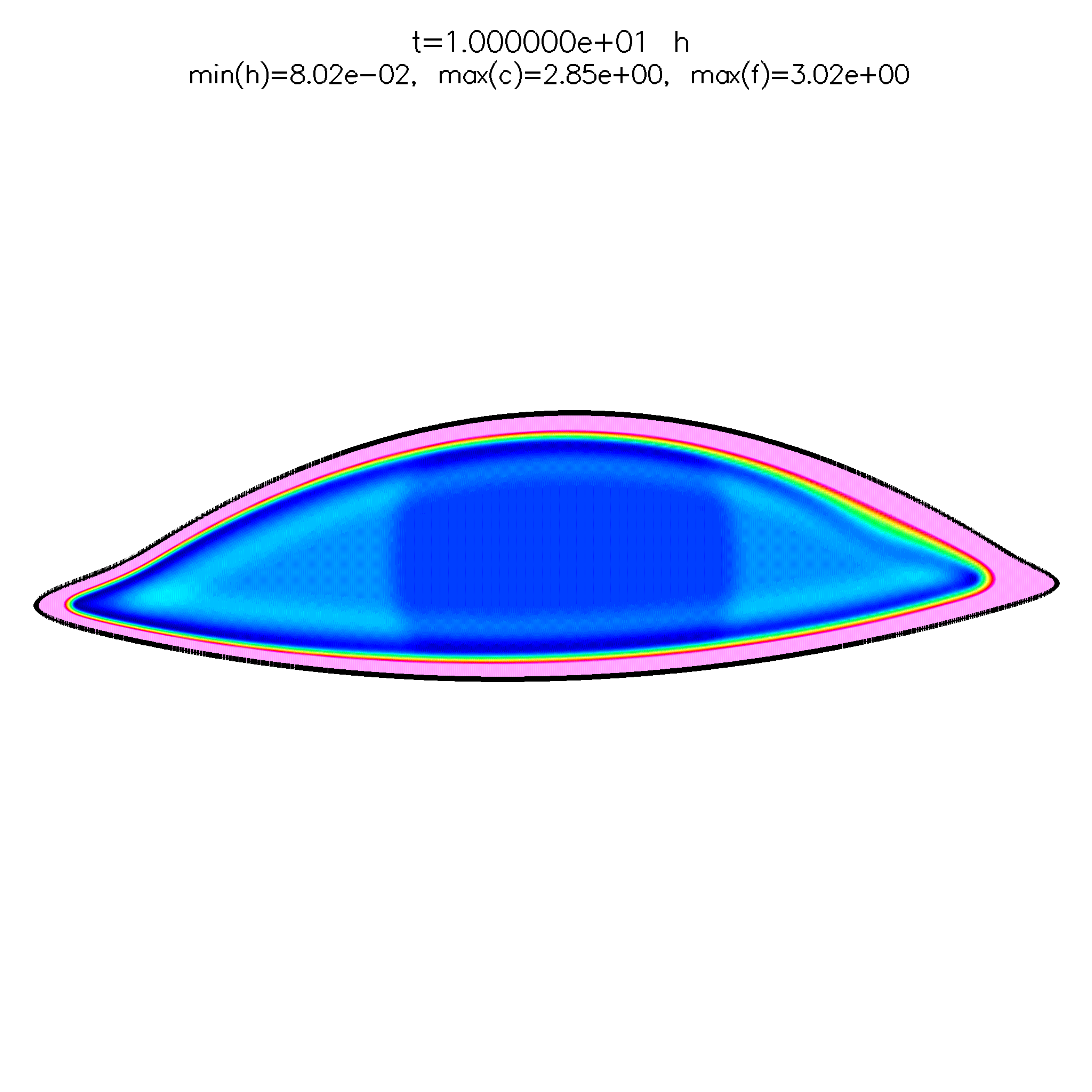}{\figWidth}};
\pgfmathparse{\xs+0.5*\figWidth} \let\xx\pgfmathresult
\draw(\xx,\row) node[anchor=north]{$\min(h)=$ 8.02e-2};
\pgfmathparse{1.*\figWidth} \let\xf\pgfmathresult
\draw(\xf,\row) node[anchor=south west,xshift=\xs cm,yshift=\ys cm] {\trimfig{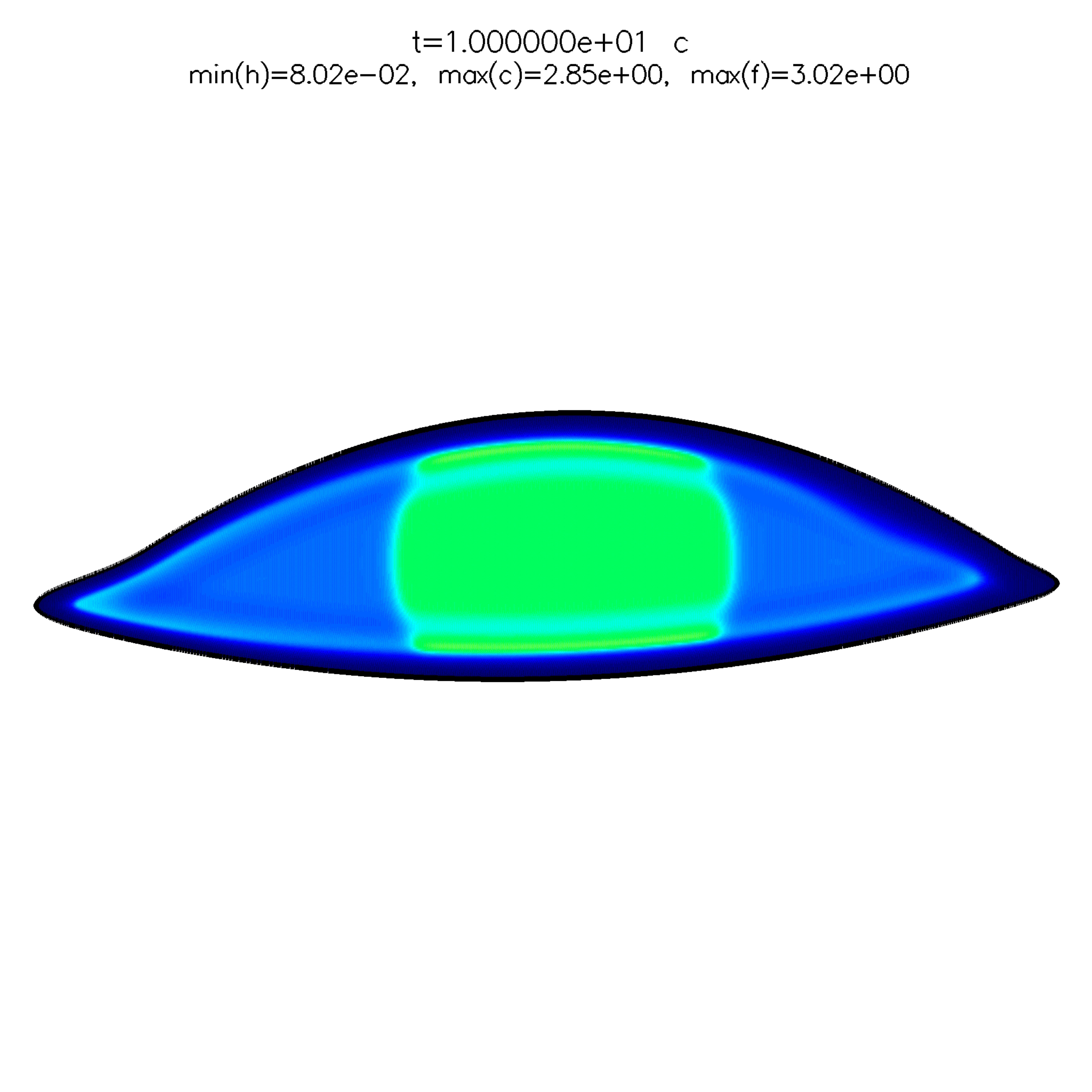}{\figWidth}};
\pgfmathparse{\xs+1.5*\figWidth} \let\xx\pgfmathresult
\draw(\xx,\row) node[anchor=north]{$\max(c)=$ 2.85};
\pgfmathparse{2.*\figWidth} \let\xf\pgfmathresult
\draw(\xf,\row) node[anchor=south west,xshift=\xs cm,yshift=\ys cm] {\trimfig{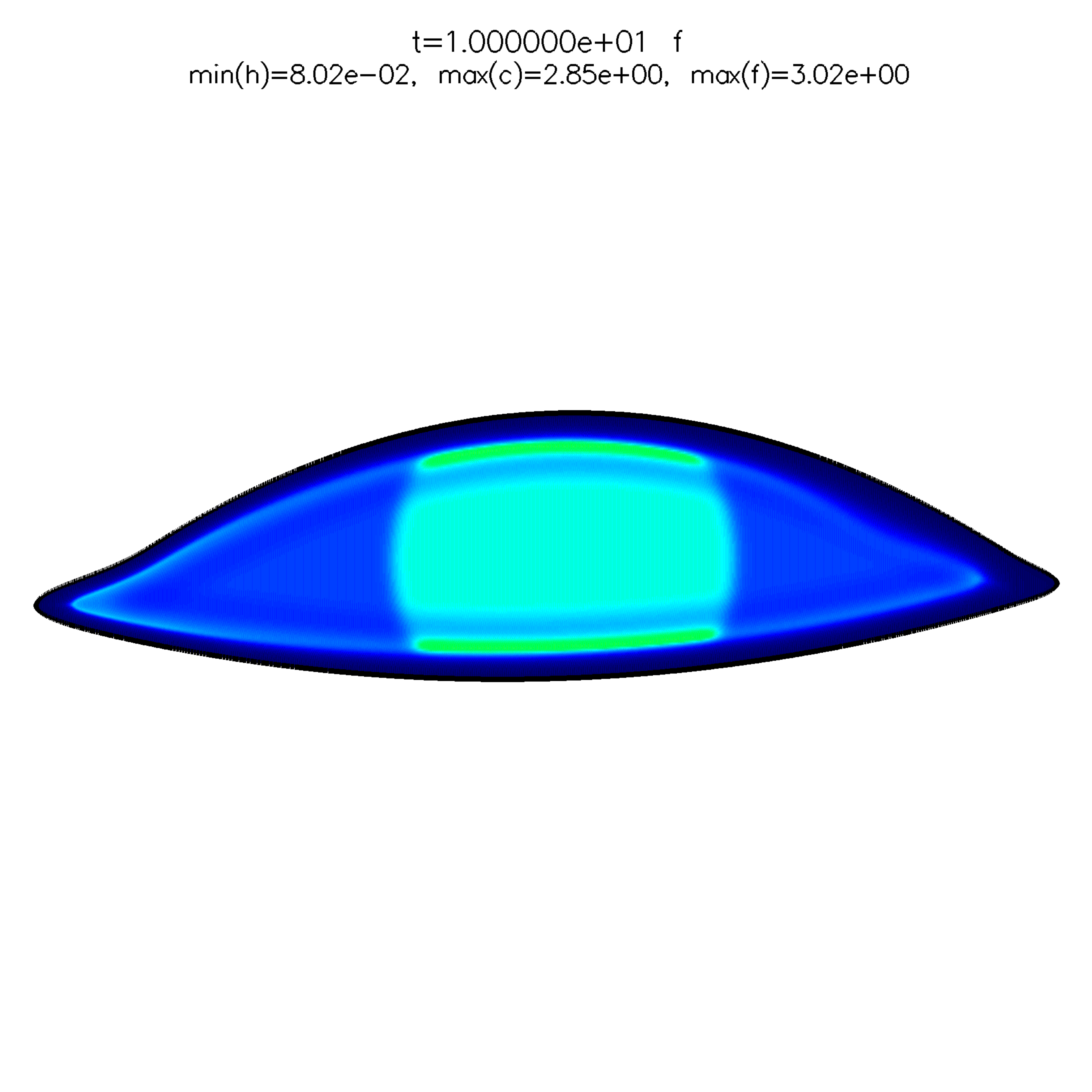}{\figWidth}};
\pgfmathparse{\xs+2.5*\figWidth} \let\xx\pgfmathresult
\draw(\xx,\row) node[anchor=north]{$\max(f)=$ 3.02};
\pgfmathparse{\row+\rowHeight} \let\row\pgfmathresult
\pgfmathparse{\row+1} \let\tP\pgfmathresult
\draw(0,\tP) node[anchor=west,xshift=-0.1cm] {$t=5$};
\pgfmathparse{0.*\figWidth} \let\xf\pgfmathresult
\draw(\xf,\row) node[anchor=south west,xshift=\xs cm,yshift=\ys cm] {\trimfig{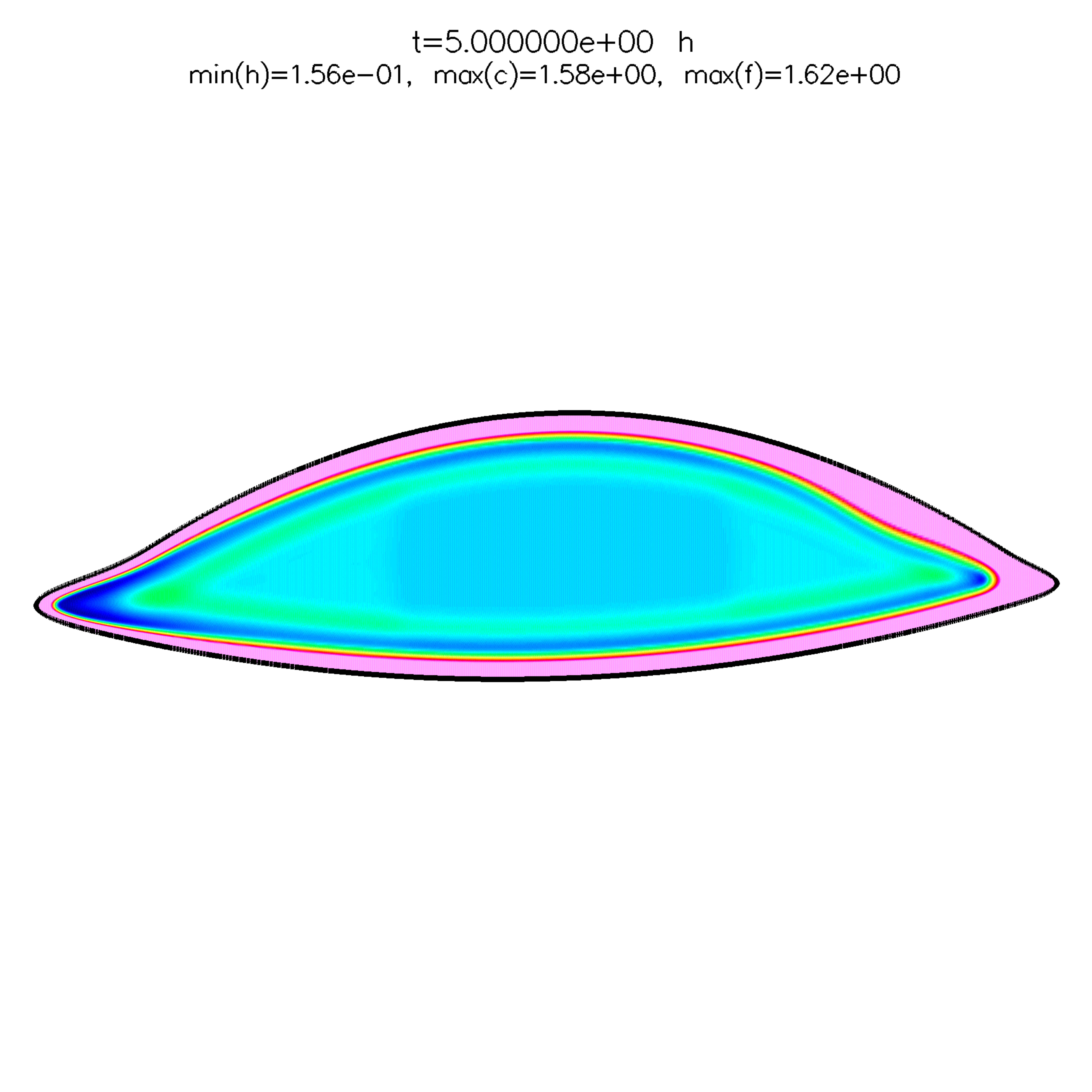}{\figWidth}};
\pgfmathparse{\xs+0.5*\figWidth} \let\xx\pgfmathresult
\draw(\xx,\row) node[anchor=north]{$\min(h)=$ 1.56e-1};
\pgfmathparse{1.*\figWidth} \let\xf\pgfmathresult
\draw(\xf,\row) node[anchor=south west,xshift=\xs cm,yshift=\ys cm] {\trimfig{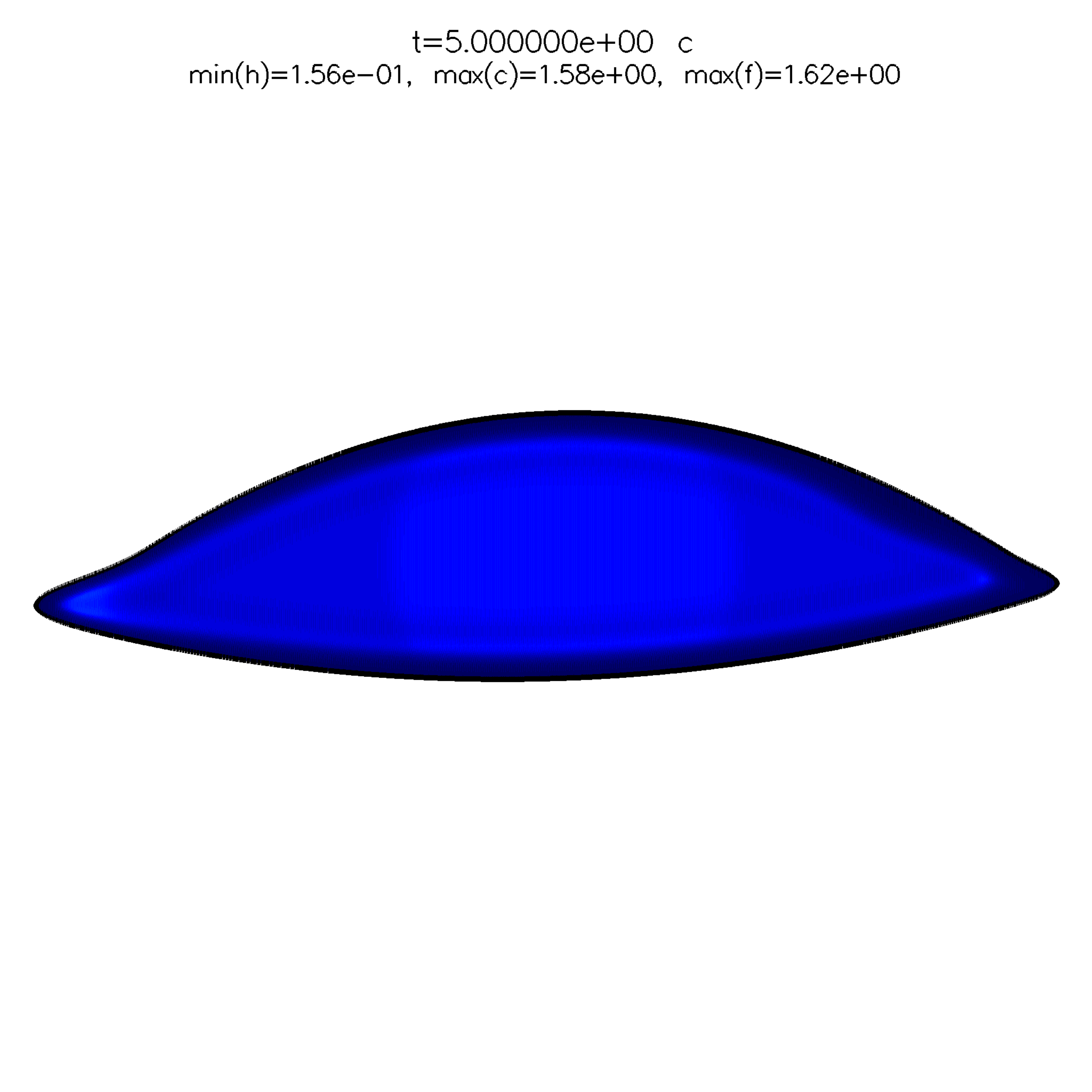}{\figWidth}};
\pgfmathparse{\xs+1.5*\figWidth} \let\xx\pgfmathresult
\draw(\xx,\row) node[anchor=north]{$\max(c)=$ 1.58};
\pgfmathparse{2.*\figWidth} \let\xf\pgfmathresult
\draw(\xf,\row) node[anchor=south west,xshift=\xs cm,yshift=\ys cm] {\trimfig{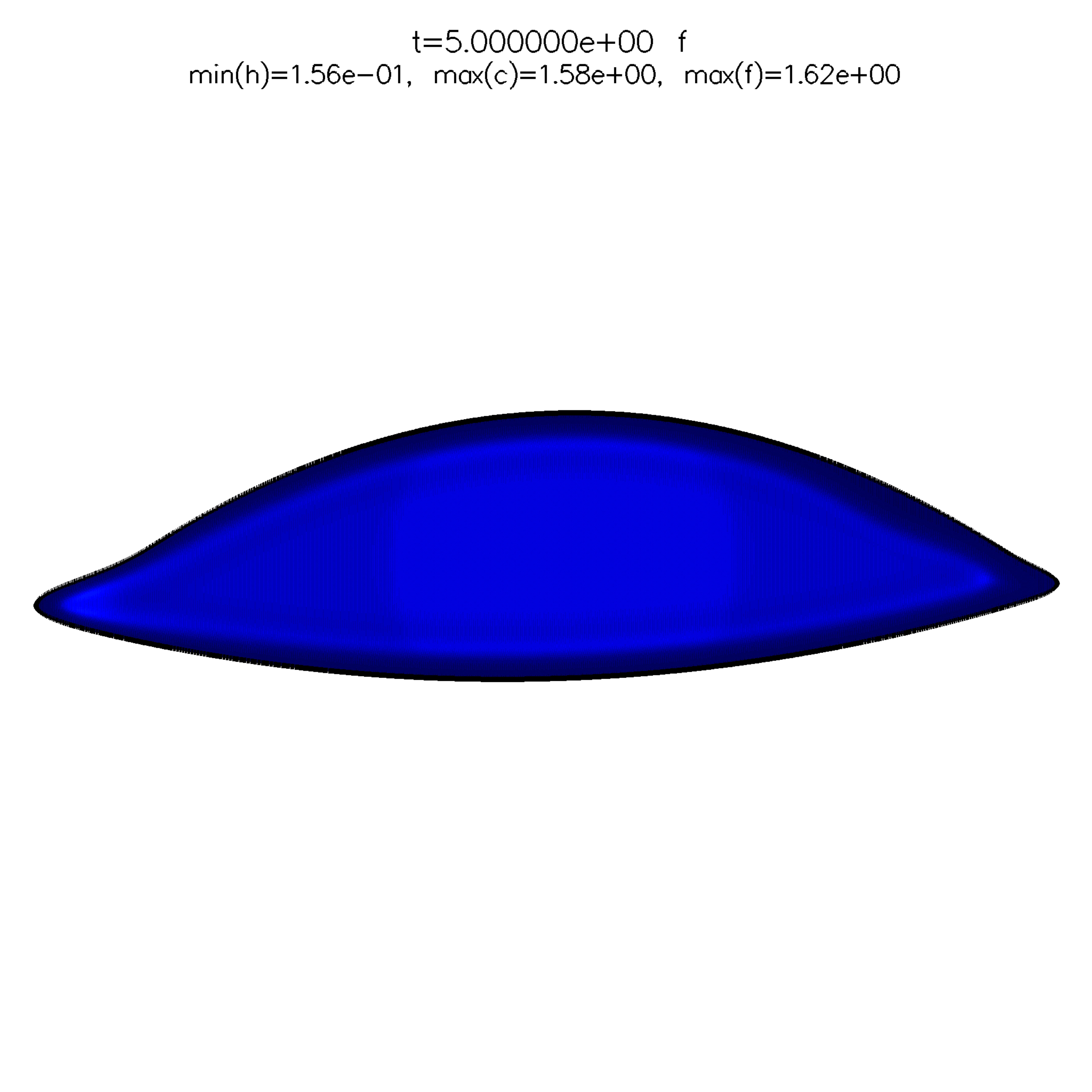}{\figWidth}};
\pgfmathparse{\xs+2.5*\figWidth} \let\xx\pgfmathresult
\draw(\xx,\row) node[anchor=north]{$\max(f)=$ 1.62};
\pgfmathparse{\row+\rowHeight} \let\row\pgfmathresult
%
\end{tikzpicture}
\end{center}
\caption{Variable permeability and evaporation is 20 micron/min, $h$,$c$ and $f$}
\label{fig:hcf_Pc(x,y)_20mpm}
\end{sidewaysfigure}
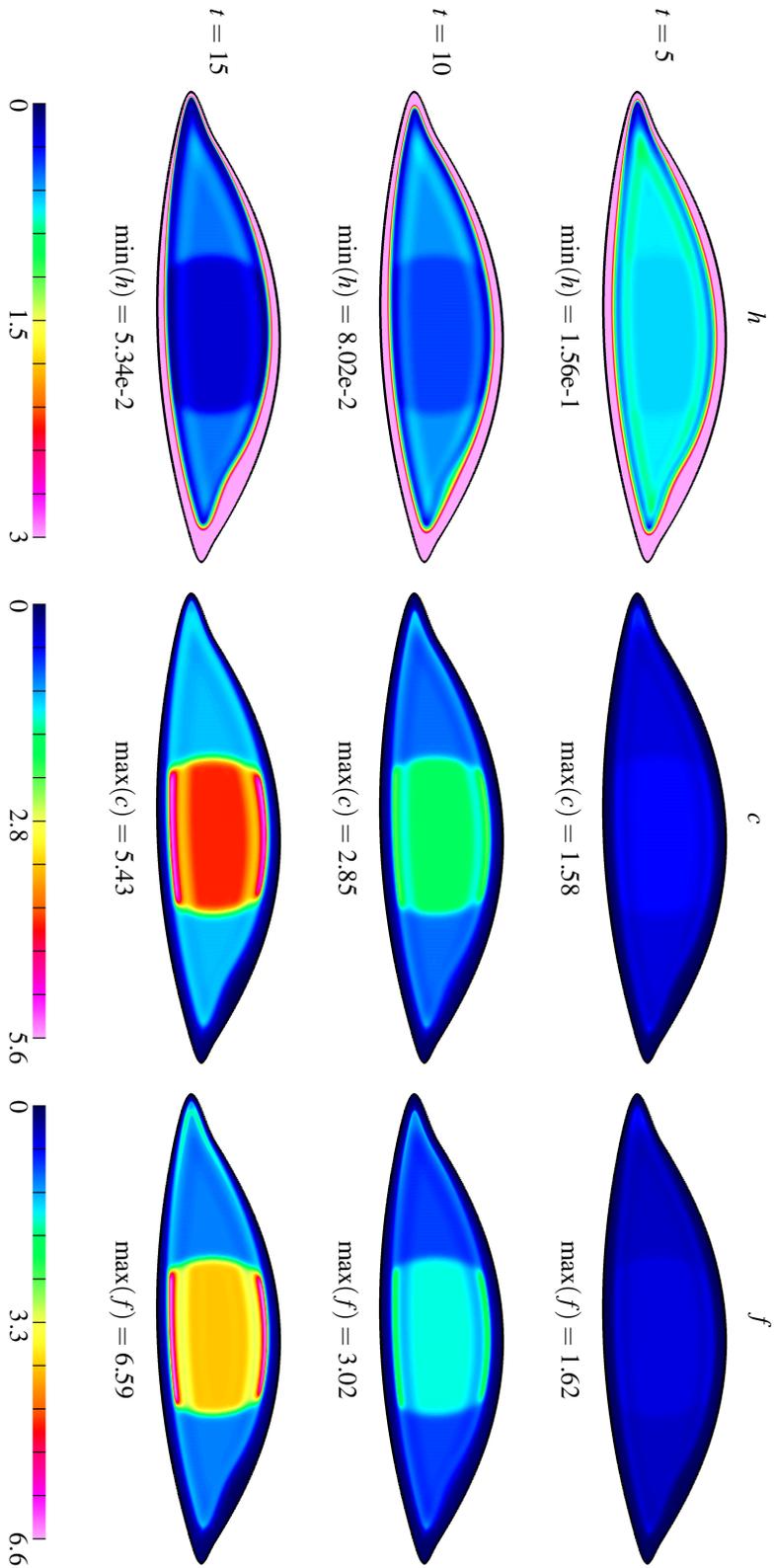
}

Fig.~\ref{fig:Intensity_Pc(x,y)_20mpm} shows the distribution of fluorescent intensity over the
exposed ocular surface after using (\ref{eqn:I-of-f}) together with the computed $f$ and
$h$. In either case, the meniscus stays at essentially constant intensity because $h$ and $f$ change little.
At this high evaporation rate, both the dilute (left) and self-quenching (right) cases show overall
dimming, but it
is much more dramatic in the latter case.  In both cases, the dimming of the intensity is most over the cornea
where the permeability is lowest and the tear film thins the most.  The dilute case (left) starts out there,
but $f$ increases sufficiently, for the initial condition we used, to also enter the self-quenching regime
later in the computation \citep{BraunEtal14}.  The self-quenching case begins there, and the entire black
line and central cornea become quite
dark.  It is not uncommon in quantitative experiments to enhance the FL images for processing so that better
results can be obtained from images that are difficult to distinguish with the naked eye.
The previously seen changes in the width of the meniscus is again made visible by the fluorescence \citep{MakiBraun10b,LiBraun12}.
{
\def\nRows{3}  
\def\nColumns{2}  
\begin{figure}[htb]
\begin{center}
\begin{tikzpicture}[scale=1]
\pgfmathparse{\columnWidth*\nColumns} \let\xb\pgfmathresult
\pgfmathparse{\rowHeight*\nRows+1} \let\yb\pgfmathresult 
\useasboundingbox (0.0,0.0) rectangle (\xb,\yb);  
\def\row{1}
\def\xs{1.1} 
\def\ys{0.3}
\pgfmathparse{\xs+0.5*\figWidth} \let\xcb\pgfmathresult 
\pgfmathparse{\xcb+0.*\cbWidth} \let\xx\pgfmathresult
\draw(\xx,\ys) node[anchor=south west] {\getCB{colorbar_green.pdf}{\cbWidth}};
\draw(\xx,\ys) node[anchor= west]{$0$};
\pgfmathparse{\xcb+0.47*\cbWidth} \let\xx\pgfmathresult
\draw(\xx,\ys) node[anchor= west]{$50$};
\pgfmathparse{\xcb+0.97*\cbWidth} \let\xx\pgfmathresult
\draw(\xx,\ys) node[anchor= west]{$100$};
\def\xs{0.8}
\def\ys{0.}
\pgfmathparse{\xs+0.5*\figWidth} \let\xx\pgfmathresult
\pgfmathparse{\yb-1} \let\yy\pgfmathresult
\draw(\xx,\yy) node[anchor=south]{Intensity $I$ with $f_0=\frac{1}{3}$};
\pgfmathparse{\xs+1.5*\figWidth} \let\xx\pgfmathresult
\draw(\xx,\yy) node[anchor=south]{Intensity $I$ with $f_0=2$};
\pgfmathparse{\row+1} \let\tP\pgfmathresult
\draw(0,\tP) node[anchor=west,xshift=-0.1cm] {$t=15$};
\draw(0,\row) node[anchor=south west,xshift=\xs cm,yshift=\ys cm] {\trimfig{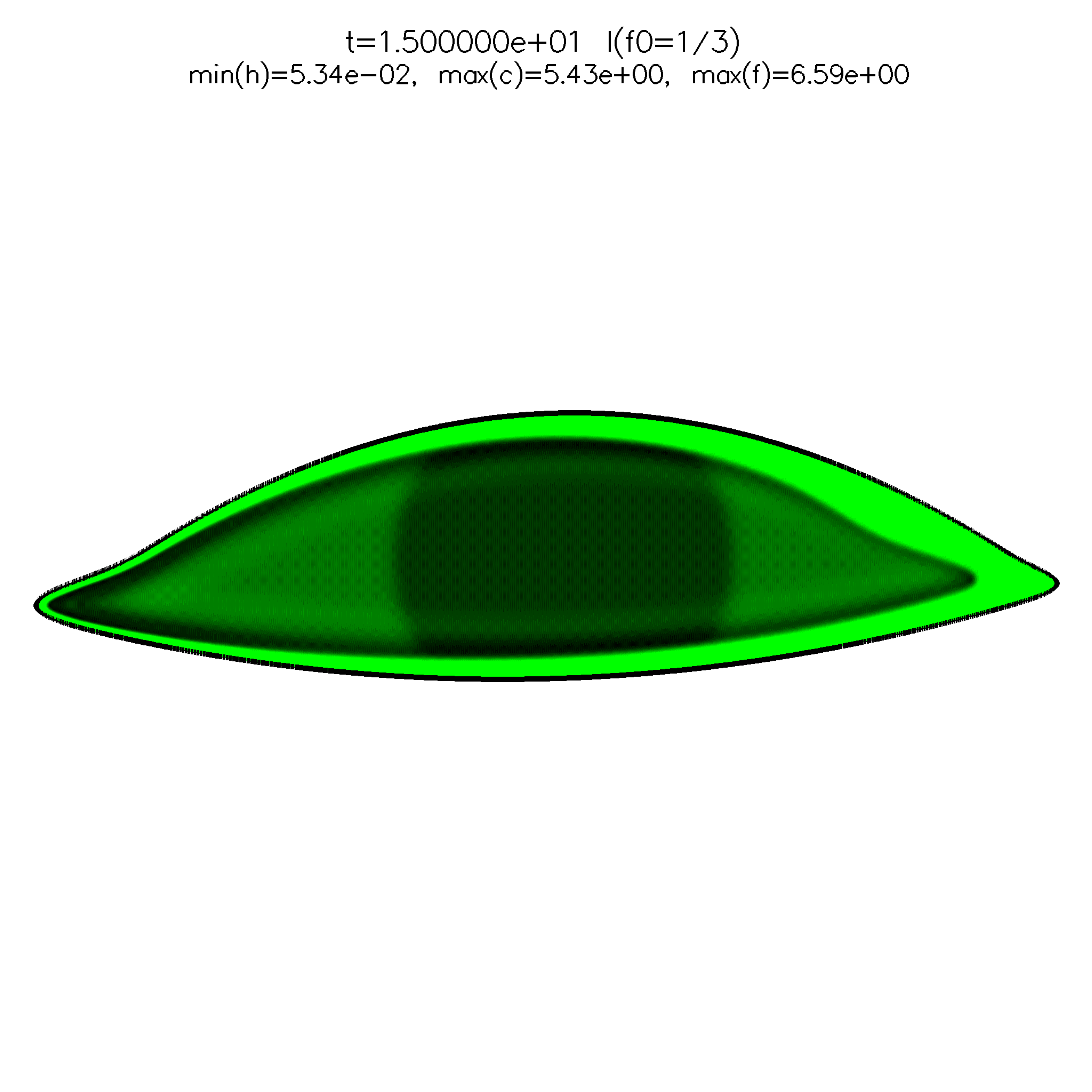}{\figWidth}};
\draw(\figWidth,\row) node[anchor=south west,xshift=\xs cm,yshift=\ys cm] {\trimfig{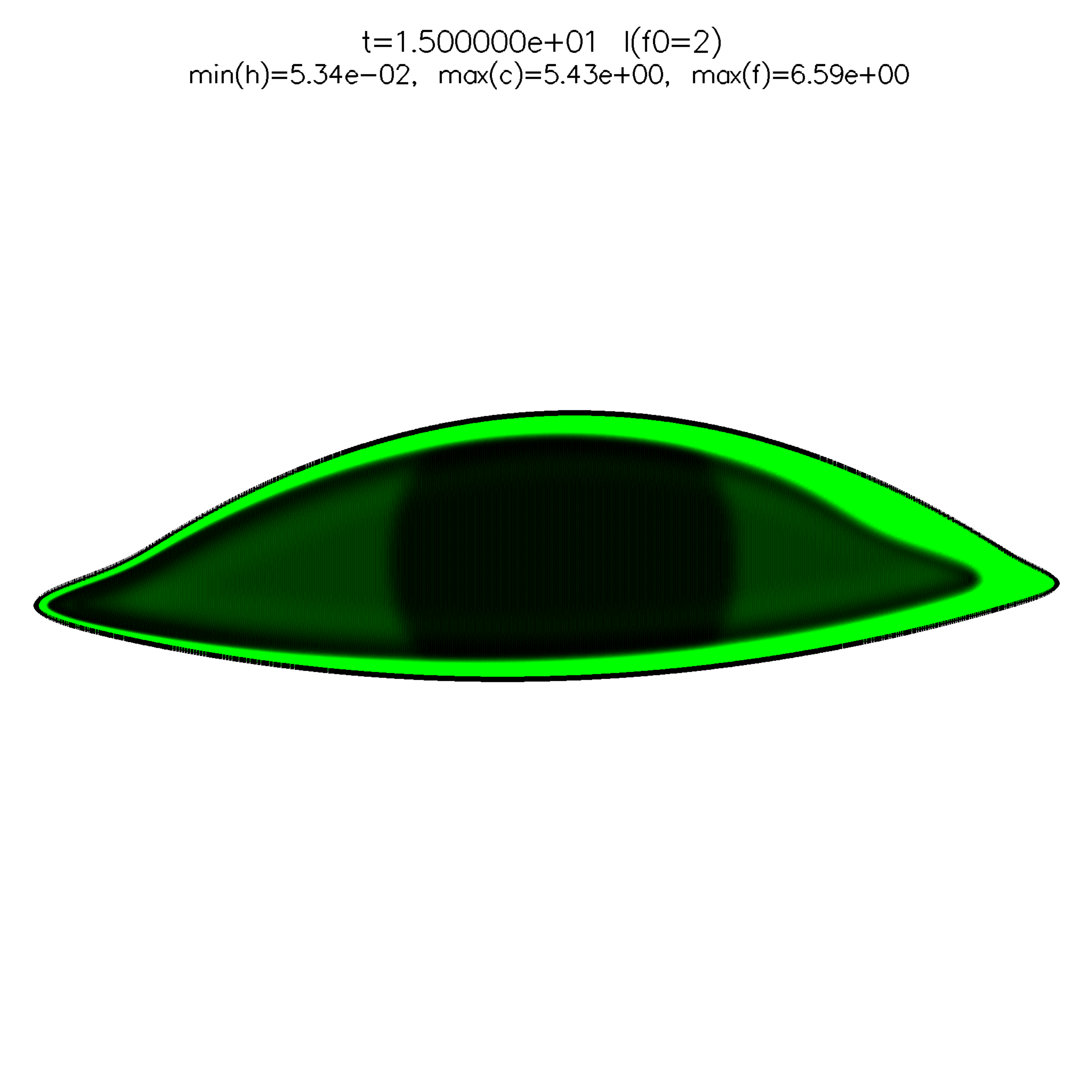}{\figWidth}};
\pgfmathparse{\row+\rowHeight} \let\row\pgfmathresult
\pgfmathparse{\row+1} \let\tP\pgfmathresult
\draw(0,\tP) node[anchor=west,xshift=-0.1cm] {$t=10$};
\draw(0,\row) node[anchor=south west,xshift=\xs cm,yshift=\ys cm] {\trimfig{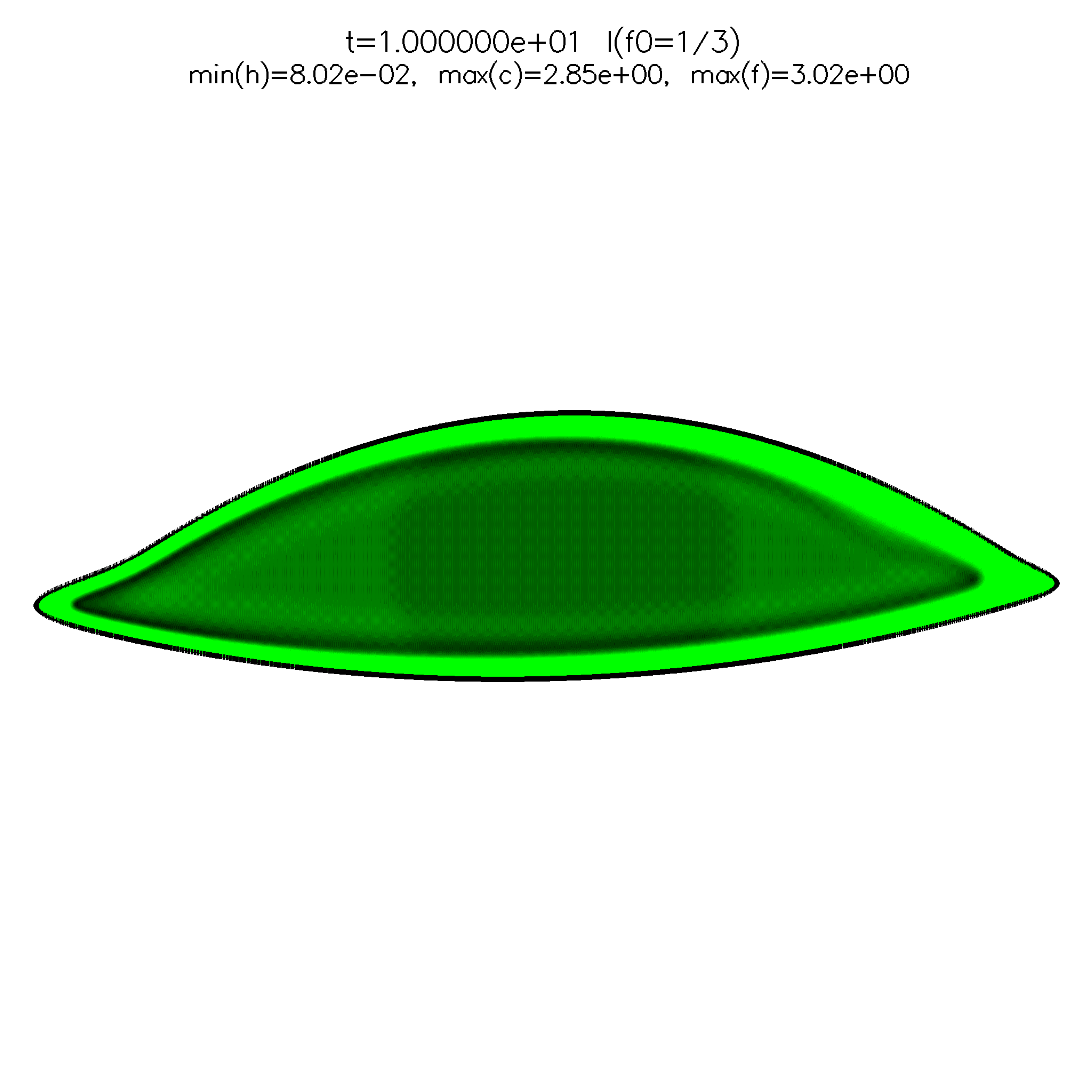}{\figWidth}};
\draw(\figWidth,\row) node[anchor=south west,xshift=\xs cm,yshift=\ys cm] {\trimfig{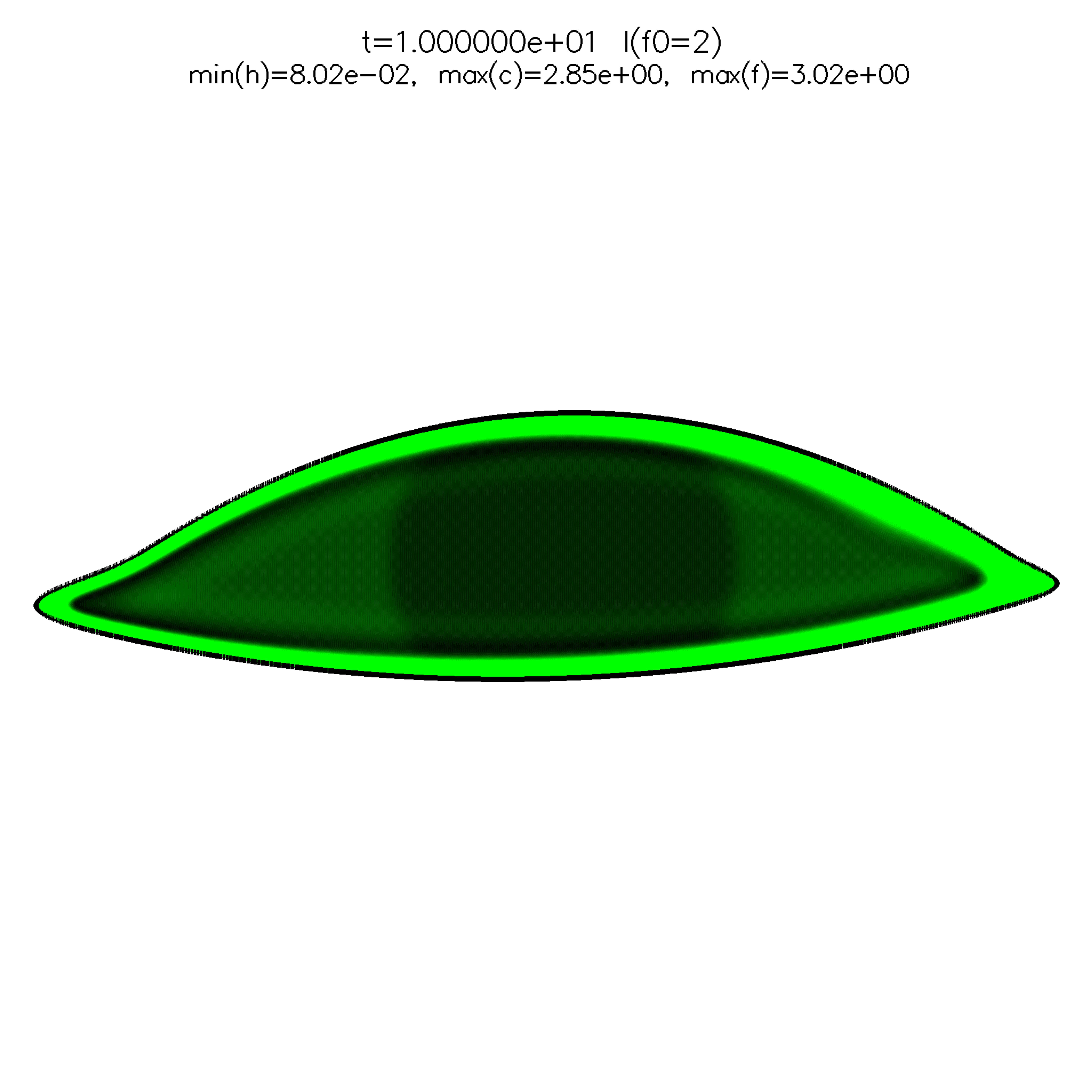}{\figWidth}};
\pgfmathparse{\row+\rowHeight} \let\row\pgfmathresult
\pgfmathparse{\row+1} \let\tP\pgfmathresult
\draw(0,\tP) node[anchor=west,xshift=-0.1cm] {$t=5$};
\draw(0,\row) node[anchor=south west,xshift=\xs cm,yshift=\ys cm] {\trimfig{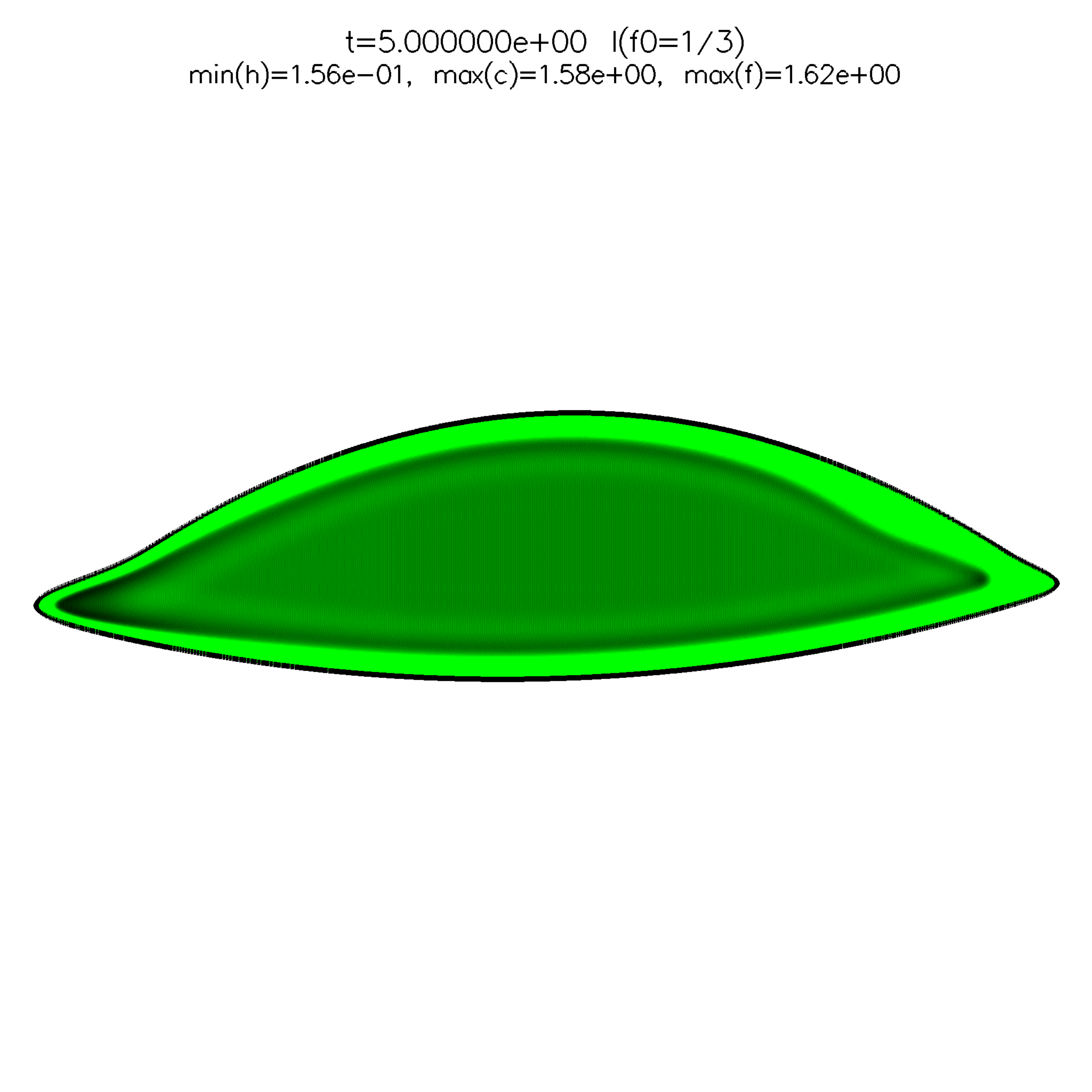}{\figWidth}};
\draw(\figWidth,\row) node[anchor=south west,xshift=\xs cm,yshift=\ys cm] {\trimfig{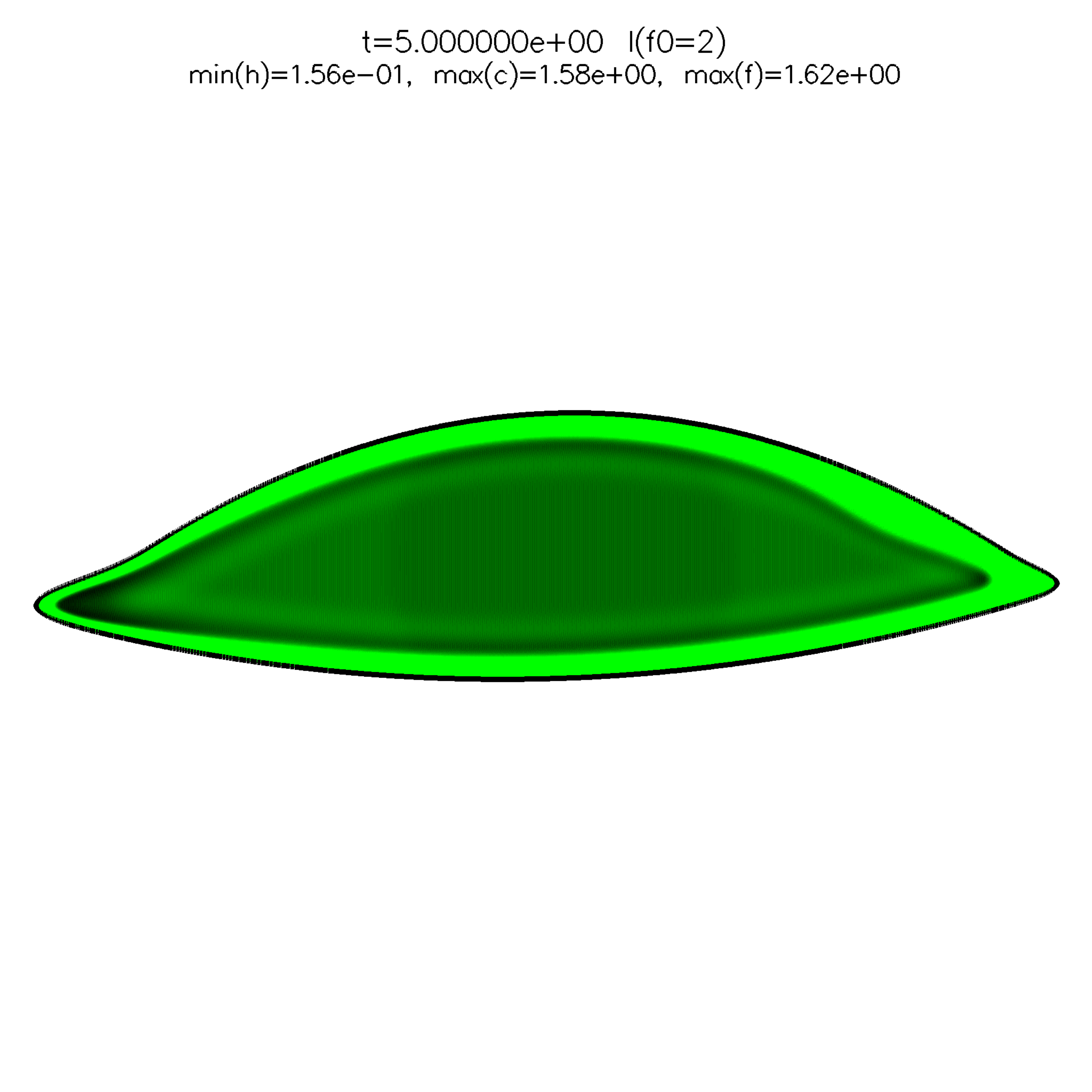}{\figWidth}};
\pgfmathparse{\row+\rowHeight} \let\row\pgfmathresult
%
\end{tikzpicture}
\end{center}
\caption{Variable permeability and evaporation is 20 micron/min, $I$ with $f_0=1/3$ and $f_0=2$}
\label{fig:Intensity_Pc(x,y)_20mpm}
\end{figure}
}

Results for the two different initial fluorescein concentrations are shown in
Figure~\ref{fig:CompareMethods_Pc(x,y)_20mpm} for $|h-a\sqrt{I}|$ with the high evaporation rate.
In either the dilute case (left) or the quenching regime (right), the menisci have a large difference
$|h-a\sqrt{I}|$ because $a$ was chosen to work in the initially flat region outside the mensici.  In the
dilute case (left), the difference between computed and estimated values at essentially any location is
large as well.  In the self-quenching side, the difference between the two is smallest at $t=5$ but
remains relatively small particularly in the central region over the cornea.
Thus the approximation to the thickness appears to be a good estimate of the thickness
over the cornea in the high evaporation rate case \cite{NicholsEtal12}.
{
\def\nRows{3}  
\def\nColumns{2}  
\begin{figure}[htb]
\begin{center}
\begin{tikzpicture}[scale=1]
\pgfmathparse{\columnWidth*\nColumns} \let\xb\pgfmathresult
\pgfmathparse{\rowHeight*\nRows+1} \let\yb\pgfmathresult 
\useasboundingbox (0.0,0.0) rectangle (\xb,\yb);  
\def\row{1}
\def\xs{1.1} 
\def\ys{0.3}
\pgfmathparse{\xs+0.5*\figWidth} \let\xcb\pgfmathresult 
\pgfmathparse{\xcb+0.*\cbWidth} \let\xx\pgfmathresult
\draw(\xx,\ys) node[anchor=south west] {\getCB{colorbar_rainbow.pdf}{\cbWidth}};
\draw(\xx,\ys) node[anchor= west]{$0$};
\pgfmathparse{\xcb+0.47*\cbWidth} \let\xx\pgfmathresult
\draw(\xx,\ys) node[anchor= west]{$0.05$};
\pgfmathparse{\xcb+0.97*\cbWidth} \let\xx\pgfmathresult
\draw(\xx,\ys) node[anchor= west]{$0.1$};
\def\xs{0.8}
\def\ys{0.}
\pgfmathparse{\xs+0.5*\figWidth} \let\xx\pgfmathresult
\pgfmathparse{\yb-1} \let\yy\pgfmathresult
\draw(\xx,\yy) node[anchor=south]{$|h-a\sqrt{I}|$ with $f_0=\frac{1}{3}$};
\pgfmathparse{\xs+1.5*\figWidth} \let\xx\pgfmathresult
\draw(\xx,\yy) node[anchor=south]{$|h-a\sqrt{I}|$ with $f_0=2$};
\pgfmathparse{\row+1} \let\tP\pgfmathresult
\draw(0,\tP) node[anchor=west,xshift=-0.1cm] {$t=15$};
\draw(0,\row) node[anchor=south west,xshift=\xs cm,yshift=\ys cm] {\trimfig{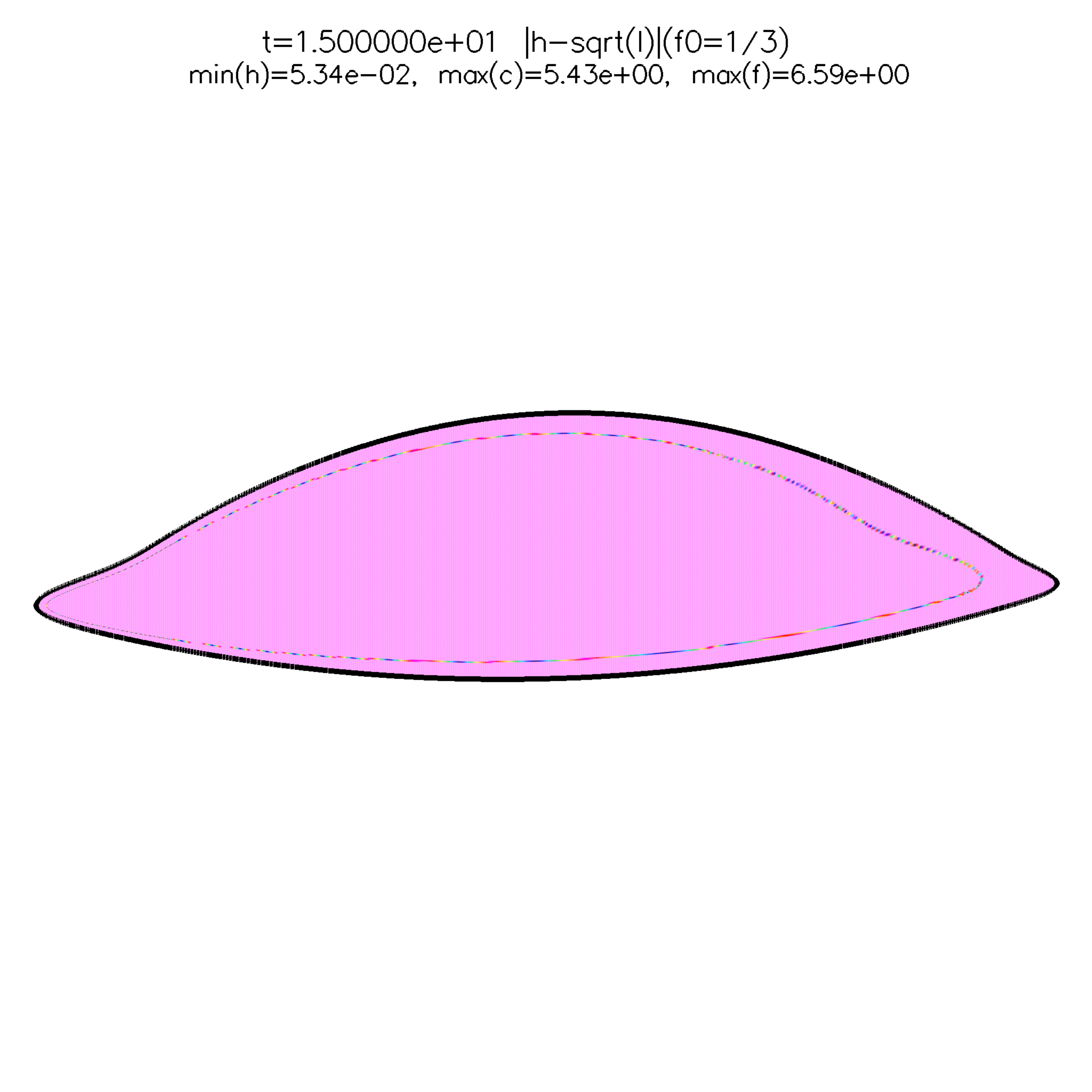}{\figWidth}};
\draw(\figWidth,\row) node[anchor=south west,xshift=\xs cm,yshift=\ys cm] {\trimfig{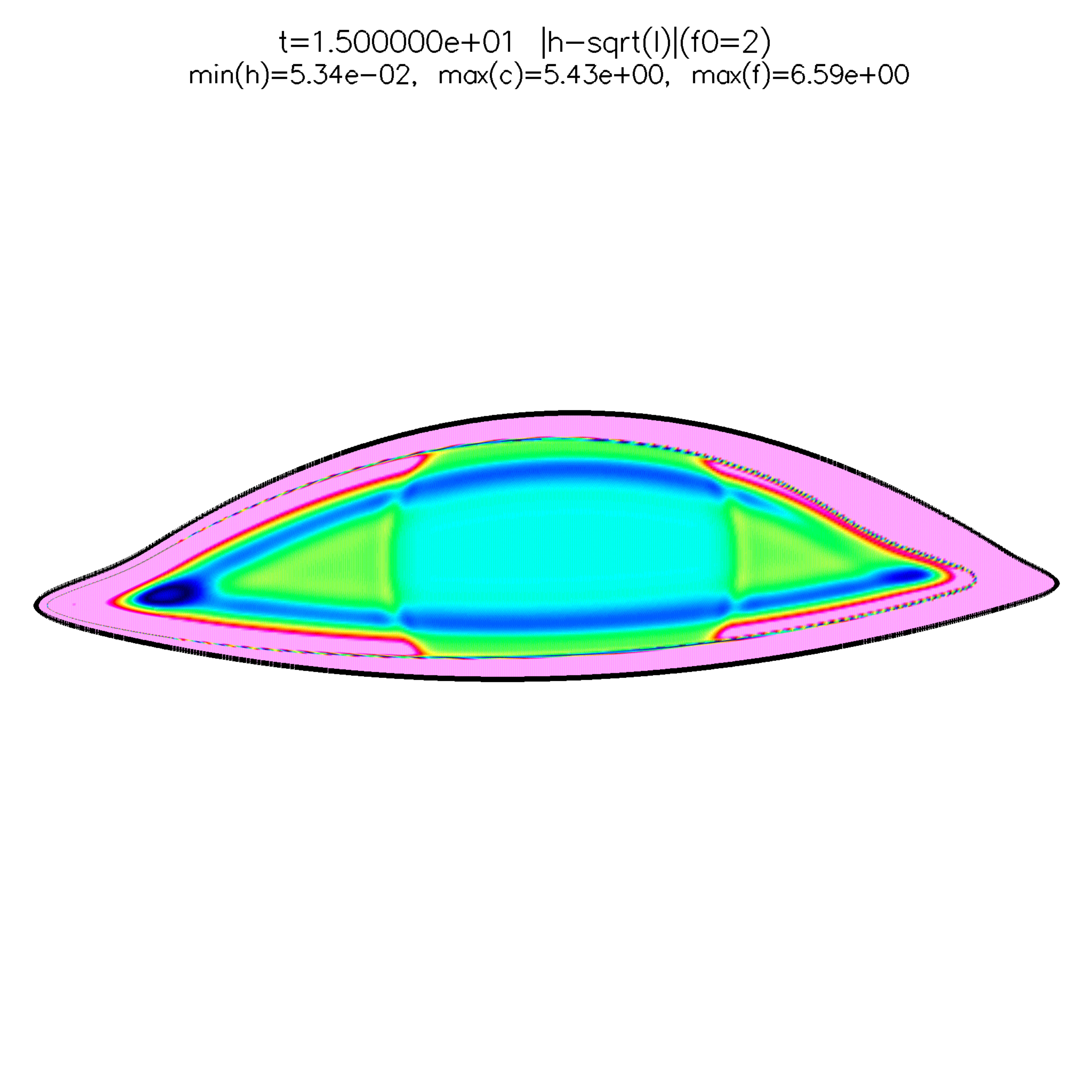}{\figWidth}};
\pgfmathparse{\row+\rowHeight} \let\row\pgfmathresult
\pgfmathparse{\row+1} \let\tP\pgfmathresult
\draw(0,\tP) node[anchor=west,xshift=-0.1cm] {$t=10$};
\draw(0,\row) node[anchor=south west,xshift=\xs cm,yshift=\ys cm] {\trimfig{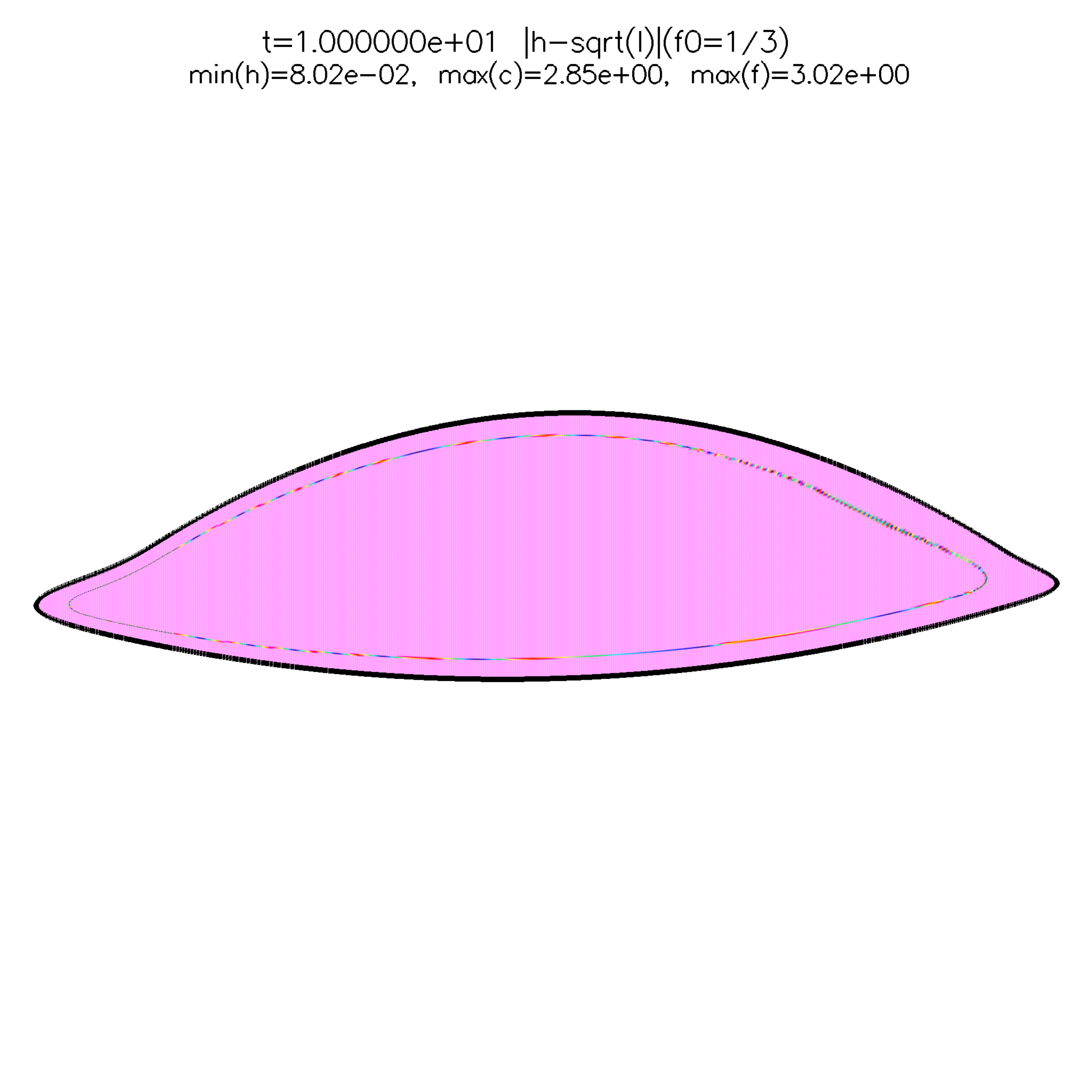}{\figWidth}};
\draw(\figWidth,\row) node[anchor=south west,xshift=\xs cm,yshift=\ys cm] {\trimfig{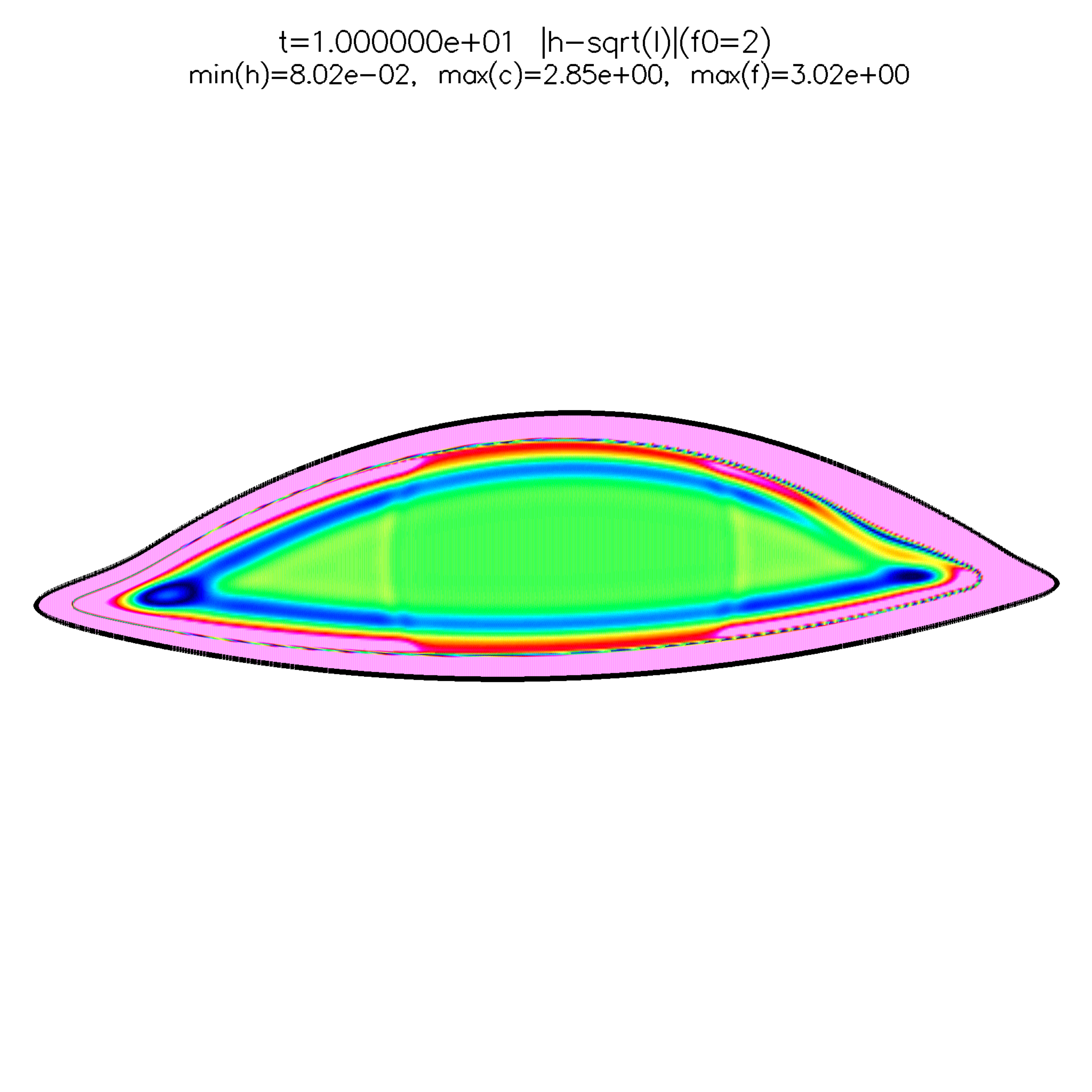}{\figWidth}};
\pgfmathparse{\row+\rowHeight} \let\row\pgfmathresult
\pgfmathparse{\row+1} \let\tP\pgfmathresult
\draw(0,\tP) node[anchor=west,xshift=-0.1cm] {$t=5$};
\draw(0,\row) node[anchor=south west,xshift=\xs cm,yshift=\ys cm] {\trimfig{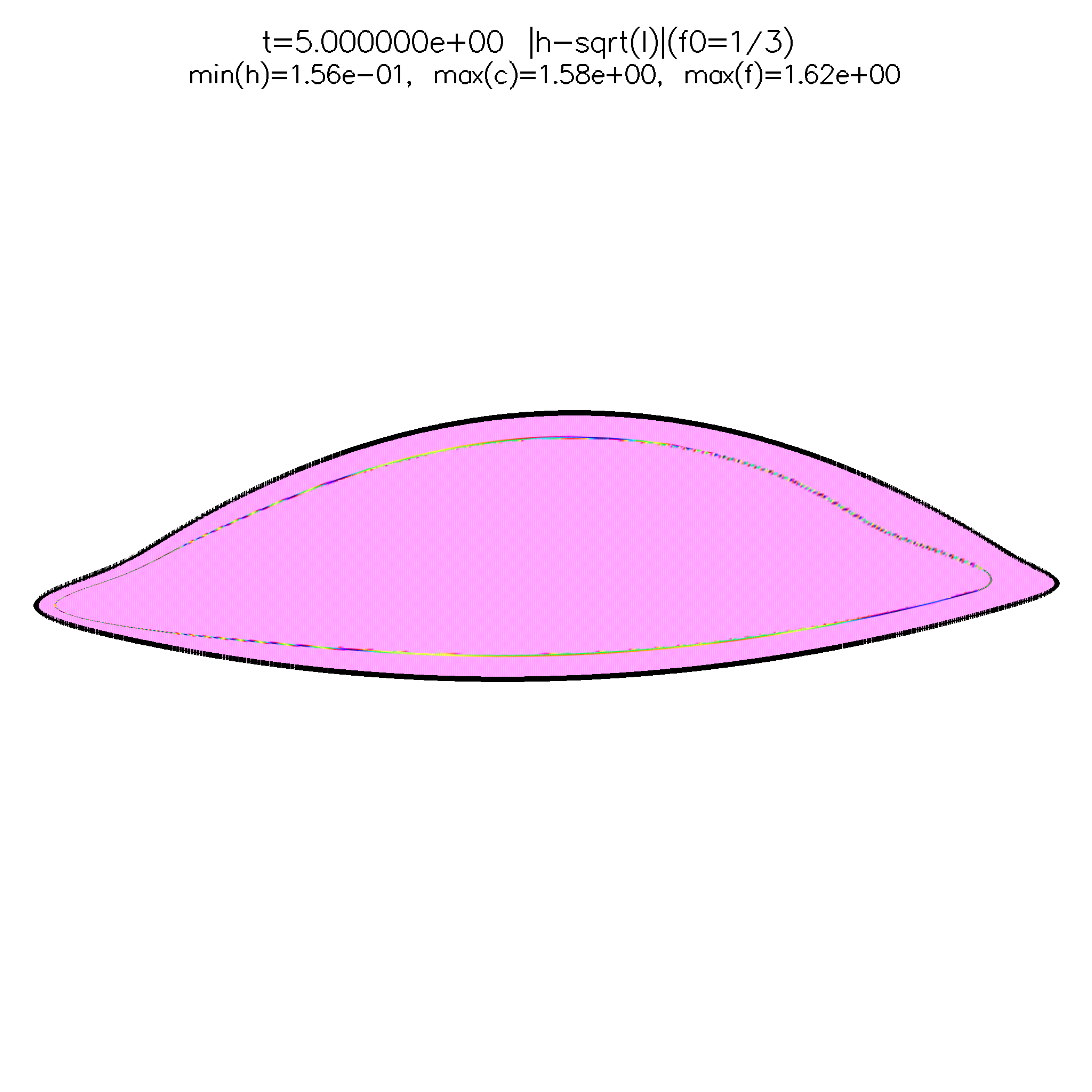}{\figWidth}};
\draw(\figWidth,\row) node[anchor=south west,xshift=\xs cm,yshift=\ys cm] {\trimfig{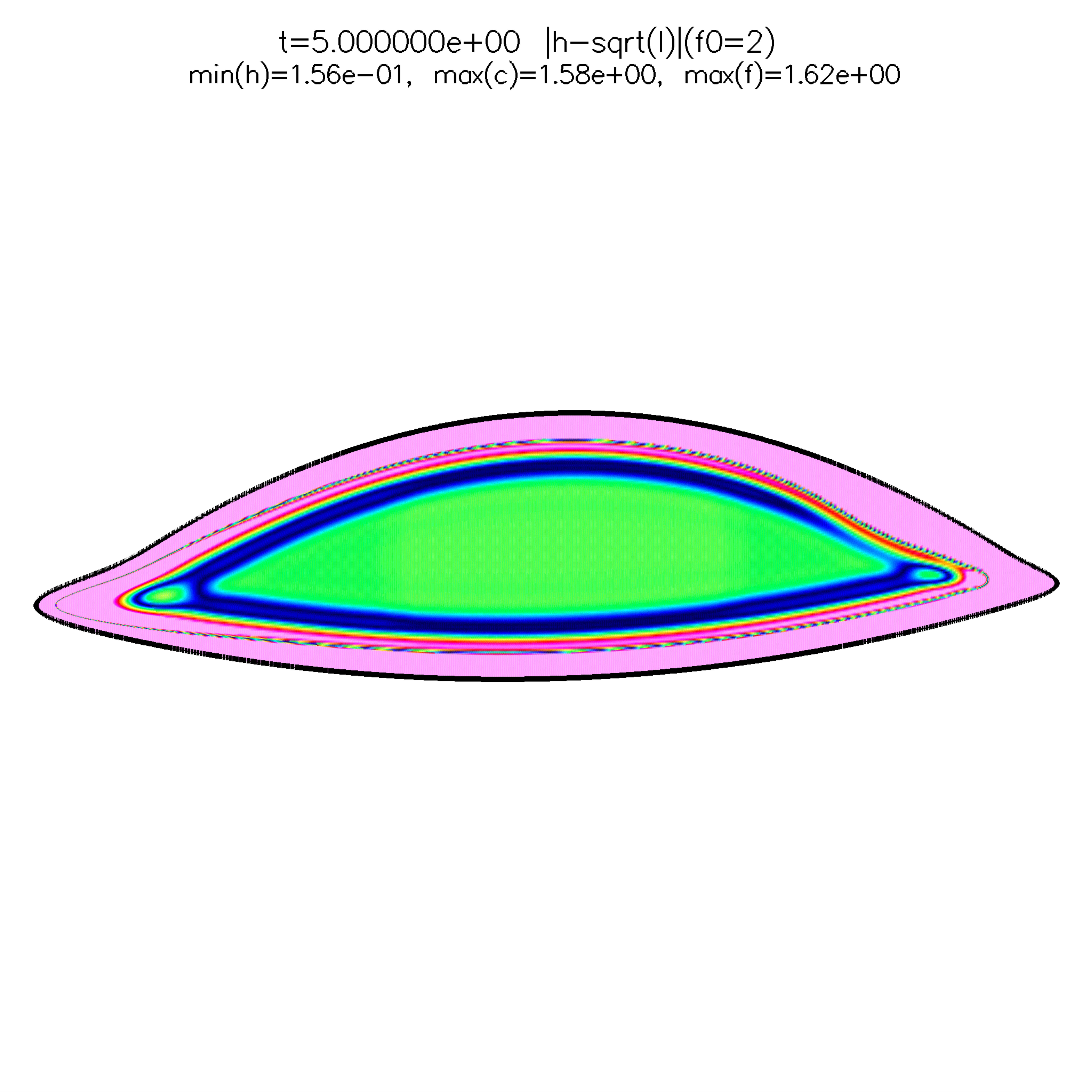}{\figWidth}};
\pgfmathparse{\row+\rowHeight} \let\row\pgfmathresult
%
\end{tikzpicture}
\end{center}
\caption{Variable permeability and evaporation is 20 micron/min, $|h-a\sqrt{I}|$ with $f_0=1/3$ and $f_0=2$}
\label{fig:CompareMethods_Pc(x,y)_20mpm}
\end{figure}
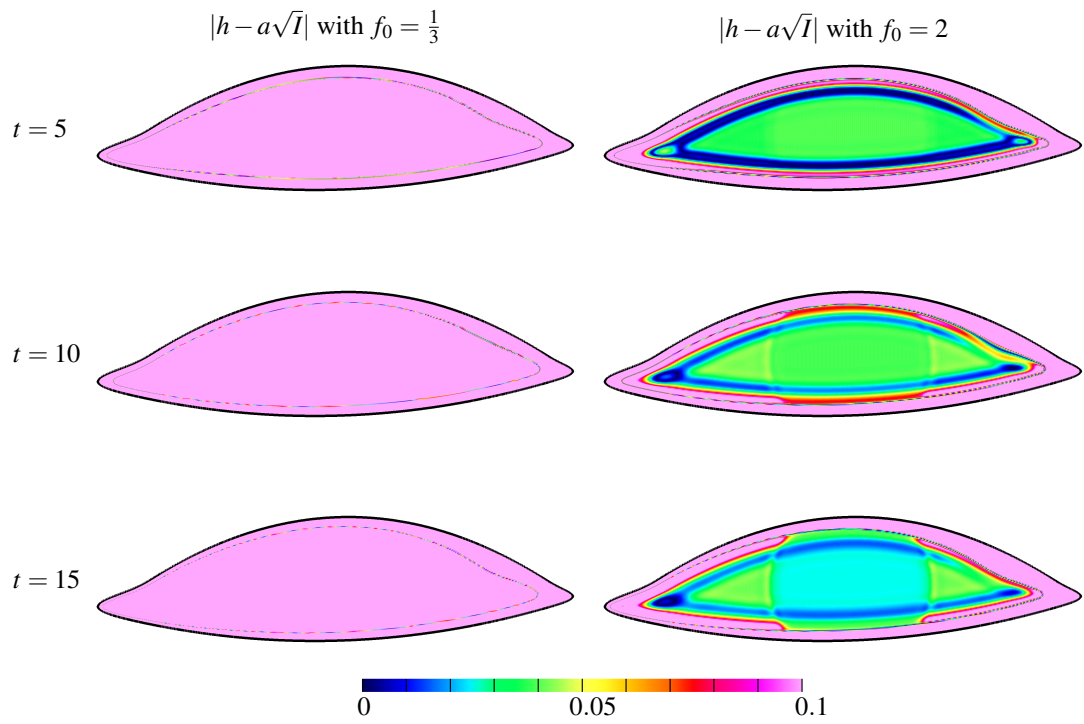
}

\clearpage
\section{Discussion}

Our computations can capture a number of observed phenomena related to fluorescence imaging and tear film dynamics.  The black line was named from its appearance in FL imaging studies, and the model captures its appearance.  Many previous tear film models have captured the phenomenon from a fluid dynamics perspective, but we believe that we are the first to do so with a mathematical model combining fluid flow and fluorescence.
In the menisci, the image remains bright because the tear film thins little and the FL concentration changes little.  This seems to agree well with
a number of experiments \citep{NicholsEtal12,King-SmithIOVS13a,King-SmithIOVS13b}.

In the central region inside the black line, the tear film has relatively little tangential flow, and thins primarily by evaporation.
The appearance of the thinning depends on the concentration of fluorescein and the rate of thinning.  In the dilute regime, the evaporation causes
the concentration to increase, and because the tangential flow is relatively small, mass conservation lead to $hf$ being nearly constant.  Then,
the intensity changes little during thinning in dilute regime because the numerator of (\ref{eqn:I-of-f}) is dominant \citep{BraunEtal14}.   In the self-quenching regime, thinning again leads to increasing $f$, and while $hf$ is nearly constant, the denominator of (\ref{eqn:I-of-f}) is dominant so that the intensity decreases during thinning roughly proportional to $f^{-2}$ \citep{BraunEtal14}.  When the evaporation rate is high, the thinning
may be dramatic, resulting in large changes in $f$, so much so that one may possibly switch from the dilute to the self-quenching regime during an experiment.  Generally in that case, the self-quenching regime can be reliably used to estimate the thinning of the tear film from the intensity
measurement.

From comparing the estimate for the thickness from $a\sqrt{I}$ with the computed thickness $h$, that the estimate worked best in
the self-quenching regime \citep[(where the approximation was derived by][]{NicholsEtal12}.  Our work shows that this works well in the central region
over the cornea for both slower and faster evaporaton.  For spot measurement in the central region and fitting
of the intensity measurements, it is possible for this match between thickness change and intensity to be very good \citep{BraunEtal15}.
For quantitative measurement of thickness change in localized TBU, the situation is less clear if the initial fluorescein concentration
is unknown (and it usually is not).  Depending on the spot size of elevated evaporation, the relative rate of evaporation in the spot and
initial fluorescein concentration, it is even possible for the intensity to increase for a time in the TBU region \citep{BraunEtalTBU16}.
The complication is that the tangential flow near TBU can rearrange the distribution of $f$ because it doesn't diffuse as well as some other
solutes in the tear film.  However, dimming does occur when in the self-quenching regime, and so the approximation to determine the thickness
from the intensity works at least qualitatively for TBU \citep{BraunEtalTBU16}.

\section{Conclusion}\label{sect:Conclusion}

The mathematical model in this paper combines tear film flow, evaporation, transport of solutes (osmolarity and fluorescein), and osmosis on an eye-shaped domain representing the exposed ocular surface.  We thus added fluorescein transport to our previous model \citep{LiBraun15}.  Most of the increase in osmolarity still occurs
over the cornea and in the black lines there.  The results show that the location and value of the minimum tear film thickness and maximum solute concentrations are found to be sensitive to the permeability at the tear/eye  {interface}.

There are some limitations to the current study for comparison with {\it in vivo}
experiment; they are similar to those list
for linking the osmolarity measured {\it in vivo} with the computed results in \citet{LiBraun15}.
We repeat those items here and add to them.
(i) There is no lid motion to mix the tear fluid as occurs {\it in vivo}, which
contributes to the variability observed {\it in vivo}.
(ii)  Once the black line is formed,
there is little exchange between the tear film in the interior with the
meniscus.  The supply and drainage of tear fluid occurs in and affects primarily the meniscus,
and together with a small relative thickness change, the meniscal values of the fluorescein concentration
stay close to the initial value.
(iii)  The volume of tear fluid is probably large compared to aqueous-deficient DES subjects, and could affect the
value of fluorescein concentration obtained, and thus the image observed.
(iv)  The initial value of the fluorescein
concentration is not usually known for {\it in vivo} experiments.  This quantity is clearly important
for quantitative interpretation of fluorescence experiments \citep{Webber86,NicholsEtal12,BraunEtal14}.
(v) Similarly, for most experiments using fluorescein to measure relative thickness changes, the initial thickness
is not known quantitatively.  Knowing this quantity experimentally would greatly aid interpretation.
(vi)  The model does not include
tear break up  (TBU) per se, and there have been results that suggest that the osmolarity could be quite
high in these localized regions
\citep[e.g.,][]{LiuBegEtal09,Braun12,PengEtal14,BraunEtal15,BraunEtalTBU16}.
(vii) There is not staining of the ocular surface in this model, but it certainly can happen
in the clinic \citep{EfronStainRev13,BronEtal15}

These points suggest directions for future research.  Additional directions would include an
dynamic lipid layer that affects the evaporation rate;  possible models are those of
of \citet{BruBrew14} and \citet{PengEtal14}, as well as \citet{Stapf16}.  Further directions would
include blinking, where mixing of the tear fluid could occur in the model; this would allow closer comparison
for both osmolarity measurement and imaging.  Localized TBU in the model would also be
desirable, but being able to resolve small TBU areas in the domain would almost certainly require
adaptive mesh refinement \citep{HenshawSchwed08}.

\section*{Acknowledgments}
This work was supported by National Science Foundation grants
DMS 1022706 and 1412085.  The content is solely the
responsibility of the authors and does not necessarily represent the
official views of the NSF.

\bibliographystyle{agsm}
\bibliography{./RefLibFL}

\appendix
\section*{Appendices}
\section{Flow and Solute Transport Model Derivation}
\label{sect:FlowModelDerivation}
We show detailed derivation of the model system (\ref{hPDE})--(\ref{pPDE}) below.
Inside the tear fluid, we model the tear film fluid with the incompressible Navier-Stokes equations and convection-diffusion equations for the energy and solutes.  For the fluid in
$0<z'<h'(x',y',t')$, we have:
\begin{align*}
& \nabla\cdot \u'=0, &\rho\left(\partial_{t'}\u'+\u'\cdot\nabla\u'\right)=-\nabla p'+\mu\Delta\u'-\rho g \mathbf{j}, \\
& \rho c_p(\partial_{t'} T'+\u'\cdot\nabla T')=k\Delta T', 
& \partial_t c' + \nabla\cdot\left(c'\u' \right) = D_c \Delta c', \\
& \partial_t f' + \nabla\cdot\left( f'\u' \right) = D_f \Delta f'. &
\end{align*}
Here $g$ is the acceleration due to gravity; henceforth we neglect it in this paper. $c'$ is the volumetric concentration of osmotically active physiological salts in the aqueous layer excluding those from fluorescein.  It is measured in units of Osmoles per m$^3$, corresponding to mOsM.  $f'$ is the concentration in moles/liter, or M.  $D_c$ is the diffusion coefficient of osmolarity; $D_f$ is the diffusivity of fluorescein. ($\mathbf{i},\mathbf{j}, \mathbf{k})$ are the  standard basis vectors in the $(x',y',z')$-directions, respectively.

At the free surface, $z'=h'$, we have the equations to balance fluid mass and energy:
 \begin{align*}
&J'_e=\rho(\u'-\u'_I)\cdot \n', & L_mJ'_e + k\n'\cdot\nabla T'=0.
\end{align*}
Here
$
\u'_I  = \partial_t' h' \mathbf{k}
$ is the interfacial velocity and $\n'$ is the normal vector to the tear film surface \citep{EdBrenWasan91}. The jump of the velocity at the free surface is due to evaporation. We also assume tangential immobility and we balance normal stress with the conjoining pressure under consideration:
 \begin{align*}
 &\u'\cdot\t'_1=\u'\cdot\t'_2=0,&-p'_v-\n'\cdot \T'\cdot\n'=\sigma\nabla\cdot\n'-\Pi'.
\end{align*}
Here $\t'_1$ and $\t'_2$ are a pair of orthogonal tangential vectors of the tear film surface,  $\T' = -p'\mathbf{I}+\mu(\nabla \u'+\nabla \u'^{T})$  is the Newtonian stress tensor, and $\Pi'=A^*/h'^3$ is the conjoining pressure \citep{EdBrenWasan91}.

Finally, we relate the interfacial temperature to the mass flux and pressure jump by the nonequilibrium condition, and we impose a no-flux condition for the osmolarity,
 \begin{align*}
& K J'_e = \alpha(p'-p'_v)+T'-T'_s, & (\u'-\u'_I)c'\cdot \n'  =  D_c\nabla c' \cdot \n'.
\end{align*}
Since we model the evaporative mass flux $J'$ as
$$
J'_e= \rho(\u'-\u'_I) \cdot \n',
$$
the no-flux condition for osmolarity  at the free surface becomes
\begin{equation*}\label{SaltNoFluxAtFreeSurface}
D_c\nabla c' \cdot \n' = \frac{c'J'_e}{\rho}.
\end{equation*}
For flourescein, we have
\begin{equation*}\label{FLNoFluxAtFreeSurface}
D_f \nabla f' \cdot \n' = \frac{f'J'_e}{\rho}.
\end{equation*}

At the cornea-tear film interface, $z' = 0'$, in addition to the specification of  no-slip conditions and the prescription of body temperature, we allow water to go through the ocular surface by osmosis, but keep the physiological salt from penetrating the ocular surface, thus we have
\begin{align*}
u'=v'=0, ~~T'=T'_{B}, ~~ w' = W'_o/\rho,~~w'c' = D_c \partial_z' c',~~w'f' = D_f \partial_z' f'.
\end{align*}
Here
\begin{align*}
W'_o = P'_c(c'-c'_0-3 f')
\end{align*}
is the osmotic volume flux per unit area and
$c'_0=302 $ Osm/m$^3$ is the isotonic physiological salt concentration, which is used to scale $c'$.
Fluorescein, when discussing the tear film, is typically referring to the use of the salt Na$_2$FL, which
ionizes to 2Na$^+$ and FL$^{2-}$ at physiological pH values of 7.2-7.6 for
tears \cite{DoughtyOPO10,EfronStainRev13,BronEtal15}.  Thus, the ion concentration is three times the Na$_2$FL
concentration, and that is the source of the factor of three in the last term of $J'_o$.  We have assumed that the
permeability is the same for these ions as naturally occuring components of osmolarity, but they possess different diffusivities
(Table~\ref{tab:DimensionalParameters}).

The following scales are used to non-dimensionalize the equations:
\begin{align*}
& x'=L'x,~y'=L'y,~z'=d'z,~h'=d'h,~ c' = c'_0\bar{c},~ f' = f'_{cr}\bar{f},~u'=U_0u,~v'=U_0v,~t'=\frac{L'}{U_0}t, \\ & w'=\frac{d'U_0}{L'}w,~
p'-p'_v=\frac{\mu U_0}{L'\epsilon^2}p,~T=\frac{T'-T'_s}{T'_B-T'_s},~
J'_e=\frac{k}{d' \mathcal{L}_m}(T'_B-T'_s)J.
\end{align*}
Here $\epsilon = {d'}/{L'}\ll1$ indicates the separation of length scales. After non-dimensionalization and neglecting gravitational acceleration,
we have,
in $0<z<h(x,y,t)$,
\begin{align*}
& \epsilon^2\mathrm{Re}\left(\partial_tu+u\partial_xu+v\partial_yu +w\partial_zu\right) = -\partial_x p +\left(\epsilon^2\partial^2_x u +\epsilon^2 \partial^2_y u +\partial^2_z u\right),\\
& \epsilon^2\mathrm{Re}\left(\partial_tv+u\partial_xv+v\partial_yv +w\partial_zv\right) = -\partial_y p +\left(\epsilon^2\partial^2_x v +\epsilon^2 \partial^2_y v +\partial^2_z v\right),\\
& \epsilon^4\mathrm{Re}\left(\partial_tw+u\partial_xw+v\partial_yw +w\partial_zw\right) =-\partial_z p+\epsilon^2\left(\epsilon^2\partial^2_x w +\epsilon^2 \partial^2_y w +\partial^2_z w\right),\\
& \epsilon^2\mathrm{Re}\mathrm{Pr}\left(\partial_tT+u\partial_xT+v\partial_yT +w\partial_zT\right) = \epsilon^2\left(\partial^2_xT+\partial^2_yT \right) +\partial^2_zT,\\
& \partial_xu +\partial_y v+\partial_z w = 0,\\
&\epsilon^2 \mathrm{Pe}_c \left[\partial_t\bar{c} +\left(u\partial_x\bar{c}+v\partial_y\bar{c}+w\partial_z\bar{c} \right) \right] = \epsilon^2\partial^2_x\bar{c} +\epsilon^2\partial^2_y\bar{c} +\partial^2_z\bar{c},\\
&\epsilon^2 \mathrm{Pe}_f \left[\partial_t\bar{f} +\left(u\partial_x\bar{f}+v\partial_y\bar{f}+w\partial_z\bar{f} \right) \right] = \epsilon^2\partial^2_x\bar{f} +\epsilon^2\partial^2_y\bar{f} +\partial^2_z\bar{f}.
\end{align*}

At $z=h(x,y,t)$,
\begin{align*}
&EJ_e = \frac{-u\partial_xh -v\partial_yh+w-\partial_th}{\sqrt{1+\epsilon^2\left(\partial_xh\right)^2+\epsilon^2\left(\partial_yh\right)^2}},\\
& J_e +\frac{-\epsilon^2\partial_xh\partial_xT-\epsilon^2\partial_yh\partial_yT+\partial_zT}{\sqrt{1+\epsilon^2\left(\partial_xh\right)^2+\epsilon^2\left(\partial_yh\right)^2}}= 0,\\
& \frac{v+\epsilon^2w\partial_yh}{\sqrt{1+\epsilon^2\left(\partial_yh\right)^2}} = \frac{u+\epsilon^2w\partial_xh}{\sqrt{1+\epsilon^2\left(\partial_xh\right)^2}} =0,
\end{align*}
\begin{align*}
&p=-\frac{2\epsilon^2\big[\epsilon^2\left(\partial_x^2h\partial_xu+\partial_y^2h\partial_yv +\partial_xh\partial_yh(\partial_yu +\partial_x v)-\partial_xh\partial_xw-\partial_yh\partial_yw\right)+\partial_zw-\partial_xh\partial_zu - \partial_yh\partial_zv\big]}{\sqrt{1+\epsilon^2\left(\partial_xh\right)^2+\epsilon^2\left(\partial_yh\right)^2}}\\
&=-S\left[\partial_x\left(\frac{\partial_xh}{\sqrt{1+\epsilon^2\left(\partial_xh\right)^2+\epsilon^2\left(\partial_yh\right)^2}}\right)+\partial_y\left(\frac{\partial_yh}{\sqrt{1+\epsilon^2\left(\partial_xh\right)^2+\epsilon^2\left(\partial_yh\right)^2}}\right)\right]-\frac{A}{h^3},
\end{align*}
\begin{equation*}
\bar{K} J_e  = \delta p+T,
\end{equation*}
\begin{equation*}
-\epsilon^2\partial_xh\partial_x\bar{c}-\epsilon^2\partial_yh\partial_y\bar{c}+\partial_z\bar{c} = E\mathrm{Pe}_c \epsilon^2\bar{c}J_e\sqrt{1+\epsilon^2\left(\partial_xh\right)^2+\epsilon^2\left(\partial_yh\right)^2}.
\end{equation*}

At $z= 0$,
\begin{align*}
& u=v=0, ~~T=1,\\
& w = W_o,\\
& W_o = P_c(c-1-Bf) \\
& \epsilon^2 \mathrm{Pe}_c w\bar{c}  =  \partial_z \bar{c}, \\
& \epsilon^2 \mathrm{Pe}_f w\bar{f}  =  \partial_z \bar{f}.
\end{align*}
$\displaystyle \mathrm{Pe}_c = \frac{U_0L'}{D_c}$ is the P\`eclet number for salt, $ \displaystyle P_c = \frac{P^{\text{tiss}}v_wc'_0}{\epsilon U_0}$ is the nondimensional permeability of the ocular surface and
$B=3f'_{cr}/c'_0$ is the ratio of osmotic components. The tissue permeability $P^{\text{tiss}}$ will take on different values as described in Section~\ref{s:results}.

We now examine the relative sizes of the osmotic suppy terms in $W_o$.
The critical concentration of fluorescein is 0.2\% in water;
this is a mass fraction of 0.002 g FL/g water.  We assume that the aqueous component of the tear
film has the same properties as water.
There are roughly 1000 g/l of water, so that $f'_{cr}= 2$ g /liter
of water.  The molar mass of Na$_2$FL is
376 g/mole, so the concentration of the undissociated molecule
$f'_{cr}= 0.0053$ mole/liter or 5.3 millimolar (mM).  The concentration of ions is then three times the concentration of FL$^{2-}$, hence the factor of three in $B$.  Substitution yields $B=0.0525$.  Because $B$ is so small and $f$ is at most only a few times the size of $c$, we have neglected the contribution of $f$ to $W_o$.  Thus, we will use
\begin{align*}
W_o = P_c(c-1)
\end{align*}
in this paper.

Proceeding to other non-dimensional parameters, we find:
$$
\epsilon = \frac{d'}{L'} = 1\times10^{-3}, ~~ \Real = \frac{U_0L'}{\mu/\rho}\approx   19.23,~~\Pran=\frac{c_p\mu}{k} \approx 8.01,
$$
where $\Real$ is the Reynolds number and $\Pran$ is the  Prandtl number. Terms involving the following parameters are regarded as small:
$$
\epsilon^2  = 1\times10^{-6},~~ \epsilon^2\Real \approx  1.92\times 10^{-5}, ~~  \epsilon^2\Real\Pran\approx 1.54\times10^{-4}.
$$
Applying  lubrication theory by  neglecting all the small terms for the fluid equations, we then have the following leading order approximations
for the water, momentum and energy conservation.
\begin{itemize}
\item[]
In $0<z<h(x,y,t)$:
\begin{align}
& \partial_xu +\partial_y v+\partial_z w = 0,\\
& 0 = -\partial_x p +\partial^2_z u,\\
& 0 = -\partial_y p +\partial^2_z v-G,\\
& 0 =-\partial_z p,\\
& 0 = \partial^2_zT
\end{align}
For the osmolarity, we expand $\bar{c}(x,y,z,t)$ as
$$
\bar{c} = \bar{c}_0 +\epsilon^2 \bar{c}_1 +O(\epsilon^4),
$$
and for fluorescein $\bar{f}(x,y,z,t)$ as
$$
\bar{f} = \bar{f}_0 +\epsilon^2 \bar{f}_1 +O(\epsilon^4),
$$

The leading order equation for the osmolarity is
\begin{equation}
\partial_z \bar{c}_0 = 0,
\end{equation}
which implies $\bar{c}_0$ is independent of $z$, i.e. $\bar{c}_0 = \bar{c}_0(x,y,t)$.  We proceed to the next order so as to find an equation for $\bar{c}_0$,  and we obtain
\begin{equation}
\partial^2_z\bar{c}_1 = \mathrm{Pe}_c \left[\partial_t\bar{c}_0 +\left(u\partial_x\bar{c}_0+v\partial_y\bar{c}_0 \right) \right] -\partial^2_x\bar{c}_0 -\partial^2_y\bar{c}_0. \label{PDEc1}
\end{equation}
The equation resulting from $\bar{f}$ is similar.

At $z=h(x,y,t)$:
\begin{align}
&EJ_e = -u\partial_xh -v\partial_yh+w-\partial_th,\\
& J_e +\partial_zT= 0,\\
& u=v=0,\\
& p=-S\left(\partial_x^2h+\partial_y^2h\right)- \frac{A}{h^3}, \label{eqn:pressure} \\
&\bar{K} J  = \delta p+T,\\
&\partial_z\bar{c}_1 =  E\mathrm{Pe}_c \bar{c}_0 J_e+\nabla h\cdot\nabla\bar{c}_0. \label{cBCath}
\end{align}
We  use $\nabla = (\partial_x, \partial_y)$  and $\Delta = (\partial^2_x+\partial^2_y)$ to represent the differential operators applied on them for convenience since $h$ and $\bar{c}_0$ are all independent on $z$.  We note that since $p$ is independent of $z$, the leading order value throughout the film is specified by (\ref{eqn:pressure}).

At $z=0$, using the simplified expression for $W_o$ yields:
\begin{align}
& u=v=0, ~~T=1,\\
& w = W_o = P_c (\bar{c}_0-1),\\
&   \partial_z \bar{c}_1=  \mathrm{Pe}_c P_c(\bar{c}_0 -1)\bar{c}_0.\label{cBCat0}
\end{align}
\end{itemize}

For the tear film, we solve for the velocity and temperature fields,
integrate the mass conservation equation and use the kinematic condition to derive  a  PDE for $h(x, y, t)$:
$$
\partial_t h +EJ_e +\nabla \cdot \mathbf{Q} - P_c(\bar{c}_0-1) = 0,
$$
with
$$
J_e =\frac{1+\delta p}{\bar{K}+h} ~~\mathrm{and}
~~\mathbf{Q} =\left(\int_0^hu\,dz,\int_0^hv\,dz\right)= \frac{h^3}{12}\nabla\left(S\Delta h +Ah^{-3}-Gy\right).
$$

For the osmolarity, we integrate Equation (\ref{PDEc1}) with respect to $z$ from $0$ to $h$. Noting that $\bar{c}_0 = \bar{c}_0(x,y,t)$, we then have
$$
\partial_z\bar{c}_1(x,y,h,t) -  \partial_z\bar{c}_1(x,y,0,t) = \mathrm{Pe}_c \left[h\partial_t\bar{c}_0 +\nabla \bar{c}_0\cdot \mathbf{Q} \right] -h\Delta \bar{c}_0.
$$
According to the boundary conditions (\ref{cBCath}) and (\ref{cBCat0}),  we derive a PDE for $\bar{c}_0(x,y,t)$:
$$
\mathrm{Pe}_c \left[h\partial_t\bar{c}_0 +\nabla \bar{c}_0\cdot \mathbf{Q} \right] -h\Delta  \bar{c}_0
=E\mathrm{Pe}_c \bar{c}_0J+\nabla h\cdot\nabla \bar{c}_0 -  \mathrm{Pe}_cP_c (\bar{c}_0-1)\bar{c}_0.
$$
The fluorescein concentration follows similar steps, and results in the following:
$$
\mathrm{Pe}_f \left[h\partial_t \bar{f}_0 +\nabla \bar{f}_0\cdot \mathbf{Q} \right] -h\Delta  \bar{f}_0
=E \mathrm{Pe}_f \bar{f}_0 J_e+\nabla h \cdot \nabla \bar{f}_0 -  \mathrm{Pe}_f P_c (\bar{c}_0-1)\bar{f}_0.
$$

For  convenience, we use $c$ in the equations instead of $\bar{c}_0$ and $f$ instead of $\bar{f}_0$.  Therefore, we have derived the governing equations (\ref{hPDE}), (\ref{cPDE}) and (\ref{fPDE}):

\begin{align}
&\partial_t h +E\frac{1+\delta p}{\bar{K}+h} +\nabla \cdot \left( -\frac{h^3}{12}\nabla p \right) - P_c(c-1) = 0,  \label{DAE1}\\
 & p+S\Delta h+Ah^{-3} = 0,  \label{DAE2}\\
 &h\partial_tc +\nabla c\cdot \left( -\frac{h^3}{12}\nabla p \right)   = E c \frac{1+\delta p}{\bar{K}+h} +\frac{1}{\mathrm{Pe}_c}\nabla\cdot (h\nabla c)  -  P_c (c-1)c,  \label{DAE3}\\
  &h\partial_tf +\nabla f\cdot \left( -\frac{h^3}{12}\nabla p \right)   = E f \frac{1+\delta p}{\bar{K}+h} +\frac{1}{\mathrm{Pe}_c}\nabla\cdot (h\nabla f)  - P_c (c-1)f.  \label{DAE4}
\end{align}
here $f_{cr}$ is the scale of $f$ in units of M.

\section{Time-dependent Fluid Flux Boundary Condition}\label{sect:Time-dependentBC}
We define
$$
Q_{lg}(s,t) = f_{lg}(t)\hat{Q}_{ls}(s), ~~Q_{p}(s,t) = f_{p}(t)\hat{Q}_{p}(s)
$$ in the time-dependent fluid flux BC (\ref{FluxBC}). The formulations of  $f_{lg}(t),f_{p}(t),\hat{Q}_{lg}(s)$, and $\hat{Q}_{p}(s)$ are listed below:
\begin{equation}\label{eqn:lacrimaltimelg}
f_{lg}(t) = \begin{cases}
\frac{1}{2}\left[ \cos{\left( \frac{\pi}{2}\frac{t - t_{lg,on}}{\Delta t_{lg}}  - \frac{\pi}{2} \right)} +1 \right], & \text{if $|t-t_{lg,on}|\leq \Delta t_{lg} $}; \\
1, & \text{if $t_{lg,on}+\Delta t_{lg} \leq t \leq t_{lg,off}-\Delta t_{lg}$}; \\
\frac{1}{2} \left[ \cos{\left( \frac{\pi}{2}\frac{t - t_{lg,off}}{\Delta t_{lg}}  + \frac{\pi}{2} \right)} +1 \right], & \text{if $|t-t_{lg,off}|\leq \Delta t_{lg}$}; \\
0, & \text{otherwise.}
\end{cases}
\end{equation}
\begin{equation}\label{eqn:lacrimaltimep}
f_{p}(t) = \begin{cases}
\frac{1}{2}\left[ \cos{\left( \frac{\pi}{2}\frac{t - t_{p,on}}{\Delta t_{p}}  - \frac{\pi}{2} \right)} +1 \right], & \text{if $|t-t_{p,on}|\leq \Delta t_{p} $}; \\
1, & \text{if $t_{p,on}+\Delta t_{p} \leq t \leq t_{p,off}-\Delta t_{p}$}; \\
\frac{1}{2} \left[ \cos{\left( \frac{\pi}{2}\frac{t - t_{p,off}}{\Delta t_{p}}  + \frac{\pi}{2} \right)} +1 \right], & \text{if $|t-t_{p,off}|\leq \Delta t_{p}$}; \\
0, & \text{otherwise.}
\end{cases}
\end{equation}
\begin{equation}\label{eqn:lacrimal}
     \hat{Q}_{lg}(s) = \begin{cases}
0, & \text{if $s<s_{lg,on} - \Delta s_{lg}$};\\
-\frac{1}{2}\hat{Q}_{0lg}\left[ \cos{\left( \frac{\pi}{2}\frac{s - s_{lg,on}}{\Delta s_{lg}}  - \frac{\pi}{2} \right)} +1 \right], & \text{if $|s-s_{lg,on}|\leq \Delta s_{lg} $}; \\
-\hat{Q}_{0lg}, & \text{if $s_{lg,on}+\Delta s_{lg} \leq s \leq s_{lg,off}-\Delta s_{lg}$}; \\
-\frac{1}{2}\hat{Q}_{0lg}\left[ \cos{\left( \frac{\pi}{2}\frac{s - s_{lg,off}}{\Delta s_{lg}}  + \frac{\pi}{2} \right)} +1 \right], & \text{if $|s-s_{lg,off}|\leq \Delta s_{lg}$}; \\
0, & \text{otherwise.}
\end{cases}
\end{equation}
\begin{equation}\label{eqn:puncta}
     \hat{Q}_{p}(s) = \begin{cases}
0, & \text{if $s<s_{p,lo} - \Delta s_{p}$};\\
-\frac{\hat{Q}_{0p}}{2}(1-p_{out})\left[ \cos{\left( \pi \frac{s - s_{p,lo}}{\Delta s_{p}}  - \pi \right)} - 1 \right], & \text{if $|s-s_{p,lo}|\leq \Delta s_{p}$};\\
0, & \text{if $s_{p,lo}+\Delta s_{p} \leq s \leq s_{p,up} - \Delta s_{p}$};\\
-\frac{\hat{Q}_{0p}}{2}(p_{out})\left[ \cos{\left( \pi \frac{s - s_{p,up}}{\Delta s_{p}}  - \pi \right)} - 1 \right], & \text{if $|s-s_{p,up}|\leq \Delta s_{p}$};\\
0, & \text{otherwise}.
\end{cases}
\end{equation}
Parameters  in the formulations are tabulated in Table \ref{tab:fluxparameterstimedep}.

\begin{table}[h]
\begin{center}
\parbox{5.5in}{
\caption{Parameters appearing in the flux boundary condition.} \label{tab:fluxparameterstimedep} }
\begin{tabular}{ccc}
\hline\hline
\text{Parameter} & \text{Description} & \text{Value} \\
\hline
$t_{lg,on}$ & On  time for lacrimal gland supply& $0.2$ \\
$t_{lg,off}$ & Off time for lacrimal gland supply & $5.2$ \\
$\Delta t_{lg}$ & Transition time of lacrimal gland supply & $0.2$ \\
$t_{p,on}$ & On time   for punctal drainage & $1.05$\\
$t_{p,off}$ & Off time    for punctal drainage  & $5.05$\\
$\Delta t_{p}$ &  Transition time of  punctal drainage& $0.05$\\
$Q_{mT}$ & Estimated steady supply from lacrimal gland&0.08\\
$\hat{Q}_{0lg}$ & Height of lacrimal gland peak & $0.4$ \\
$\hat{Q}_{0p}$ &  Height of punctal drainage peak & $4$ \\
$\Delta t_{bc}$ & Flux cycle time & $10$ \\
$s_{lg,on}$ & On-ramp location for lacrimal gland peak & $4.2$\\
$s_{lg,off}$ & Off-ramp location for lacrimal gland peak & $4.6$ \\
$\Delta s_{lg}$ & On-ramp and off-ramp width  of  lacrimal peak & $0.2$ \\
$p_{out}$ &Fraction of drainage from upper punctum & $0.5$\\
$s_{p,lo}$ &Lower punctal drainage peak location & $11.16$\\
$s_{p,up}$ &Upper punctal drainage peak location & $11.76$\\
$\Delta s_p$ &Punctal drainage peak width& $0.05$\\
\hline
\end{tabular}
\end{center}
\end{table}

\section{Fluorescence Model}
\label{sect:IntensityModelDerivation}

Fluorescein, when discussing the tear film, is typically referring to the use of the salt Na$_2$FL, which
ionizes to 2Na$^+$ and FL$^{2-}$ at physiological pH values of 7.2-7.6 for
tears \citep{DoughtyOPO10,EfronStainRev13,BronEtal15}.
It glows green when exposed to blue
light \citep{Webber86}, and has been used in a variety of ways for the tear \citep{EfronStainRev13,BronEtal15}, including as a first indication of TBU \citep{Norn69}, as well as a variety of biological applications \citep{Lakowicz}.

The fluorescent intensity observed in the tear film is a product of the efficiency of fluorescence  $E_f(f')$ and the
absorptance $A_f(h',f')$ \citep{NicholsEtal12}.  Here $h'$ is the tear film thickness and $f'$ is the fluorescein concentration. For the efficiency,
\begin{equation}
E(f') = \frac{C_1}{1+(f'/f_{cr})^2},
\end{equation}
where $C_1$ is a constant.  For the absorptance,
\begin{equation}
A(h',f') = C_2 \left( 1-e^{-\kappa h' f'} \right);
\end{equation}
$C_2$ is a constant and $\kappa$ is
the Naperian extinction coefficient.
The FL intensity $I$ is then given by their product \citep{Webber86,NicholsEtal12}:
\begin{equation} \label{e:dimensionalI}
I = I_0 \frac{1-e^{-\kappa h' f'}}{1+(f'/f_{cr})^2}.
\end{equation}
$I_0$ is a normalization factor that takes into
account a number of factors including the optical system.  For fixed $h'$, expanding for
small $f'$ yields a leading term proportional to $f'$ and $h'$; this is the dilute regime.
Expanding for large $f'$ shows that the intensity decreases proportional to $1/(f')^{2}$;
this is the self-quenching regime \citep{NicholsEtal12}.  When the spatially-uniform tear
film thins by evaporation, mass conservation requires that $h'f'=f'_0h'_0$ where the
subscript zeros indicate initial values.  Thus, for the
spatially uniform case in the self-quenching regime, the thickness is proportional to the
square root of the intensity \citep{NicholsEtal12,BraunEtal14}.  This approximation was
used to estimate the tear film thickness from the FL images
in previous work \citep{King-SmithIOVS13a,BraunEtal15}.

Using $h'=hd$ and $f'=f_{cr}f$, we obtain
the nondimensional version of (\ref{e:dimensionalI}) \citep{Webber86,NicholsEtal12},
one obtains (\ref{eqn:I-of-f}). The constant
$\phi=\epsilon_f d f_{cr}$ is the nondimensional Naperian extinction coefficient; a representative value is given
in Table~\ref{tab:DimensionlessParameters}.

\end{document}